\documentclass[journal,twoside,web]{ieeecolor}
\usepackage{tmi}
\usepackage{cite}
\usepackage{amsmath,amssymb,amsfonts}
\usepackage{graphicx}
\usepackage{textcomp}
\usepackage{multirow}
\usepackage{float}
\usepackage{booktabs}
\usepackage{tabularx}
\usepackage{amsmath}
\usepackage{adjustbox}
\usepackage{url}
\usepackage{comment}
\usepackage{stfloats}      % or: \usepackage{dblfloatfix}
\usepackage{placeins}      % gives \FloatBarrier
\usepackage[T1]{fontenc}
\usepackage{algorithm}
\usepackage{algpseudocode}
\usepackage{tcolorbox} % Add in your preamble
\tcbuselibrary{listingsutf8}

\usepackage{array}
\usepackage{booktabs}

\usepackage{xcolor}

\PassOptionsToPackage{table}{xcolor}

\renewcommand{\arraystretch}{1.2}
\usepackage{subcaption}

\usepackage[hidelinks]{hyperref}

\newcommand{\tool}[1]{\mbox{\texttt{#1}}}
\newcommand{\etal}{\textit{et al.}}

\def\BibTeX{{\rm B\kern-.05em{\sc i\kern-.025em b}\kern-.08em
    T\kern-.1667em\lower.7ex\hbox{E}\kern-.125emX}}
\begin{document}
\title{MRI Super-Resolution with Deep Learning: A Comprehensive Survey}

\author{
\fontsize{10.6 pt}{12.5 pt}\selectfont
Mohammad Khateri, Serge Vasylechko, Morteza Ghahremani, Liam Timms, Deniz Kocanaogullari,
\\
Simon K. Warfield, Camilo Jaimes, Davood Karimi, Alejandra Sierra, Jussi Tohka, Sila Kurugol, Onur Afacan%
\thanks{This work was supported in part by the Research Council of Finland under Grants 323385 and 358944; by the Flagship of Advanced Mathematics for Sensing Imaging and Modelling; by the saastamoinen foundation; by the KAUTE Foundation; and by Finnish Cultural Foundation.}
\thanks{Mohammad Khateri, Alejandra Sierra, and Jussi Tohka are with the A. I. Virtanen Institute for Molecular Sciences, Faculty of Health Sciences, University of Eastern Finland, 70210 Kuopio, Finland (e-mail: \{mohammad.khateri, alejandra.sierralopez, jussi.tohka\}@uef.fi).}
\thanks{Serge Vasylechko, Liam Timms, Deniz Kocanaogullari, Simon Warfield, Davood Karimi, Sila Kurugol, and Onur Afacan are with Harvard Medical School and Boston Children’s Hospital, Boston, MA 02115, USA (e-mail: \{serge.vasylechko, liam.timms, deniz.kocanaogullari, simon.warfield, davood.karimi, sila.kurugol, onur.afacan\}@childrens.harvard.edu).
}
\thanks{Morteza Ghahremani is with the Artificial Intelligence in Medical Imaging group at the Department of Radiology, Technical University of Munich, 80333 München, Germany (e-mail:morteza.ghahremani@tum.de).}
\thanks{Camilo Jaimes is with the Department of Radiology, Massachusetts
General Hospital, Boston, MA 02114 USA (e-mail: cjaimescobos@mgb.org).
}
}

\markboth{\journalname, VOL. XX, NO. XX, XXXX 2025}
{Khateri \MakeLowercase{\textit{et al.}}: MRI Super-Resolution with Deep Learning: A Comprehensive Survey}

\maketitle
\hypersetup{pdfborder={0 0 0}}

\begin{abstract}
High-resolution (HR) magnetic resonance imaging (MRI) is crucial for many clinical and research applications. However, achieving it remains costly and constrained by technical trade-offs and experimental limitations. Super-resolution (SR) presents a promising computational approach to overcome these challenges by generating HR images from more affordable low-resolution (LR) scans, potentially improving diagnostic accuracy and efficiency without requiring additional hardware. This survey reviews recent advances in MRI SR techniques, with a focus on deep learning (DL) approaches. It examines DL-based MRI SR methods from the perspectives of computer vision, computational imaging, inverse problems, and MR physics, covering theoretical foundations, architectural designs, learning strategies, benchmark datasets, and performance metrics. We propose a systematic taxonomy to categorize these methods and present an in-depth study of both established and emerging SR techniques applicable to MRI, considering unique challenges in clinical and research contexts. We also highlight open challenges and directions that the community needs to address. Additionally, we provide a collection of essential open-access resources, tools, and tutorials, available on our GitHub: {\fontsize{7.3}{9}\selectfont
\textcolor{blue}{\href{https://github.com/mkhateri/Awesome-MRI-Super-Resolution}{\texttt{https://github.com/mkhateri/Awesome-MRI-Super-Resolution}.}}}
\end{abstract}


\section{Introduction}
\label{sec:introduction}

\IEEEPARstart{H}{igh}-resolution (HR) magnetic resonance imaging (MRI) is essential for capturing fine anatomical details and accurately assessing physiological and functional processes. However, acquiring HR images comes with inherent trade-offs among spatial resolution, signal-to-noise ratio (SNR), and scan time--three tightly coupled factors where improving one often degrades the others. For example, achieving higher spatial resolution typically requires longer scans, increasing both imaging costs and the risk of motion artifacts. Low-resolution (LR) imaging is faster and may boost SNR, but often lacks the detail needed for high-quality diagnostics. Balancing this trade-off remains a central challenge in MRI \cite{afacan2016evaluation, brown2014magnetic, sui2021gradient}. Although advanced acquisition techniques, e.g., parallel imaging \cite{pruessmann1999sense} and compressed sensing \cite{lazarus20203d}, offer partial solutions, super-resolution (SR) techniques have shown a more favorable trade-off among these parameters than direct HR acquisition \cite{plenge2012super}.

SR techniques reconstruct HR images from more accessible LR scans using computational, hardware-independent methods. By enhancing resolution post-acquisition, SR can help to bridge the gap between fast imaging and diagnostic precision, making it an increasingly valuable tool in MRI \cite{fiat2005method, kang20243d, hokamura2024exploring, chaudhari2018super, zhao2019applications, qiu2023medical, vis2021accuracy, chen2022spatial, feng2021brain, sui2021fast}.

Given the pressing need for HR MRI, there is a clear demand for a comprehensive survey focused on SR techniques tailored to this modality. While most SR surveys address general computer vision tasks \cite{zhang2025deep, wang2020deep, chen2022real, liu2022blind, chauhan2023deep, li2021beginner, moser2024diffusion, gendy2025diffusion, moser2023hitchhiker, lepcha2023image, tian2022generative, ooi2021deep}, relatively few focus on medical applications \cite{umirzakova2023medical, li2021review, yang2023deep, kaji2019overview, yu2025general}, and even fewer specifically target MRI SR \cite{van2012super, ji2024deep, muhammad2024brain}. Existing surveys often overlook MRI-specific challenges and recent advances, such as clinical relevance, diverse MRI applications, and modern developments from computer vision, computational imaging, and inverse problems—including denoising diffusion models, implicit neural representations, foundation models, and advanced learning methods. A dedicated survey on MRI SR is essential to advance technology, benchmark methods, improve clinical diagnostics, and guide future research by highlighting emerging trends and gaps. To meet this need, our study presents a comprehensive survey of MRI SR techniques with the following contributions:

\begin{enumerate}
\item A thorough review of existing and potentially applicable DL-based approaches to MRI SR, ranging from theoretical foundations to research and clinical applications.

\item A comprehensive review of key aspects of DL-based MRI SR, including architectural design, performance evaluation,  benchmark datasets, practical applications, and learning strategies, with an emphasis on developing a systematic taxonomy to categorize these techniques.

\item A review of recent advances in computer vision for MRI SR, including vision Transformers (ViTs), diffusion models, foundation models, implicit neural representations, and Gaussian splatting.

\item A discussion of current challenges and future directions in DL-based MRI SR, along with a GitHub repository with open-access codes, datasets, tools, and tutorials.
\end{enumerate}

\begin{figure*}[!t] 
\label{Fig_taxonomy}
\centering
\includegraphics[width=0.99\textwidth, height=0.65\textwidth ]{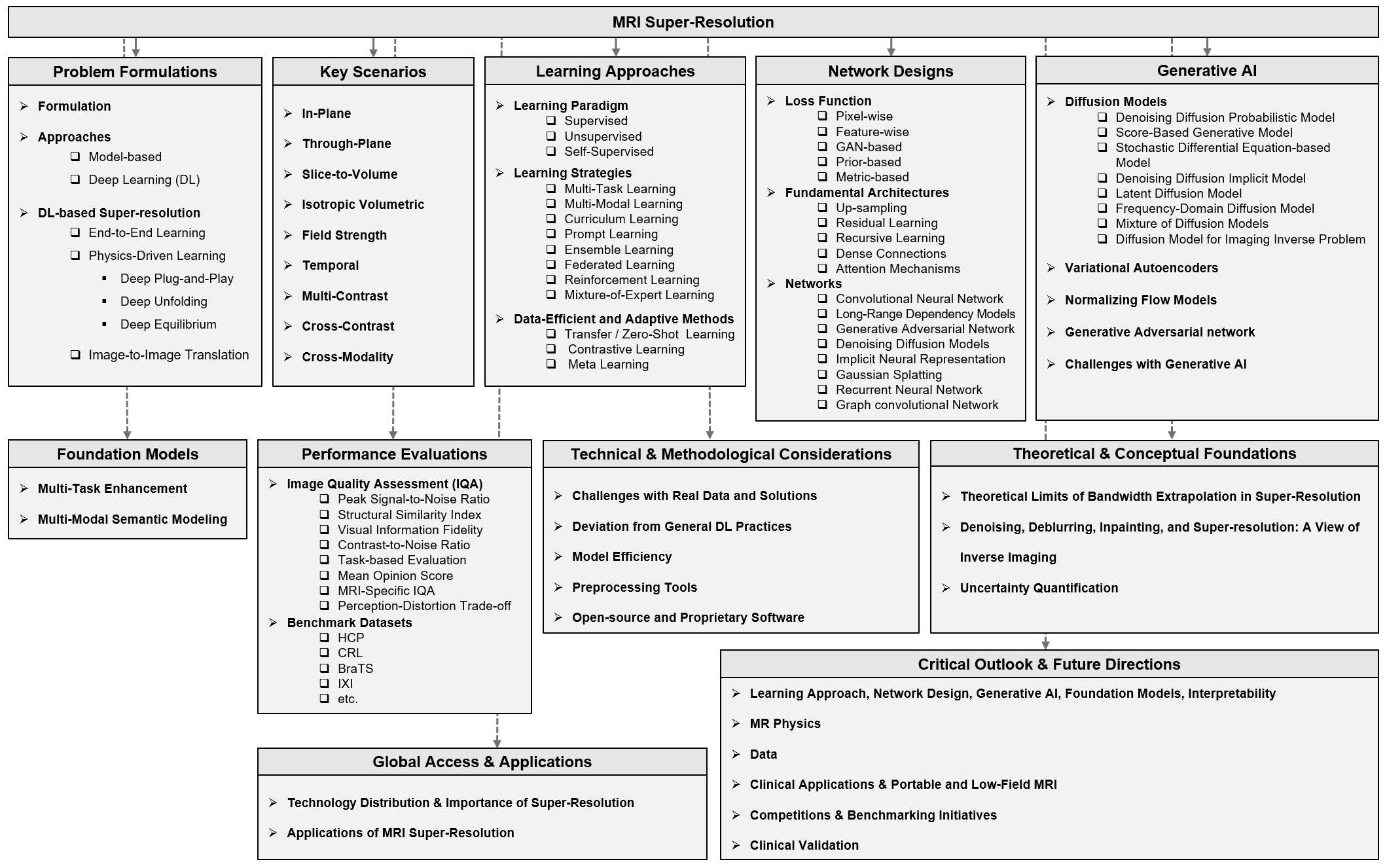}
\caption{Taxonomy of MRI super-resolution methods.}
\end{figure*}

\section{Problem Formulation and Terminology}
\label{Section_SR_formulation}

\subsection{Super-Resolution Formulation} 

SR is a low-level vision task that aims to recover the unknown HR image $ x \in \mathbb{R}^{H \times W \times D}$  from a degraded observation $y \in \mathbb{R}^{h \times w \times d}$:
\begin{equation}
\label{eq_1}
y = \mathcal{H}_{\delta}(x),
\end{equation}
where, $\mathcal{H}_{\delta}(\cdot): \mathbb{R}^{H \times W \times D} \rightarrow \mathbb{R}^{h \times w \times d}$ is the degradation operator arising from imaging system. We assume throughout that the MRI data have already been reconstructed from k-space into images, and that SR is applied in the image domain. SR is formulated as an ill-posed inverse problem because multiple HR images can correspond to the same LR observation. The degradation operator $\mathcal{H}_{\delta}(\cdot)$ is parameterized by $\delta$ primarily includes a geometric transformation $\Gamma(\cdot)$, a blurring operator, which is a convolution $\circledast$ with the blur kernel $\kappa$, an s-fold under-sampling operator $\downarrow_s$, and noise $n$, i.e., $\delta = \{\Gamma, \kappa, \downarrow_s, n\}$ \cite{plenge2012super, van2012super,lyu2020mri, zhao2023generative}. In practice, $\delta$ is not directly available; only the LR observation exists, and assumptions on $\delta$ are based on estimates or a subset of realistic degradation parameters. The ratios $ \tau_{x} = {H}/{h} $, $ \tau_{y} = {W}/{w} $, and $ \tau_{z} = {D}/{d} $  represent the undersampling factors in the $x$, $y$, and $z$ directions, respectively.

Theoretically, the SR process is to recover unknown $x$ through  $\mathcal{H}_{\delta}^{-1}(y)$, which requires determining the inverse mapping $\mathcal{H}_{\delta}^{-1}(\cdot) : \mathbb{R}^{h \times w \times d} \rightarrow \mathbb{R}^{H \times W \times D}$. If this mapping can be determined, it would enable the practical recovery of the HR image from the LR observation. SR techniques can generally be categorized into two groups based on their approach to this mapping: model-based and learning-based methods.

%\section{Model-Based vs. Learning-Based Super-Resolution}
%\subsection{Model-Based Super-Resolution} 

\subsection{Super-Resolution Approaches} 
\subsubsection{Model-Based} 
\label{model_based_sr}
Model-based SR methods aim to explicitly model the degradation function in Eq.~\ref{eq_1} as a combination of several operations, commonly referred to as the forward model. Assuming that the geometric transformation, blurring kernel, and under-sampling operator are known, and considering noise as additive--an assumption valid when the SNR exceeds three dB\cite{gudbjartsson1995rician}--the degradation operator is expressed as:
\begin{equation}
\label{eq_2}
\mathcal{H}_{\delta}(x) = (\Gamma(x) \circledast \kappa) \downarrow_s + n, \hspace{2mm}   \{\Gamma, \kappa, s, n\} \subset \delta  
\end{equation}
The solution $\hat{x}$ can then be obtained by optimizing maximum a posteriori (MAP) formulation as below:
\begin{equation}
\label{eq_MAP}
\hat{x} = \arg\min_{x}\{ {\Vert y - (\Gamma(x) \circledast \kappa) \downarrow_s  \Vert}^q_p + \lambda\mathcal{R}(x)\}.
\end{equation}
The first term represents the likelihood, calculated as the $p$-norm distance between the observation $y$ and the degraded latent image $x$, where $0 < p, q \leq 2$ are dictated by the noise distribution \cite{meng2013robust, bouman2022foundations, ren2019simultaneous}. When the SNR exceeds three dB, the noise can be assumed to be identical Gaussian \cite{gudbjartsson1995rician}, leading to $p = q = 2$. $R(\cdot)$ is the regularization (or prior) term, penalizing the latent image $x$ based on prior knowledge of the data. The parameter $\lambda$ controls the balance between the likelihood and prior term. To mitigate the ill-posedness of SR problems, various regularization techniques have been developed, including Tikhonov \cite{
%plenge2012super,
zhang2008application, brudfors2019tool}, total variation (TV) \cite{rudin1992nonlinear, tourbier2015efficient, shi2015lrtv}, self-similarity \cite{glasner2009super,manjon2010mri, bustin2018isotropic}, low-rankness \cite{shi2015lrtv, cherukuri2019deep, li2022motion, zhang2022accelerated}, sparse representation \cite{zhang2015image, wang2014sparse, zhang2012hierarchical}, non-local mean \cite{protter2008generalizing, manjon2010non, jafari2014mri}, and gradient guidance \cite{farsiu2004fast, sui2019isotropic, sui2020learning}. Handcrafted priors can improve SR but have limited performance compared to data-driven approaches. Building effective SR models typically requires the optimization of multiple priors, which is time- and memory-intensive and involves tuning trade-off parameters. Furthermore, SR models are often tailored to specific degradation scenarios, necessitating different models for varying conditions. If the degradation in the LR images does not match the assumed model, significant artifacts may occur due to domain gaps \cite{liu2024unsupervised, liang2022efficient, chen2022real}.

\subsubsection{Learning-Based} 
Learning-based SR methods establish a mapping between LR and HR image spaces, enabling the restoration of HR images from LR inputs. This mapping can be viewed as a regression problem, where the goal is to predict HR images based on LR inputs. Early research by \cite{freeman2000learning}, introduced co-occurrence priors between LR and HR image patches. Since then, numerous patch-based approaches have emerged, techniques like manifold learning \cite{lu2015mr}, and sparse representation \cite{ velasco2017sparse, zhang2015mr,jia2017new}. Recently, deep learning (DL)-based SR methods have demonstrated exceptional performance \cite{yuan2018unsupervised, wang2024knowledge, guerreiro2023super, zhao2019channel, yu2023rirgan, zhi2023coarse, wang2023disgan, lei2023decomposition, huang2021mri}. DL-based methods with end-to-end training automatically learn priors and degradation models directly from the data, eliminating the need for manual design. These methods optimize the objective function below over a set of LR-HR pairs $\{(y_i, x_i)\}_{i=1}^N$: 
\begin{equation} \label{eq_33} 
\hat{\theta} = \arg\min_{\theta} \sum_{i=1}^{N} \mathcal{L}(f_{\theta}(y_i), x_i) + \lambda \Phi(\theta),
\end{equation}
where $\mathcal{L}(f_{\theta}(y_i), x_i)$ represents the loss function between the network's prediction and the ground truth, $\Phi(\cdot)$ represents the regularization term that supports model generalization, and $\lambda$ is the trade-off parameter that balances the optimization problem. The trained model is then applied to an unseen LR image, $f_{\theta}(y)$, to generate the super-resolved output $\hat{x}$. Hereinafter, our discussion will center around DL-based methods.

\subsection{Perspectives on DL-based Super-Resolution} 
\begin{figure}[!t]
    \centering
    \includegraphics[width=0.99\linewidth]{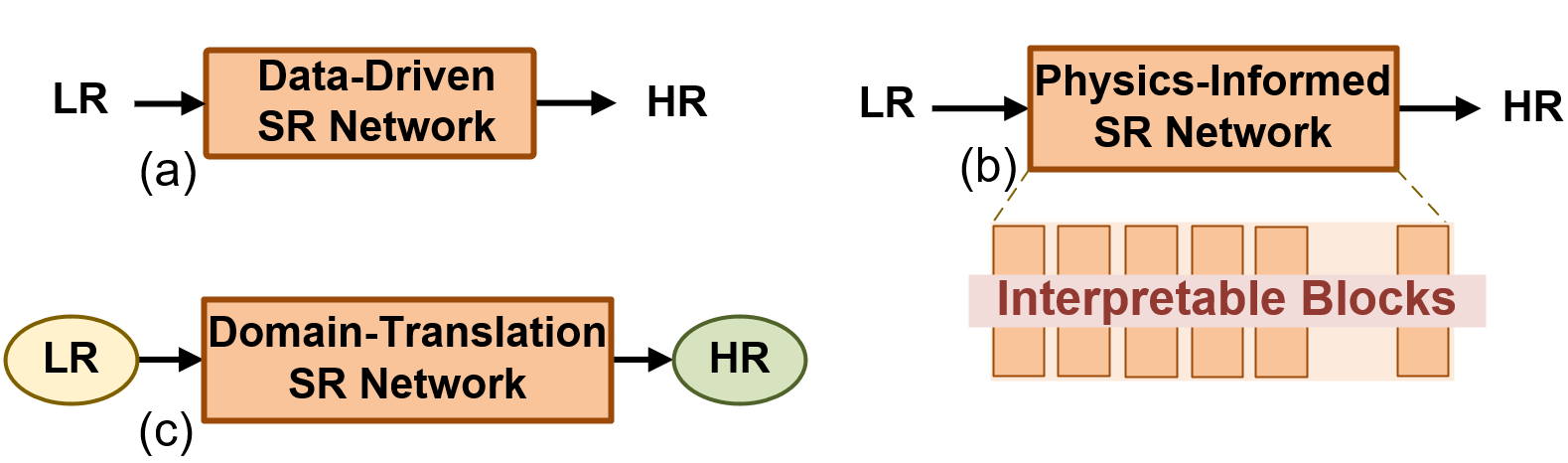}
    \caption{Perspectives on MRI super-resolution methods: 
    (a) Data-driven approach that learns LR-to-HR mapping purely from data; 
    (b) Physics-informed approach that incorporates the underlying imaging physics while mapping the LR-to-HR; 
    and (c) Image-to-image translation approach that translates across LR and HR domains, such as modalities or contrasts.}
    \label{Fig_sr_perspectives}
\end{figure}

DL-based SR techniques can be approached from different perspectives, each providing unique insights and methodologies for addressing the inverse problem of recovering HR images from LR acquisitions. These perspectives differ in how they incorporate prior knowledge, the degradation model, and apply learning strategies. This section explores these perspectives, including end-to-end data-driven, physics-informed techniques, image-to-image translation (see Fig. \ref{Fig_sr_perspectives}).

\subsubsection{End-to-End Data-Driven Perspective} 
In the plain end-to-end learning paradigm, the SR task is treated as a data-driven problem, where a neural network directly maps LR images to HR counterparts. The success of this approach hinges on the availability and quality of large-scale datasets, comprising either real-world LR-HR pairs or synthetic pairs generated by downsampling HR images. Dataset diversity and realism are critical to ensuring robust model performance. Various learning paradigms--including supervised, unsupervised, and self-supervised--can be applied depending on the data context and availability \cite{yu2023rirgan, zhao2019channel, zhi2023coarse}.

\subsubsection{Physics-Driven Perspective} The physics-driven or physics-informed approach--also referred to as the interpretable data-driven approach--combines traditional model-based methods with data-driven techniques to ensure that the SR solution adheres to the physical constraints of the MRI acquisition process. By integrating domain knowledge about imaging physics, these methods achieve more reliable and interpretable results than plain end-to-end data-driven methods~\cite{ chen2023physics, hammernik2023physics, karl2023foundations, yang2023mgdun,zhu2023physics, kazemi2024pi}. This approach is linked to various imaging inverse problems, grounded in strong mathematical principles~\cite{song2023solving, qu2024solving, gilton2021deep, alkan2023variational, chand2024multi, chung2024deep}. Physics-informed SR approaches are classified into I) deep plug-and-play, II) deep unfolding, and III) deep equilibrium methods, which will be elaborated in section \ref{physics_driven_sr}.

\subsubsection{Image-to-Image Translation Perspective} The image-to-image translation perspective frames SR as a task where the model learns to map images from the LR domain to the HR domain. This approach is widely used in medical image analysis, where variations in resolution, contrast, or modality are treated as distinct domains. The objective is to ``translate'' LR images into HR counterparts with enhanced anatomical detail and visibility. These tasks often involve challenges such as multiple modalities, varying contrast, or differences in imaging techniques \cite{wang2023ct, zeng2023mmnet, wang2023spatial}. Generative models, such as generative adversarial networks (GANs) and denoising diffusion probabilistic models (DDPMs), are frequently employed within this framework to produce high-quality content \cite{safayani2025unpaired}, which will be further explored in Section \ref{generative_ai}.

%\subsubsection{Low-Level Vision Enhancement Problem} This perspective focuses on enhancing image quality by first upsampling the image, followed by image enhancement techniques like sharpening and contrast adjustments to improve clarity. It prioritizes visual quality without modeling the MRI acquisition process, making it useful for tasks that emphasize sharpness and aesthetics \cite{he2024diffusion, ahmadian2021self}. (will be removed/modify)

\section{MRI Super-Resolution}
\label{sec: MRISR_scenarios}
\subsection{Key Scenarios in MRI Super-Resolution}
MRI SR addresses the fundamental trade-offs of the MRI triangle—spatial resolution, SNR, and scan time—as well as hardware limitations, e.g., magnet strength, gradient performance, and coil design. To overcome these challenges, various imaging strategies have been developed, including fast acquisition protocols, motion-robust techniques, and cost-effective low-field MRI. SR techniques offer computational support across these approaches, enabling high-quality imaging and broader accessibility without requiring modifications to scanner hardware or acquisition protocols. These capabilities make SR valuable in several key scenarios. % (see Fig. \ref{MRISR_scenarios}).

% \begin{figure*}[t]
% \centering
% \includegraphics[width=0.99\textwidth]{figs/SR_paradigm.png}
% \caption{ (updating..) Key scenarios in MRI super-resolution: 
% (a) In-plane SR: enhancing resolution within the imaging plane; 
% (b) Through-plane SR: improving resolution between slices; 
% (c) Slice-to-volume reconstruction: combining multiple stacks into an HR isotropic volume; 
% (d) Isotropic volumetric SR: uniformly enhancing resolution across all spatial dimensions; 
% (e) Temporal SR: increasing temporal resolution in dynamic imaging; 
% (f) Field-strength SR: enhancing resolution across different magnet strengths; 
% (g) Multi-contrast SR: integrating multiple contrasts to boost resolution; 
% (h) Cross-contrast SR: generating one contrast modality from another at higher resolution; 
% (i) Cross-modality SR: leveraging different imaging modalities (e.g., CT, PET) to enhance MRI resolution.}
% \label{MRISR_scenarios}
% \end{figure*}

\subsubsection{In-Plane Super-Resolution} 
In-plane SR enhances spatial resolution within the imaging plane $(x, y)$, enabling finer anatomical detail in individual slices. Physically, this often involves increasing the matrix size (e.g., from $128 \times 128$ to $256 \times 256$), which reduces voxel size and consequently lowers SNR. To preserve image quality--typically defined by adequate SNR and tissue contrast (sequence-dependent intensity differences; e.g., T1/T2/proton-density weighting)--longer scan times are usually required. In-plane SR (e.g., \cite{xue2019progressive}) mitigates these limitations by computationally boosting resolution without markedly compromising SNR or increasing acquisition time.

\subsubsection{Through-Plane Super-Resolution} 
Through-plane SR targets the resolution anisotropy in MRI data, where the in-plane resolution $(x, y)$ is higher than the through-plane resolution $(z)$. This anisotropy can hinder 3D analysis. Achieving isotropic resolution, where the through-plane resolution matches the in-plane resolution, requires reducing slice thickness. However, thinner slices reduce voxel size, decreasing SNR as well as increasing scan time as more slices are needed to cover the same volume. SR techniques mitigate these trade-offs by computationally enhancing through-plane resolution, allowing for isotropic reconstruction without sacrificing SNR or extending scan time. A variety of methods have been proposed to achieve isotropic MRI from a single anisotropic scan \cite{jog2016self, zhao2020smore, Beni2024SIMPLE, liu2024unsupervised} or from multiple anisotropic stacks acquired in orthogonal planes \cite{sui2022scan, gholipour2010robust, sui2021gradient, xu2021stress, lin2023high}. Many of these methods build on the foundational idea introduced by Elad and Feuer \cite{elad1997restoration}, who first formulated single-image SR reconstruction from multiple measurements.

\subsubsection{Slice-to-Volume Reconstruction}
Slice-to-volume reconstruction (SVR) is a form of through-plane SR in which multiple anisotropic data--2D slice stacks acquired under motion and from varying orientations--are combined to reconstruct an HR, isotropic volume. SVR jointly performs motion correction and SR by leveraging complementary information across stacks. This is especially valuable in fetal and neonatal imaging, where subject motion and limited acquisition time compromise 3D image quality. Each stack contributes partial, anisotropic observations of the anatomy, and reconstruction is further complicated by both intra- and inter-stack motion. SVR addresses these challenges by exploiting cross-stack redundancy and complementary spatial coverage across orientations—each stack provides high in-plane detail along its native slice axis and supplies through-plane information missing in the others \cite{uus2020deformable, xu2023nesvor, dannecker2025meta}. While SVR relies on multiple 2D stacks, a recent study \cite{young2024fully} demonstrates the feasibility of single-stack SVR, utilizing DL to directly infer HR isotropic volumes from a single anisotropic input.

\subsubsection{Isotropic Volumetric Super-Resolution}
In certain scenarios, such as fetal diffusion MRI, volumes are acquired with uniformly low resolution across all spatial dimensions $(x, y, z)$ due to the scan time constraints. The goal in this setting is to enhance resolution isotropically in all directions. Unlike through-plane SR, where high in-plane detail serves as a structural prior, isotropic SR is a more ill-posed, poorly conditioned inverse problem because high-frequency content is missing simultaneously in \(x\), \(y\), and \(z\), leaving no directional cues. Consequently, most approaches depend on fully supervised learning with external HR isotropic datasets. Recent work has shown promising results in isotropically enhancing LR volumes using such strategies \cite{wu2022arbitrary}.

\subsubsection{Super-Resolution Across Field Strengths}
MRI field strength is a key determinant of image quality, influencing SNR, spatial resolution, and diagnostic utility. At lower fields, image quality is degraded. SR methods address this limitation by computationally enhancing resolution, narrowing the gap to high-field performance and enabling diagnostic-quality imaging at lower fields. This is particularly relevant for low-field and portable systems \cite{islam2023improving}, where SR helps democratize MRI and approximate high-field imaging. SR also offers a cost-effective alternative in resource-limited or point-of-care settings. Its clinical viability is increasingly supported, with studies showing improved diagnostic quality across standard to high-field MRI \cite{bahrami20177t, bahrami2016reconstruction, siam2024improving, eidex2024high, jha2023trganet, zhang2019dual, cui20247t}, as well as in ultra-low-field \cite{dayarathna2024ultra, jiang2022super} and portable low-field systems \cite{islam2023improving, lin2023low}.

Several low-field MRI systems have received FDA clearance for clinical use, including the ultra-low-field, portable Hyperfine Swoop (0.064 T) for bedside neuroimaging \cite{sheth2021portable}. Recent reviews also discuss the regulatory landscape and clinical potential of other low-field systems, such as Siemens MAGNETOM Free.Max (0.55 T) and Promaxo (0.066 T) \cite{cooley2023lowfield, arnold2023lowfield}.

\subsubsection{Temporal Super-Resolution} 
Temporal SR aims to enhance the temporal resolution of dynamic MRI--that is, the frequency at which images are acquired over time. Higher temporal resolution allows for more precise capture of rapid physiological changes, which is critical in applications such as cardiac imaging, functional MRI, and blood flow assessment. Conventionally, increasing temporal resolution requires faster acquisitions, often at the cost of reduced spatial resolution and lower SNR due to limited sampling. SR techniques address this trade-off by reconstructing temporally enhanced images from low-temporal-resolution data, thereby improving the fidelity of dynamic processes without requiring faster acquisition protocols. Applications of temporal SR include cardiac and functional imaging \cite{li2017novel, nie2020super}, time-resolved flow imaging such as 4D flow MRI \cite{kazemi2024pi, shone2023deep, ericsson2024generalized}, and respiration-resolved 4D MRI \cite{chilla2017deformable}.

\subsubsection{Multi-Contrast MRI Super-Resolution} 
In clinical practice, multiple MRI contrasts (e.g., T1, T2, FLAIR) are routinely employed to capture complementary anatomical and pathological information. Multi-contrast SR leverages the synergy among these modalities by integrating diverse contrast information to enhance image resolution beyond what is achievable with a single modality. By incorporating multiple contrasts, it offers a more comprehensive depiction of tissue properties—effectively performing image fusion—and supports more accurate diagnosis and informed treatment planning. Recent studies highlight the increasing importance of multi-contrast SR, showcasing progress in both algorithmic innovation and clinical applicability \cite{lyu2020multi, feng2021multi, feng2024exploring, liu2024mapanet, huang2023ddformer, yang2022model, mcginnis2023single, zou2023multi, kong2023dual, lei2024joint, li2024rethinking, zheng2024sgsr, lei2025robust, li2025high}. %Multi-contrast SR largely builds on tools and insights from modality synthesis and domain translation.

\subsubsection{Cross-contrast MRI Super-Resolution}
Cross-contrast SR focuses on translating or synthesizing one MRI modality (e.g., T2, FLAIR) from another (e.g., T1), often within an image-to-image translation framework. This approach is particularly valuable in scenarios where certain contrasts are missing, corrupted, or impractical to acquire due to time limitations or patient-specific constraints. By learning mappings between modalities, SR models can generate HR representations of unavailable contrasts, supporting diagnosis and multi-modal analysis. Unlike multi-contrast SR, which fuses multiple available contrasts to enhance a target modality, cross-contrast SR predicts a distinct contrast, potentially at higher resolution, from a single input. Recent deep learning approaches have shown promise in this area, particularly those based on GANs~\cite{armanious2020medgan, feng2024bridging, yang2020mri}, and diffusion models \cite{ozbey2023unsupervised}. This task intersects with broader research on modality synthesis and domain translation. 

\subsubsection{Cross-Modality Super-Resolution}
Cross-modality SR leverages complementary information across different imaging modalities--such as MRI, CT, and PET--to enhance resolution beyond the limits of a single modality. By mapping between modalities, it reconstructs HR images in one domain (e.g., MRI) using data from another with higher resolution or distinct characteristics (e.g., PET for functional insights). This approach is valuable when direct HR imaging is impractical due to risks (e.g., radiation in CT) or limited accessibility. DL methods based on domain translation \cite{armanious2020medgan}, cross-modal synthesis, and fusion networks have driven progress in this area. Applications include enhancing PET resolution with MRI priors, boosting MRI resolution using CT guidance, and synthesizing detailed anatomical images from LR functional scans \cite{liang2024medical, dayarathna2024deep}.

\section{Physics-Driven Super-Resolution}
\label{physics_driven_sr}
Physics-driven SR integrates knowledge of imaging physics with learning approaches to solve inverse imaging problems. Notable strategies include deep image priors \cite{ulyanov2018deep, zhang2021plug, yoo2021time} and generative priors \cite{sun2024provable, huang2023transmrsr, zach2023stable}, which serve as implicit regularizer. This section examines three major frameworks: (i) deep plug-and-play (PnP), (ii) deep unfolding, and (iii) deep equilibrium models.
We highlight how these frameworks integrate physical modeling with deep learning.

\subsection{Deep Plug-and-Play} 
The PnP framework integrates DL-based priors into traditional model-based algorithms, with applications across a broad spectrum of inverse imaging problems \cite{venkatakrishnan2013plug, zhang2019deep, karl2023foundations, shoushtari2023prior, kamilov2023plug, ahmad2020plug, gao2024plug, terris2025fire}. The central idea is to replace the handcrafted regularization term in the iterative solution of the MAP formulation (Eq.~\ref{eq_MAP}) with a pre-trained deep denoiser \cite{kamilov2023plug}, which is "plugged" into the iterative optimization process using methods such as alternating direction method of multipliers (ADMM) \cite{boyd2011distributed}, half-quadratic splitting (HQS) \cite{nikolova2005analysis}, or majorization-minimization (MM) \cite{sun2016majorization}. The general procedure is summarized in Algorithm~\ref{alg:pnp}. PnP combines flexibility and interpretability, adapting to various tasks without needing ground-truth data. Its main drawback is the computational and memory overhead incurred by the repeated execution of the forward model and denoising network, which can hinder scalability. In addition, similar to other model-based methods, it requires manual tuning of the regularization parameter~$\lambda$ and the number of iterations~$K$.

\begin{algorithm}[!t]
\caption{Plug-and-Play for Image Super-Resolution}
\label{alg:pnp}
\begin{algorithmic}[1]
\Require Observed image $y$, degradation operator $\mathcal{H}_\delta(\cdot) = (\Gamma(\cdot) \circledast \kappa)\downarrow_s$, pre-trained denoiser $\mathcal{D}_\sigma(\cdot)$, regularization parameter $\lambda$, number of iterations $K$
\State Initialize $x_0$ by upsampling $y$
\For{$k = 1$ to $K$}
    \State \textbf{Data Consistency Step:}
    \[
    z_k \gets \arg\min_{z} \left\| y - \mathcal{H}_\delta(z) \right\|_2^2 + \lambda \left\| z - x_{k-1} \right\|_2^2
    \]
    \State \textbf{Denoising Step:}
    \[
    x_k \gets \mathcal{D}_\sigma\left(z_k\right)
    \]
\EndFor
\State \Return $x_K$
\end{algorithmic}
\end{algorithm}

\subsection{Deep Unfolding}
\begin{figure}[!h]
\centering
\includegraphics[width=0.49\textwidth]{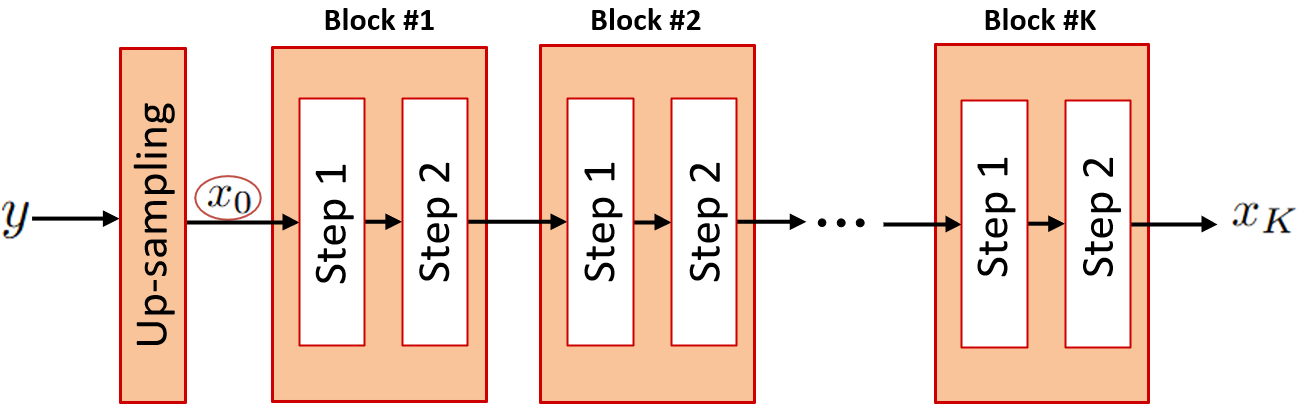}
\caption{Deep unfolding models provide a physics-informed approach to imaging inverse problems by integrating the forward model with a neural network within an iterative reconstruction framework, as illustrated in Algorithm~\ref{alg:deep_unfolding}. Each unrolled layer corresponds to a single iteration of an optimization algorithm~\cite{zhang2020deep, monga2021algorithm}. }
\label{DU}
\end{figure}

Deep unfolding, or unrolling, provides a systematic bridge between iterative model-based algorithms and deep learning. In this approach, each iteration of an optimization algorithm is unrolled into the neural network layers, allowing for the end-to-end training of the network. This offers the advantage of efficient parameter learning and faster inference compared to traditional iterative methods \cite{yang2022model, monga2021algorithm}. A key benefit of deep unfolding is its interpretability: it retains the underlying mathematical structure of the optimization problem, ensuring that each layer has a clear role corresponding to the steps of an iterative method. This combination of interpretability and efficiency makes deep unfolding particularly powerful for applications like image super-resolution \cite{yang2023mgdun, zhao2023generative, yamato2023super, wang2025diff}.
Following this approach, the MRI SR method is structured as an iterative two-step process comprising data consistency enforcement and prior-based refinement, as detailed in Algorithm~\ref{alg:deep_unfolding}. This recursive formulation mirrors the architecture of deep networks, where each layer progressively refines the reconstruction by alternating between fidelity to the observed data and the incorporation of prior knowledge. A conceptual overview of deep unfolding appears in Fig.~\ref{DU}.

\begin{algorithm}[!ht]
\caption{Deep Unfolding for Image Super-Resolution}
\label{alg:deep_unfolding}
\begin{algorithmic}[1]
\Require Observed image $y$, degradation operator $\mathcal{H}_\delta(\cdot) = (\Gamma(\cdot) \circledast \kappa)\downarrow_s$, regularization $\mathcal{R}(\cdot)$,  parameters $\mu, \lambda$, number of iterations $K$
\State Initialize $x_0$ by upsampling $y$
\For{$k = 1$ to $K$}
    \State \textbf{Step 1:}
    \[
    z_k \gets \arg\min_{z} \left\| y - \mathcal{H}_\delta(z) \right\|_2^2 + \mu \left\| z - x_{k-1} \right\|_2^2
    \]
    \State \textbf{Step 2:}
    \[
    x_k \gets \arg\min_{x} \frac{\mu}{2} \left\| z_k - x \right\|_2^2 + \lambda \mathcal{R}(x)
    \]
\EndFor
\State \Return $x_K$
\end{algorithmic}
\end{algorithm}

\subsection{Deep Equilibrium}
\begin{figure}[!h] 
\centering
\includegraphics[width=0.45\textwidth]{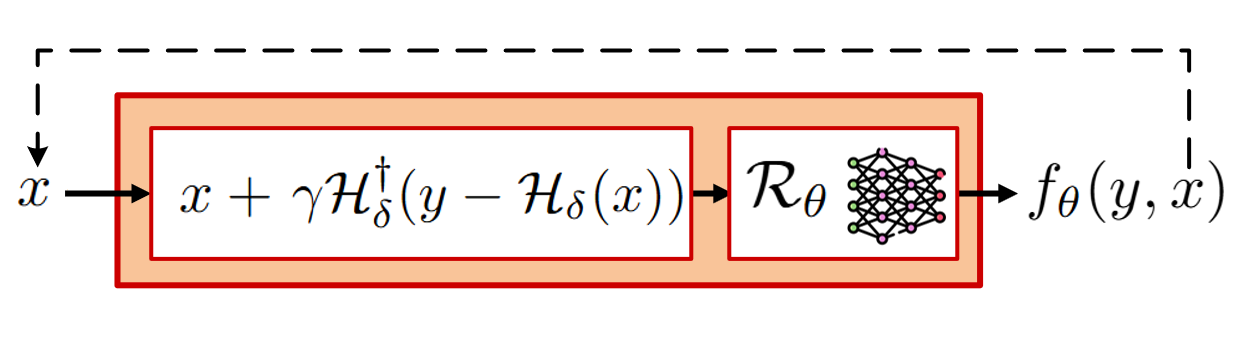}
\caption{Deep Equilibrium models provide a physics-driven approach to imaging inverse problems by incorporating the forward model and a neural network within a fixed-point formulation \cite{gilton2021deep}.} 
\label{Fig_DEQ}
\end{figure}

Deep equilibrium (DEQ) networks extend iterative reconstruction methods by directly solving for the fixed point of a parameterized update rule. Instead of explicitly unrolling a fixed number of iterations, DEQ models define an implicit mapping and compute its equilibrium state \(x^*\) satisfying,
\begin{equation}
    x^* = f_{\theta}(y, x^*),
\end{equation}
where \(f_{\theta}\) denotes a learned refinement operator. A common formulation for imaging inverse problems is $f_\theta(y,x) = \mathcal{R}_\theta\!\left( x +  \gamma \mathcal{H}_\delta^\dagger\!\left(y - \mathcal{H}_\delta(x)\right) \right)$, where \(\mathcal{H}_\delta\) and \(\mathcal{H}_\delta^\dagger\) represent the forward degradation operator and its adjoint, \(\mathcal{R}_\theta\) is a learned prior, and $\gamma$ are tunable parameters~\cite{gilton2021deep}. 
Unlike deep unfolding, which unrolls \(K\) iterations into network layers, DEQ models compute the equilibrium point implicitly using fixed-point solvers such as Anderson acceleration \cite{walker2011anderson} or Broyden's method \cite{bai2019deep}. This enables \emph{constant memory usage}, since only the equilibrium state is required for both forward and backward passes. As a result, DEQ models can scale to much deeper implicit architectures compared to unrolled networks. However, convergence to a stable fixed point is not guaranteed and may depend on architectural choices, initialization, and the learned operator~\cite{zou2023deep, buzzard2018plug, cao2024deep}. Additionally, implicit differentiation introduces extra computational overhead. A conceptual illustration of the DEQ formulation is shown in Fig.~\ref{Fig_DEQ}.

% % Later in the document
% \begin{table*}[ht]
% \centering
% \caption{Comparison of Deep Plug-and-Play, Deep Unfolding, and Deep Equilibrium}
% \label{tab:comparison_methods}
% \small
% \renewcommand{\arraystretch}{1.1} % tighter row spacing
% \rowcolors{2}{gray!10}{white}
% \begin{tabularx}{\textwidth}{p{3.6cm}X X X}
% \toprule
% \textbf{Method} & \textbf{Key Idea} & \textbf{Pros} & \textbf{Cons} \\ 
% \midrule

% \textbf{Deep Plug-and-Play (PnP)} & 
% Uses pre-trained denoisers as priors in model-based optimization. & 
% • Flexible with pre-trained priors \newline • No paired data needed & 
% • Computationally expensive \newline • Memory-intensive \\

% \textbf{Deep Unfolding} & 
% Unrolls iterative algorithms into network layers for end-to-end learning. & 
% • Efficient parameter learning \newline • Faster inference \newline • Interpretable & 
% • Requires paired data \newline • Less flexible than PnP \\

% \textbf{Deep Equilibrium (DEQ)} & 
% Solves for the fixed point of a dynamical system instead of fixed iterations. & 
% • Reduced memory \newline • Flexible in depth & 
% • Slower convergence \newline • Convergence not always guaranteed \\
% \bottomrule
% \end{tabularx}
% \end{table*}

\section{Learning Approaches in MRI SR}
\label{learning_approach}

DL-based MRI SR approaches can be broadly categorized into three main groups: 
(1) \textit{learning paradigms}, 
(2) \textit{learning strategies}, and 
(3) \textit{data-efficient and adaptive methods}. 

These groups represent complementary dimensions rather than mutually exclusive categories; an MRI SR method may simultaneously fall under multiple dimensions (e.g., a self-supervised multi-task learning method).

\subsection{Learning Paradigms}
\begin{figure*}[!t]
\centering
\includegraphics[width=0.99\textwidth,height=0.2\textwidth]{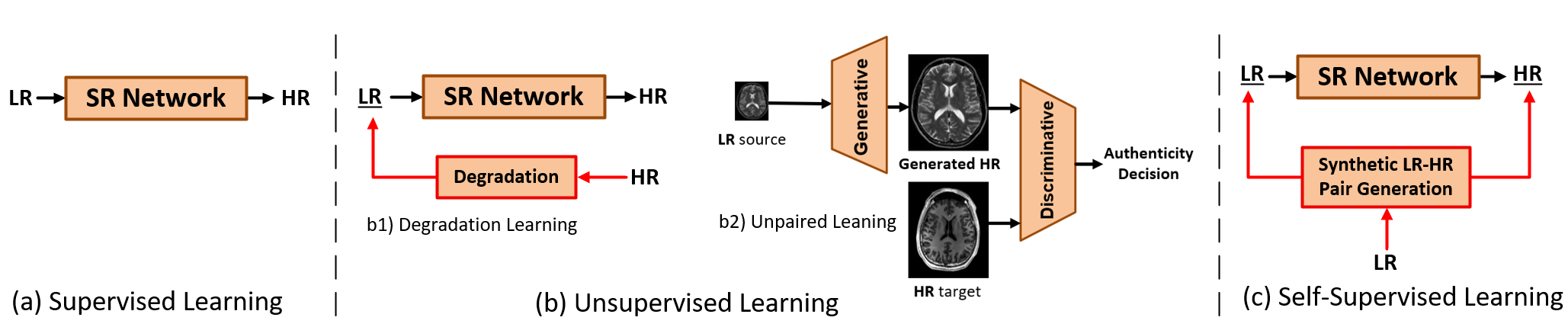}
\caption{Learning paradigms for MRI super-resolution: (a) Supervised learning, where the network is trained on physically acquired and well-aligned pairs of LR-HR scans; (b) Unsupervised learning, where no LR-HR pairs are available: 1) generating synthetic LR images from HR scans to create LR-HR training pairs, and 2) scenarios where only LR scans and unpaired examples from the HR target domain are accessible; and (c) Self-supervised learning, where training pairs are generated directly from the available LR scans.}
\label{Fig_Learning_paradigm}
\end{figure*}

MRI SR methods can be classified into three learning paradigms based on the level of supervision:
\textit{(1) Supervised learning}, which uses physically acquired and spatially aligned LR-HR image pairs;
\textit{(2) Unsupervised learning}, which removes the need for paired data by leveraging degradation learning, or unpaired distributions, domain adaptation;
\textit{(3) Self-supervised learning}, which generates supervision directly from the input image without external training data (see Fig. \ref{Fig_Learning_paradigm}).

\begin{figure}[!t]
\centering
\includegraphics[width=0.45\textwidth,height=0.13\textwidth]{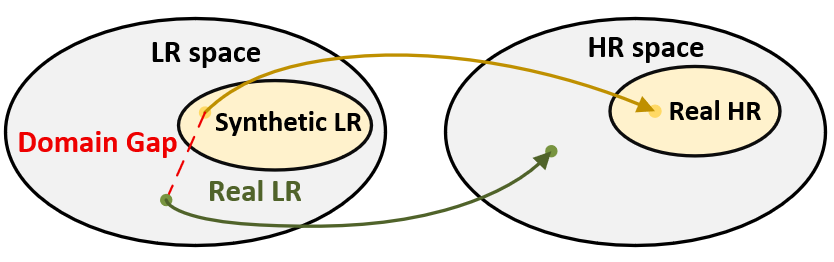}
\caption{Supervised vs. unsupervised super-resolution: Domain gaps between synthetic LR training data and real acquired LR images lead to discrepancies in the super-resolved outputs when models trained on synthetic data are applied to real-world inputs.}
\label{Fig_supervised_vs_unsupervised}
\end{figure}

\subsubsection{Supervised Learning}
Supervised learning trains models on physically acquired and registered LR-HR image pairs, which has been studied in natural image SR \cite{chen2022real}. However, applying this paradigm to MRI presents key challenges:
(i) Acquiring matched LR-HR pairs is difficult due to motion artifacts, scanner variability, and the high cost and time required for scans—issues exacerbated in dynamic or susceptible regions (e.g., the abdomen) and populations such as fetuses and neonates.
(ii) Even when LR and HR scans are available, they often exhibit non-rigid misalignments caused by patient motion or anatomical deformation. Although image registration can reduce misalignment, it often fails to correct complex non-linear changes, particularly in soft tissue \cite{zakeri2023dragnet}.
Because of these limitations, few studies have utilized truly acquired LR–HR MRI pairs for supervised training. Notable examples include a paired multi-modal MRI dataset at 3T and 7T~\cite{chu2025paired}, and the work by Chun et al.~\cite{chun2019mri}, which used a small set of real LR–HR pairs to supervise a downsampling network for learning realistic degradation. 
Due to the practical difficulties in collecting well-aligned paired data, there has been a growing shift toward unsupervised and self-supervised SR methods that better accommodate the constraints of clinical MRI acquisition.

\subsubsection{Unsupervised}
Unsupervised learning addresses the challenge of unavailable or misaligned LR-HR image pairs by training models on unpaired data or synthetically degraded images. This approach is especially suited to MRI, where motion, scanner variability, and acquisition costs often hinder the collection of paired datasets. Common strategies include:

\textbf{Degradation-based learning:}
Unsupervised SR methods often simulate LR images from HR scans using predefined degradation models, such as down-sampling~\cite{shi2018mr, zhang2021mr}. In MRI, k-space truncation is the most commonly used strategy, where HR images are transformed into the frequency domain, partially truncated, and then reconstructed into synthetic LR volumes~\cite{zhao2019channel, sui2022scan, li2021volumenet, chen2018brain}. Additionally, Iglesias et al.~\cite{iglesias2023synthsr} proposed SynthSR, a framework that synthetically generates HR images from segmentation maps and produces paired LR images via down-sampling.
While such down-sampling techniques are widely used, they tend to oversimplify the complexity of real-world clinical acquisitions. In practice, LR and HR images are often acquired using distinct under-sampling protocols—such as partial Fourier, elliptical k-space, or parallel imaging, which are difficult to replicate accurately due to scanner-specific parameters and unknown acquisition settings. Moreover, differences in SNR, coil sensitivity, and system calibration further contribute to discrepancies between synthetic and real degradations~\cite{liu2024unsupervised}. These mismatches lead to a domain gap that can hinder model generalization at inference time (see Fig.~\ref{Fig_supervised_vs_unsupervised}). % In such settings, unsupervised SR approaches offer a promising solution by relaxing the need for exact degradation modeling and enabling more robust learning under degradation shifts.

\textbf{Domain Transfer (Unpaired Image-to-Image Translation):}
Degradation-based unsupervised methods reduce reliance on paired data but still depend on simplified assumptions that fail to capture the full complexity of real MRI acquisitions. This mismatch creates a domain gap between synthetic and clinical data, motivating domain transfer approaches that directly learn mappings between unpaired LR and HR domains. In MRI SR, domain transfer strategies are often inspired by CycleGAN-based frameworks~\cite{zhu2017unpaired}, which treat LR and HR images as distinct domains and learn bidirectional mappings between them. Cycle-in-cycle architectures~\cite{yuan2018unsupervised} have been proposed to enhance domain consistency, forming the basis for several subsequent approaches. 
For example, Zhou et al.~\cite{zhou2022blind} introduced an unsupervised SR framework that combines a domain correction network in the LR space with an upscaling network for HR reconstruction, enabling training without explicit knowledge of the degradation process. Building on this idea, Liu et al.~\cite{liu2024unsupervised} proposed domain adaptation in both image and latent feature spaces to improve alignment for unpaired and misaligned MRI data. Domain transfer has also been applied to multimodal MRI SR~\cite{iwamoto2022unsupervised, feng2024bridging} and for synthesizing 7T-like images from 3T scans~\cite{siam2024improving}. A comprehensive overview of unpaired image-to-image translation in medical imaging is available in~\cite{chen2023deep_image2image}.

\subsubsection{Self-Supervised Learning}
Self-supervised SR eliminates the need for external datasets by training directly on the input LR image, exploiting its internal structure. Unlike supervised methods, self-supervised approaches can adapt to variations in scanner type, acquisition protocol, or anatomical region, making them highly practical for clinical use. In this setting, regions of the input with relatively higher structural fidelity are treated as proxy HR, while corresponding synthetic LR samples are generated from them to form training pairs. A representative method is zero-shot self-supervised SR (ZSSR)~\cite{shocher2018zero}, which exploits patch recurrence across scales to synthetically generate LR–HR pairs from a single image. This principle has been extended to MRI in methods such as SMORE~\cite{zhao2020smore}, which reconstructs isotropic 3D volumes from anisotropic inputs without external supervision.
Subsequent self-supervised strategies have been applied to a range of MRI tasks, including dynamic fetal imaging~\cite{xu2021stress}, arbitrary-scale through-plane SR~\cite{zhang2023self}, and multi-contrast DWI SR with inter-scan motion correction~\cite{gundogdu2024self}. Sui et al.~\cite{sui2021mri} further advanced this direction with a scan-specific generative degradation learning approach, inspired by blind SR~\cite{liu2022blind}, which models degradation directly from the input data—enhancing robustness to scanner-specific artifacts and acquisition variability. More recently, \cite{sui2025unsupervised} extended scan-specific SR with an unsupervised transformer that captures long-range spatial structure, yielding improved patient-specific reconstructions.

\subsection{Learning Strategies}

\subsubsection{Multi-Task Learning} 
Multi-task learning is a strategy in which a model is trained to perform multiple related tasks simultaneously, leveraging shared representations to enhance overall performance. In the context of SR, this approach improves outcomes by incorporating complementary information from tasks such as image reconstruction, artifact correction, and related processes. Multi-task learning has been extensively studied in MRI. For instance, joint learning of SR and denoising has been explored in \cite{yu2023rirgan, chung2022mr, fathi2020super, patel2025super}, the combinations of SR and image reconstruction are investigated in \cite{feng2021task, ding2024cross, chen2024joint, wang2023mhan}, the integration of SR with motion correction is discussed in \cite{chen2024motion, zhang2023mocosr}, and SR combined with Gibbs artifact removal is presented in \cite{liu2023unsupervised}. Additionally, the synergy between SR and segmentation is explored in \cite{delannoy2020segsrgan}, and the combination of SR with registration and reconstruction is addressed in \cite{corona2021variational}.

\subsubsection{Multi-Modal MRI Learning}
Clinical protocols often acquire multiple MRI sequences with complementary contrasts (e.g., T1, T2, FLAIR, diffusion). Conventional SR methods usually process each modality independently, overlooking cross-modal correlations that could enhance reconstruction. Multi-modal MRI SR addresses this by fusing information across contrasts to guide reconstruction. Xu et al.~\cite{xu2025multi}, for example, proposed a multiscale fusion network that reconstructs HR T2-weighted images using high-frequency structural details from T1-weighted scans. Stimpel et al.~\cite{stimpel2019multi} developed a deep guided filtering framework in which an HR modality (e.g., T1-weighted MRI or CT) guides the SR of a LR modality (e.g., T2-weighted MRI), combining a locally linear guided filter with a learned guidance map to improve transparency and robustness. More recently, Li et al.~\cite{li2022transformer} introduced McMRSR, a Transformer-based framework that models long-range dependencies and employs multi-scale contextual matching and aggregation across contrasts.

\subsubsection{Curriculum Learning} 
Curriculum learning is a training strategy where tasks are introduced in order of increasing difficulty, enabling models to first master simpler concepts before tackling more complex ones \cite{bengio2009curriculum, gao2017demand}. In SR, this approach improves model robustness to high scaling factors, noise, and blurring by gradually exposing the network to such challenges during training, resulting in better reconstruction quality and generalization \cite{li2019feedback, wang2018fully}.

\subsubsection{Prompt Learning}
Prompt learning introduces additional contextual cues—known as prompts—to guide the model's reconstruction strategy. In image SR, prompts typically encode degradation characteristics such as blur, noise, or distortion, helping the model adapt its response accordingly. One effective approach employs degradation-conditioned prompt layers to focus on relevant contextual features, improving super-resolved output \cite{khan2024lightweight, wang2023promptrestorer}. More recently, \cite{Ai_2024_CVPR} introduced a multimodal prompt framework leveraging stable diffusion priors to enhance adaptiveness, generalization, and fidelity in image restoration tasks. Although this approach has not yet been applied to MRI, it has the potential to incorporate scan parameters and acquisition protocols as prompts to improve reconstruction performance.

\subsubsection{Ensemble Learning (Multi-Supervision)}
Ensemble learning combines predictions from multiple models to achieve improved performance compared to any individual model \cite{dong2020survey}. Classical strategies include bagging (e.g., Random Forests), boosting (e.g., AdaBoost), and stacking. In MRI SR, ensembles are typically constructed by integrating outputs from several SR networks trained on different priors, architectures, or datasets. This aggregation helps mitigate artifacts and preserve fine details, as the complementary strengths of the models lead to more robust and visually consistent reconstructions \cite{lyu2020mri}.

\subsubsection{Federated Learning}
Federated learning offers a privacy-preserving paradigm for MRI SR by enabling collaborative model training across multiple institutions without sharing raw patient data. This is particularly valuable in neuroimaging, where datasets are often small and fragmented across sites. Recently, Basira et al.~\cite{ghilea2023replica} proposed RepFL, a replica-based FL framework that creates perturbed replicas of each client’s data to improve data diversity and aggregates heterogeneous models for graph super-resolution. Applied to brain connectome reconstruction from MRI, RepFL demonstrated superior performance over conventional FL methods, highlighting the potential of federated approaches for multi-resolution MRI analysis under data scarcity and privacy constraints.

\subsubsection{Reinforcement Learning Models (RLs)} 
RLs have been explored for SR by framing it as a decision-making process where agents iteratively enhance image resolution. These models adaptively learn strategies for dynamic or stage-based upscaling, showing potential in MRI SR for handling complex data variations and improving image quality \cite{vassilo2020multi, chen2024star}. However, RL-based SR methods are less common than CNN- or GAN-based approaches, since defining suitable states, actions, and reward functions for sequential resolution enhancement is challenging and often less efficient than direct end-to-end learning.

\subsubsection{Mixture of Experts (MoE)} 
MoE framework combines a lightweight gating mechanism with a pool of specialized subnetworks—experts—activating only the most relevant subset for each input, enabling high representational capacity with efficient computation \cite{jacobs1991adaptive}. In MRI SR, experts can specialize in handling specific degradations, anatomical regions, or scanner contrasts, while the gate dynamically selects the best combination for each volume. MoE-DiffSR \cite{wang2025moediff} demonstrates this approach using a diffusion backbone with gated experts, achieving strong results on multi-scanner brain and knee datasets. Recent advances in routing and regularization further improve load balancing and scalability, making MoE a promising direction for MRI SR.

\subsection{Data-Efficient and Adaptive Methods}

\subsubsection{Transfer Learning and Zero-Shot}
Transfer learning enables models trained on one task to be adapted for related tasks with limited data. In MRI SR, it enhances efficiency by leveraging pre-trained models from other imaging modalities or existing datasets, reducing reliance on large domain-specific data \cite{park2020fast}. This approach accelerates development and improves performance, especially in clinical scenarios with limited HR MRI availability. It has been applied in brain \cite{liang2023mouse} and cardiac MRI \cite{xia2021super}. A notable subcategory, zero-shot learning, adapts models directly to test data without task-specific fine-tuning and has shown promise in medical imaging \cite{chen2023cunerf}.

\subsubsection{Contrastive Learning} 
Contrastive learning trains models by pulling together similar samples (positives) and pushing apart dissimilar ones (negatives), enabling robust feature representations. Originally developed in computer vision \cite{wu2023practical, chen2022unpaired}, it has recently been applied to MRI SR to cope with scarce paired data. Approaches contrast LR inputs with HR counterparts, synthetic with real LR images, or different contrasts from the same subject \cite{li2024unpaired}, encouraging fine-grained structural consistency and reducing domain gaps. Huang et al.~\cite{huang2024mftn}, for example, introduced a perceptual contrastive loss to improve texture realism in multi-contrast MRI SR.

\subsubsection{Meta learning} 
Meta-learning, or "learning to learn," enables models to rapidly adapt to new imaging tasks using minimal data. In natural image SR, meta-transfer learning methods (e.g., MTL-SR~\cite{soh2020meta}) have demonstrated impressive adaptability to unseen degradations. Although such approaches have been validated in natural images, their application to MRI SR is only beginning to emerge. Recent work includes Delta-WKV~\cite{lu2025delta}, a meta-in-context learner that dynamically adjusts model weights during inference to capture both local and global anatomical patterns. Evaluated on benchmark MRI datasets, it achieved state-of-the-art MRI SR quality while reducing computational cost.

%%%%%%%%%%%%%%%%%%%%%%%%%%%%%%%%%%%%%%%%%%%%%%%%%%
%%%%%%%%%%%%%%%%%%%%%%%%%%%%%%%%%%%%%%%%%%%%%%%%%%

\begin{table*}[t]
\centering
\caption{Commonly Used Loss Functions in Image Super-Resolution.}
\begin{tabularx}{0.8\textwidth}{p{2.5cm}X}
\toprule
\textbf{Loss Category} & \textbf{Loss Function (Batch-wise)} \\
\midrule

\multirow{3}{*}{\textbf{Pixel-wise}} & 
\textbf{Mean Squared Error:} 
$\mathcal{L}_{\ell_2} = \frac{1}{N} \sum_{i=1}^{N} \frac{1}{hwd} \sum_{j,k,l} (x^{(i)}_{j,k,l} - \hat{x}^{(i)}_{j,k,l})^2$ \\
& 
\textbf{Mean Absolute Error:} 
$\mathcal{L}_{\ell_1} = \frac{1}{N} \sum_{i=1}^{N} \frac{1}{hwd} \sum_{j,k,l} |x^{(i)}_{j,k,l} - \hat{x}^{(i)}_{j,k,l}|$ \\
& 
\textbf{Charbonnier Loss:} 
$\mathcal{L}_{\text{Charbonnier}} = \frac{1}{N} \sum_{i=1}^{N} \sqrt{\left(\mathcal{L}_{\ell_1}^{(i)}\right)^2 + \epsilon^2}$ \\

\midrule

\multirow{2}{*}{\textbf{Feature-wise}} & 
\textbf{Perceptual Loss:} 
$\mathcal{L}_{\text{Perceptual}} = \frac{1}{N}\sum_{i=1}^{N}\left\| \phi^{(l)}(x^{(i)}) - \phi^{(l)}(\hat{x}^{(i)}) \right\|_p^p$ \\
& 
\textbf{Style Loss:} 
$\mathcal{L}_{\text{style}} = \frac{1}{N} \sum_{i=1}^{N} \left\| G^{(l)}(\hat{x}^{(i)}) - G^{(l)}(x^{(i)}) \right\|_F^2$ \\

\midrule

\multirow{2}{*}{\textbf{GAN-based}} & 
\textbf{Adversarial Loss:} 
$\mathcal{L}_{\mathcal{G}} = -\frac{1}{N} \sum_{i=1}^{N} \log (\mathcal{D}(\mathcal{G}(y^{(i)})))$ \\
& 
\textbf{Cycle Consistency Loss:} 
$\mathcal{L}_{\text{cyc}} = \frac{1}{N} \sum_{i=1}^{N} \left( \lVert \mathcal{G}(\mathcal{H}(x^{(i)})) - x^{(i)} \rVert_1 + \lVert \mathcal{H}(\mathcal{G}(y^{(i)})) - y^{(i)} \rVert_1 \right)$ \\

\midrule

\multirow{4}{*}{\textbf{Prior-based}} & 
\textbf{Total Variation Loss:} 
$\mathcal{L}_{\text{TV}} = \frac{1}{N} \sum_{i=1}^{N} \frac{1}{hwd} \sum_{j,k,l} \sqrt{(\nabla_j \hat{x}^{(i)}_{j,k,l})^2 + (\nabla_k \hat{x}^{(i)}_{j,k,l})^2 + (\nabla_l \hat{x}^{(i)}_{j,k,l})^2}$ \\
& 
\textbf{Physics-based Loss:} 
$\mathcal{L}_{\text{Physics}} = \frac{1}{N} \sum_{i=1}^{N} \lVert y^{(i)} - \mathcal{D}(\hat{x}^{(i)}) \rVert_2^2$ \\
& 
\textbf{Histogram Loss:} 
$\mathcal{L}_{\text{Histogram}} = \frac{1}{N} \sum_{i=1}^{N} \delta(\text{Hist}(x^{(i)}), \text{Hist}(\hat{x}^{(i)}))$ \\
& 
\textbf{Edge Loss:} 
$\mathcal{L}_{\text{Edge}} = \frac{1}{N} \sum_{i=1}^{N} \lVert \mathcal{E}(x^{(i)}) - \mathcal{E}(\hat{x}^{(i)}) \rVert_p^p$, $p \in \{ 1, 2\}$ \\

\midrule

\textbf{Metric-based} & 
\textbf{Metric-Based Loss:} 
$\mathcal{L}_{\text{Metric}} = \frac{1}{N} \sum_{i=1}^{N} \lVert \mathcal{M}(x^{(i)}, x^{(i)}) - \mathcal{M}(x^{(i)}, \hat{x}^{(i)}) \rVert_2^2$ \\

\bottomrule
\end{tabularx}
\label{table:loss_functions}
\end{table*}

\section{Network Design}
Designing effective MRI SR systems requires coordinated choices across loss functions, architectural building blocks, and network paradigms. 
The following subsections review key loss formulations, core architectural components, and representative network families, highlighting their roles, trade-offs, and use in MRI SR.

\subsection{Loss Functions}
In SR, loss functions quantify reconstruction errors and guide model optimization. The most common are pixel- or voxel-wise $\ell_p$-norm losses, which measure pixel- or voxel-level intensity differences between reconstructed and reference images. While simple and effective, these losses often fail to capture perceptual quality or structural fidelity. Depending on the learning paradigm (e.g., supervised or unsupervised), complementary objectives have therefore been introduced, including feature-based, GAN-based, prior-informed, and metric-driven losses. These aim to enhance visual realism and anatomical accuracy. We review key loss functions here, where $x$ and $\hat{x}$ denote the ground-truth and predicted HR images, respectively. For clarity, all losses are formulated for a single image pair but computed over batches in practice (see Table~\ref{table:loss_functions}).

\label{sec:loss_function}

\subsubsection{Pixel-/Voxel-wise Loss} Pixel-/voxel-wise losses measure the discrepancy between the super-resolved image $\hat{x}$ and the ground-truth $x$ at the intensity level. Prominent losses in this category are $\ell_1$-norm, measuring the mean absolute error (MAE), the $\ell_2$-norm, measuring the mean square error (MSE), and Charbonnier loss \cite{charbonnier1994two}, a differentiable variation of the $\ell_1$ loss:
\begin{equation}
\label{eq_pl_l1}
\mathcal{L}_{\ell_1}(x, \hat{x}) = \frac{1}{hwd} \sum_{i,j,k} \left| x_{i,j,k} - \hat{x}_{i,j,k} \right|
\end{equation}

\begin{equation}
\label{eq_pl_l2}
\mathcal{L}_{\ell_2}(x, \hat{x}) = \frac{1}{hwd} \sum_{i,j,k} \left( x_{i,j,k} - \hat{x}_{i,j,k} \right)^2
\end{equation}

\begin{equation}
\label{eq_pl_Charbonnier}
\mathcal{L}_{\text{Charbonnier}}(x,\hat{x}) = \sqrt{(\mathcal{L}_{\ell_1}\big(x,\hat{x}) \big)^2 + \epsilon^2},
\end{equation}
where $h$, $w$, and $d$ represent the height, width, and depth of $x$, respectively, while $\epsilon$ is a small constant, e.g. $10^{-3}$, used to stabilize the optimization process. When comparing $\ell_1$ to $\ell_2$,  the $\ell_2$ loss penalizes larger errors more severely but is more lenient with smaller errors, yielding smoother results. In practice, the $\ell_1$ loss has consistently exhibited superior performance over the $\ell_2$ loss for different restoration tasks \cite{lim2017enhanced, zhao2016loss}.

Moreover, pixel-/voxel-wise loss functions exhibit the classical ``regression-to-the-mean'' effect \cite{delbracio2023inversion, delbracio2021projected}. 
Because $\ell_p$ norms minimize average per-pixel discrepancies, the optimal estimator under ill-posed conditions is the conditional expectation $x_{\mathrm{pred}} = \mathbb{E}[x_{\mathrm{HR}} \mid x_{\mathrm{LR}}]$, causing models trained with $\ell_p$ losses to average over plausible solutions and smooth out high-frequency details.

\subsubsection{Feature-wise Loss} Unlike voxel-/ pixel-wise losses, which aim for intensity-perfect matches between $\hat{x}$ and $x$, feature-based losses encourage them to share similar feature representations computed by a pre-trained classification network $\phi(\cdot)$, e.g., VGG~\cite{simonyan2014very}. The high-level representations extracted by such a network on the $l$-th layer are denoted as $\phi^{(l)}(\cdot)$. Two prominent categories of feature-based losses are perceptual~\cite{johnson2016perceptual, sajjadi2017enhancenet} and style~\cite{gatys2016image, wang2018recovering} loss functions.

\textbf{Perceptual Loss.} The perceptual loss, also known as content loss, is defined as the $\ell_p$-norm distance between the high-level representations of two images:
\begin{equation}
\label{eq_pl_l332}
\mathcal{L}_{\text{Perceptual}}(x,\hat{x}; \phi, l) = \mathcal{L}_{\ell_p}{\big( \phi^{(l)}(x), \phi^{(l)}(\hat{x}) \big)},
\end{equation}
Perceptual loss encourages the output image $\hat{x}$ to resemble to target image ${x}$ in terms of human perception, yielding visually more pleased results.

\textbf{Style Loss:} The style loss penalizes $\hat{x}$ when it deviates from the target image $x$ in terms of style-related attributes, i.e., colors, textures, and patterns \cite{gatys2016image}. This loss utilizes the Gram matrix $G^{(l)}_{ij}(x) = \text{vec}(\phi^{(l)}_i(x)) \cdot \text{vec}(\phi^{(l)}_j(x))$, where $\text{vec}(\cdot)$ denotes the vectorization operation, and $\phi^{(l)}_i(x)$ represents the $i$-th channel of feature maps at layer $l$ of image $x$. The style loss is defined as:
\begin{equation}
\label{eq_pl_l3322}
\mathcal{L}_{\text{style}}(\hat{x}, x;\phi,l) = \left\| G^{(l)}_{ij}(\hat{x}) - G^{(l)}_{ij}(x) \right\|_F^2,
\end{equation}
where $F$ denotes the Frobenius norm. This loss encourages the network to maintain style integrity, particularly in preserving textures, which can yield visually appealing results.

\subsubsection{GAN-based Loss}
Generative adversarial networks (GANs) are notable for their capacity to produce realistic high-quality data. GANs comprise a generator, responsible for data generation, and a discriminator, which evaluates the authenticity of the generated data by comparing it to target distribution samples.
%During training, GANs follow a two-step procedure: I) In the initial stage, the discriminator focuses on enhancing its capability to differentiate between real and generated data, while the generator remains fixed; II) In the subsequent phase, roles are reversed, as the generator refines its skill to produce data that closely mimics real examples, ultimately outsmarting the discriminator. These approaches have been used for image restoration in different ways.

\textbf{Adversarial Loss.}
In the context of super-resolution, the generator \(\mathcal{G}(\cdot)\) produced super-resolved image, and a discriminator $\mathcal{D}(\cdot)$ assesses whether the super-resolved image is real or not. This approach has been implemented in various frameworks as illustrated in \cite{ledig2017photo, jo2020investigating, schonfeld2020u, goodfellow2014generative}. A ``vanilla'' GAN involves the training of two networks, which entails alternating between optimizing two loss functions: 1) the generator loss \(\mathcal{L}_{\mathcal{G}}\), and 2) the discriminator loss \(\mathcal{L}_{\mathcal{D}}\):
\begin{align}
\mathcal{L}_{\mathcal{D}} &= -\mathbb{E}_{x}[\log (\mathcal{D}(x))] - \mathbb{E}_{y}[\log(1 - \mathcal{D}(\mathcal{G}(y)))], \notag \\
\mathcal{L}_{{\mathcal{G}}} &= -\mathbb{E}_{y}[\log (\mathcal{D}(\mathcal{G}(y)))].
\end{align}
While state-of-the-art GANs are capable of generating highly realistic details, they are susceptible to issues such as mode collapse and convergence problems, and sometimes require additional methods for stabilization \cite{moser2024diffusion}.

\textbf{Cycle Consistency Loss}: Introduced in the context of unpaired image-to-image translation within GAN frameworks \cite{zhu2017unpaired, yuan2018unsupervised}, cycle consistency loss ensures that mappings between two distinct domains are inverses of each other. Specifically, it enforces that the mapping \(\mathcal{G}: \mathbb{R}^{w \times h \times d} \rightarrow \mathbb{R}^{W \times H \times D}\) and its reverse \(\mathcal{H}: \mathbb{R}^{W \times H \times D} \rightarrow \mathbb{R}^{w \times h \times d}\) are both bijections. This consistency is promoted through forward consistency: \(x \approx \mathcal{G}(\mathcal{H}(x))\) and backward consistency: \(y \approx \mathcal{H}(\mathcal{G}(y))\). The encouragement of this behavior is enforced using a cycle consistency loss:
\begin{multline}
\mathcal{L}_{\text{cyc}}= \mathbb{E}_{x} \left[ \lVert \mathcal{G}(\mathcal{H}(x)) - x \rVert_1 \right] + \mathbb{E}_{y} \left[ \lVert \mathcal{H}(\mathcal{G}(y)) - y \rVert_1 \right] \hspace{5pt}
\end{multline}
By enforcing both forward and backward consistency, cycle consistency loss limits the range of possible mapping functions. This loss can maintain content fidelity between input and translated images, particularly for unpaired restoration tasks where ground truth is unavailable.

\subsubsection{Prior-based Loss} These losses integrate our prior knowledge about data into optimization, guiding models to meet predetermined expectations based on prior information. Depending on the formulation, they may operate independently of ground truth (e.g., TV) or in conjunction with ground truth data.

\textbf{Total Variation (TV):}
TV loss encourages spatial smoothness by penalizing large differences between neighboring pixels. It helps suppress spurious variations and artifacts in reconstructed images by promoting piecewise-smooth regions \cite{rudin1992nonlinear}. The loss is defined as:
\begin{equation}
    \mathcal{L}_{\text{TV}}(\hat{x}) = \frac{1}{hwd} \sum_{i,j,k} \sqrt{
    (\nabla_{i} \hat{x}_{i,j,k})^2 + 
    (\nabla_{j} \hat{x}_{i,j,k})^2 + 
    (\nabla_{k} \hat{x}_{i,j,k})^2 }
\label{eq:tv_loss}
\end{equation}
where $\nabla_i$, $\nabla_j$, and $\nabla_k$ denote the spatial gradients along the height, width, and depth axes, respectively. While stable in 2D, extending TV to volumetric MRI poses numerical stability challenges. Practical implementations of 3D TV for MRI SR have been developed, for example, in fetal brain imaging \cite{tourbier2015efficient}, multi-contrast brain MRI \cite{brudfors2018mri}, and more recently in deep self-supervised fMRI SR frameworks \cite{perez2025tv}.

\textbf{Physics-Based:}
These loss functions embed fundamental physical principles underlying the imaging process into the training objective. By enforcing physics-consistent constraints, they guide model outputs to remain faithful to the physical laws governing MRI acquisition, thereby narrowing the solution space and improving the realism of reconstructed images \cite{zhu2023physics, li2024enhancing}. A common approach is to formulate the loss based on the forward or degradation model:
\begin{equation}
    \mathcal{L}_{\text{Physics}} = \lVert {y} - \mathcal{H}(\hat{x}) \rVert_2^2
\end{equation}
where $y$ is the observed measurement, $\hat{x}$ is the reconstructed image, and $\mathcal{H}(\cdot)$ denotes the degradation operator~\cite{wu2025spatial}.

\textbf{Histogram:}
Histogram loss is useful in tasks such as SR, denoising, and contrast enhancement, where maintaining the statistical distribution of pixel intensities is essential for perceptual quality. By minimizing the difference between the histograms of the predicted and target images, this loss encourages the restoration of intensity patterns that closely match the ground truth, thereby enhancing visual realism and fidelity \cite{delbracio2021projected, xiao2019histogram, kim2022learning}. A generic formulation compares histogram distributions using a divergence or distance metric:
\begin{equation}
    \mathcal{L}_{\text{Histogram}} = \delta(\text{Hist}(x), \text{Hist}(\hat{x}))
\end{equation}
where $\delta(\cdot,\cdot)$ denotes a distance or divergence measure (e.g., $L_2$, KL divergence), and $\text{Hist}(\cdot)$ computes the normalized histogram of the image.

\textbf{Edge-Preserving:}
Edge-preserving losses aim to maintain sharp structural boundaries during image reconstruction. Methods based on multi-scale gradient field priors \cite{sui2019isotropic} penalize discrepancies in edge information between the predicted and target images. These losses are particularly valuable in medical imaging, where preserving anatomical features such as tissue interfaces or lesion boundaries is critical. A typical formulation compares edge maps extracted by an edge operator $\mathcal{E}(\cdot)$:
\begin{equation}
    \mathcal{L}_{\text{Edge}} = \lVert \mathcal{E}(x) - \mathcal{E}(\hat{x}) \rVert_1^1,
\end{equation}
where $\mathcal{E}(\cdot)$ computes edge features, e.g., gradients \cite{hua2023convolutional}.

\subsubsection{Metric-based}
Metric-based loss functions guide training based on perceptual or evaluation-driven criteria rather than simple pixel-wise differences. By optimizing image quality metrics such as SSIM \cite{li2024resolution}, these losses align better with human visual perception when produce HR outputs:
\begin{equation}
    \mathcal{L}_{\text{Metric}} = \lVert \mathcal{M}(x, x) - \mathcal{M}(x, \hat{x}) \rVert_2^2,
\end{equation}
where $\mathcal{M}(\cdot, \cdot)$ denotes a metric, and $\mathcal{M}(x, x)$ represents its ideal (reference) value; for example, $\mathcal{M}(x, x) = 1$ in the case of SSIM.

\subsection{Fundamental Network Architectures}

\subsubsection{Up-sampling Methods}   
\begin{figure*}[!th] 
\centering
\includegraphics[width=0.99\textwidth]{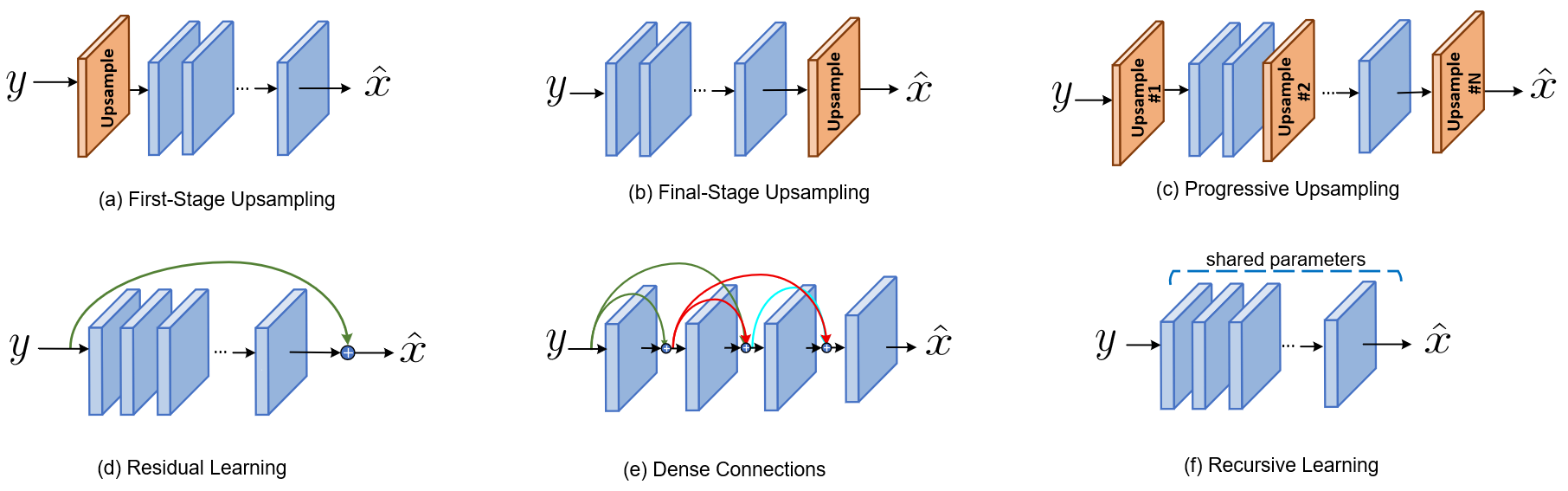}
\caption{
Overview of super-resolution architectures:  
(a) Pre-upsampling,  
(b) Final upsampling,  
(c) Progressive upsampling, 
(d) Residual learning with long skip connections,  
(e) Dense connections with efficient feature reuse,  
(f) Recursive learning with shared layers.
}
\label{fig:upsampling}
\end{figure*}

Upsampling, or upscaling, is a fundamental operation in SR tasks, aimed at increasing the spatial dimensions and enhancing the resolution of an image or feature map. Depending on the network design, upsampling can be introduced at various stages: as a first-stage pre-processing step, at the final stage to reconstruct HR outputs, or progressively within intermediate layers. An overview of these strategies is illustrated in Fig.~\ref{fig:upsampling}. Existing methods can be broadly categorized into interpolation-based techniques and learning-based approaches.

\textbf{Interpolation-Based Methods:} 
Traditional upsampling techniques such as nearest-neighbor, bilinear, bicubic, and Lanczos interpolation are widely used due to their simplicity, interpretability, and computational efficiency. Their 3D counterparts (e.g., trilinear interpolation) extend these principles to volumetric data. While easy to implement and fast, interpolation-based methods cannot recover high-frequency information and therefore tend to produce overly smooth or blurred results, particularly at large scaling factors.

% \textit{Nearest-Neighbor Interpolation:} It computes the new pixel by assigning the value of the nearest pixel to each interpolated position without considering the other pixels. This method is computationally efficient and fast; however, it typically results in blocky artifacts.

% \textit{Bilinear Interpolation:} It computes new pixel values by performing linear interpolation along both the horizontal and vertical axes, using the four nearest pixel values ($2\times2$ neighborhood). This method offers improved visual quality compared to nearest-neighbor interpolation, producing smoother transitions between pixels, while maintaining relatively fast computational performance.

% \textit{Bicubic Interpolation:} It performs cubic interpolation along both axes, utilizing a $4 \times 4$ pixel neighborhood. This method produces smoother and higher-quality results with fewer artifacts compared to bilinear interpolation. However, it is computationally slower. 

% Note that the 3D versions of these methods follow the same principles, extending the interpolation to an additional axis. For example, bilinear interpolation becomes trilinear interpolation, which operates over a $2\times2\times2$ voxel neighborhood.

% Though easy to implement and versatile, these techniques struggle to capture high-frequency details, often resulting in blurred or overly smooth outputs, especially at higher upscaling factors, limiting their performance.

\textbf{Learning-based}
Unlike interpolation-based approaches, learning-based upsampling methods leverage DL techniques to enhance feature maps. These methods are capable of recovering fine details and textures that traditional techniques struggle to capture. Standard learning-based strategies include transposed convolution, sub-pixel layers, and meta-upscaling.

\textit{Transposed Convolution:} Transposed convolutions, also referred to as deconvolutions, are commonly used in SR models to increase the spatial size of feature maps \cite{zeiler2010deconvolutional, zeiler2014visualizing}. This operation reverses the spatial effects of standard convolutions by inserting zeros between values and applying a convolution, effectively expanding feature maps to higher resolutions. Integrated into end-to-end SR models with learnable parameters, transposed convolutions enable adaptive upscaling of feature maps. However, despite their widespread use in SR tasks \cite{fu20243dattgan}, they can introduce artifacts like uneven overlaps, leading to checkerboard patterns \cite{odena2016deconvolution}.

\textit{Sub-Pixel Layer:} The sub-pixel layer upscales feature maps by generating additional channels through convolution and rearranging them to form an HR output, a process known as pixel shuffle \cite{shi2016real}. This method performs upsampling in the channel dimension, providing greater efficiency, especially with smaller kernel sizes, compared to transposed convolutions. Given a scaling factor $\tau$ and input channels $c$, the sub-pixel layer produces a feature map of size 
$h \times w \times \tau^2  c,$  which is then reshaped to  $\tau h \times \tau w \times c$ to form the upsampled output. Compared to transposed convolutions, the sub-pixel layer offers a larger receptive field, enabling the capture of more contextual information for realistic detail generation, commonly used in many SR tasks \cite{khateri2024noclean, zhao2019channel, qiu2023progressive, li2024rethinking}. However, the uneven distribution of receptive fields can introduce artifacts, such as blocky regions and unsmooth transitions.

\textit{Meta Upscale:} The meta upscale module offers flexibility by enabling arbitrary scaling factors without requiring separate models for different scales \cite{hu2019meta, deng2023meta}. Unlike traditional SR methods, which necessitate predefined scaling factors, this approach leverages meta-learning to dynamically predict convolution weights for each HR position based on small patches of the LR feature map. This allows continuous zooming and adaptive upsampling with a single model, minimizing computational overhead while maintaining high performance across various scaling factors, frequently used in MRI SR \cite{wu2022arbitrary, han2024arbitrary, tan2020arbitrary, li2023rethinking}.

Learning-based methods represent a significant improvement over traditional interpolation, providing superior performance and the ability to recover high-frequency information crucial for SR tasks.

\subsubsection{Residual Learning}
Residual learning is a key technique in SR models that focuses on learning the differences, or residuals, between the input LR and the reference HR, simplifying the task by concentrating on reconstructing high-frequency details. It can be broadly categorized into global and local residual learning. Global residual learning targets the residuals between the LR input and HR output, reducing the need to map the entire image. By focusing on residuals—most of which are near zero—this approach reduces model complexity and training difficulty \cite{kim2016accurate}, and frequently has been applied in MRI SR \cite{feng2021brain}. Local residual learning, inspired by ResNet \cite{he2016deep}, introduces shortcut connections within model layers to prevent gradient vanishing and degradation, allowing deeper models to effectively learn complex features. Both global and local residual learning typically use shortcut connections, often combined through element-wise addition, though concatenation is also possible. Global residual learning enhances high-frequency detail reconstruction, while local residual learning enables deeper models to manage more complex transformations. The integration of both has become a widely adopted approach for improving performance in DL-based SR tasks, including MRI SR \cite{han2023prostate, xue2019progressive, du2020super}.

\subsubsection{Dense Connections}
Dense connections link each layer within a dense block to all the previous layers, while simultaneously passing its output forward to subsequent layers. This creates a highly interconnected architecture that enhances feature reuse and improves gradient flow. By leveraging growth rate optimization and techniques like 1x1 convolutions for channel compression, dense connections not only reduce the model size but also help prevent overfitting. Despite some computational overhead, they significantly improve image quality by fusing low- and high-level features, leading to superior reconstruction of fine details \cite{tong2017image}. Dense connections are a highly effective strategy for optimizing SR tasks, particularly in high-precision applications like MRI SR \cite{chen2018efficient, wang2020enhanced}.

\subsubsection{Recursive Learning}
Recursive learning enables SR models to capture higher-level features while keeping the number of parameters low by reapplying the same module multiple times \cite{kim2016deeply, tai2017image}. Although recursive learning is effective in enhancing feature extraction, it can be computationally intensive and susceptible to challenges such as vanishing or exploding gradients. Techniques like residual learning are commonly employed to mitigate these issues \cite{shi2018mr}.

\subsubsection{Attention Mechanisms}
Attention modules enhance SR by steering the network toward the most informative cues across channel, spatial, temporal, and relational domains.

\textbf{Channel Attention:} 
Channel attention re-weights feature maps along the channel dimension so that informative filters contribute more to the reconstruction. Residual channel attention network (RCAN) \cite{zhang2018image} popularized this idea in SR. The squeeze-and-excitation (SE) block of Hu et al. \cite{hu2018squeeze}, later adapted to MRI SR \cite{zhang2021mr}, captures global channel statistics via pooling, while second-order channel attention (SOCA) leverages covariance information for finer discrimination \cite{dai2019second}.

\textbf{Spatial Attention:} 
Spatial attention focuses on highlighting important regions or pixels in an image, directing the model’s attention to spatially relevant areas to enhance performance \cite{woo2018cbam, liu2020residual}.

\textbf{Spatial and Channel Attention:} 
Joint spatial–channel modules let a network attend to \emph{where} (spatial mask) and \emph{what} (channel mask) simultaneously. Holistic Attention Network (HAN) \cite{niu2020single} pioneers this design in SR. Transformer-style self-attention, adopted in SwinIR \cite{liang2021swinir}, further models long-range context while preserving local detail. Recent work augments these blocks with multi-scale kernels and gating, enabling coarse-to-fine feature re-calibration and suppressing block artefacts \cite{chen2023dual, wang2024multi}.

\textbf{Spatial-Temporal Attention:} 
Spatio-temporal blocks jointly attend to \emph{where} (spatial saliency) and \emph{when} (frame-to-frame coherence), making them pivotal for video SR \cite{fu20243dattgan} and dynamic MRI. STADNet \cite{lyu2024stadnet} illustrates the concept: a location-aware spatial branch reuses neighbouring frames for detail, while a motion-aware temporal branch, guided by optical flow, aligns dynamics. The fused attention improves anatomical fidelity and temporal consistency, reducing blur in cardiac cine MRI.

\textbf{Self-Attention:}  
Self-attention models long-range dependencies by allowing each position in a feature map to interact with all others, thereby aggregating richer contextual information than local convolutions. It was first introduced to SR in natural images through Transformer-based architectures such as IPT \cite{chen2021pre} and SwinIR \cite{liang2021swinir}, which demonstrated strong capability in recovering fine textures and global structures. To address the computational burden of full attention, hierarchical and window-based variants have been proposed. In MRI SR, Transformer-based attention mechanisms have been increasingly adopted, for example, in arbitrary-scale reconstruction \cite{han2024arbitrary} and multi-contrast SR \cite{lyu2023multicontrast}, where they enhance structural fidelity by capturing non-local anatomical relationships.

\textbf{Cross-Attention:} 
Cross-attention enables one feature stream to query another, fusing complementary cues from inputs with differing resolutions or modalities \cite{jiang2022super}. In multi-contrast MRI SR, DCAMSR \cite{huang2023accurate} uses dual cross-attention to align T1- and T2-weighted features, while MS2CAM \cite{hu2025ms2cam} introduces a multi-scale self-cross-attention block that exchanges information across spatial scales, producing sharper and more consistent reconstructions.

\textbf{Graph Attention:} 
Graph attention represents voxels (or patches) as nodes and learns edge weights to aggregate non-local context, making it well-suited to multi-contrast MRI SR. GraphSR \cite{hua2024multi} builds spatially aware graphs to fuse complementary cues across contrasts, sharpening inter-regional detail. A lightweight successor, TDAFD \cite{zhao2025lightweight}, couples dual-attention feature distillation with a recursive volumetric Transformer, achieving state-of-the-art performance with only $1$ M parameters.

\subsection{Networks}
Network architectures play an important role in MRI SR by directly mapping LR inputs to HR outputs. Since the introduction of CNNs to image restoration tasks, architectural advancements such as U-Net \cite{ronneberger2015u}, ResNet \cite{he2016deep}, and DenseNet \cite{huang2017densely} have significantly improved SR performance. These models enhance the preservation of fine anatomical details, reduce artifacts, and maintain structural consistency, which is critical in clinical imaging.
CNN-based architectures offer a strong balance between flexibility and computational efficiency, making them foundational in early and current MRI SR pipelines. As deep learning evolves, innovations have extended beyond traditional CNNs. Attention-based mechanisms and Transformer architectures, such as vision Transformers (ViTs) and SwinIR \cite{liang2021swinir}, improve the modeling of long-range dependencies and spatial precision. In parallel, generative AI approaches—particularly those based on GANs \cite{delannoy2020segsrgan} and diffusion models \cite{saharia2022image}—have further advanced perceptual quality and anatomical plausibility. These models synthesize realistic HR outputs from degraded inputs, often guided by structural or modality-specific priors.
The following sections provide an overview of key network paradigms that have driven progress in MRI SR, highlighting their unique contributions and trade-offs.

\subsubsection{Convolutional Neural Networks (CNNs)}
The use of CNNs in SR was first introduced with SRCNN \cite{dong2015image}, which established an end-to-end learning framework that outperformed traditional interpolation and sparse-coding approaches in both reconstruction accuracy and scalability. This marked a turning point for data-driven SR methods, including in medical imaging. Early applications to MRI SR followed soon after. Pham et al.~\cite{pham2017brain} proposed one of the first deep residual CNNs for enhancing brain MRI, while Bahrami et al.~\cite{bahrami2016convolutional} employed a patch-based 3D CNN to reconstruct 7T-like brain MRI from 3T scans.
Building on these foundations, more advanced CNN architectures were adapted for MRI SR. U-Net \cite{ronneberger2015u}, originally proposed for biomedical segmentation, has been widely adopted for SR tasks, where its encoder–decoder design and skip connections enable effective multi-scale feature extraction and detail preservation \cite{zhou2022super}. ResNet \cite{he2016deep} addressed the vanishing gradient problem and allowed deeper, more stable networks, which were leveraged in models such as DeepResolve \cite{chaudhari2018super} for musculoskeletal MRI. DenseNet \cite{huang2017densely} improved parameter efficiency through dense connectivity and has inspired residual-dense architectures for brain MRI SR \cite{zhu2021residual}. 
These architectural innovations advanced MRI SR by improving tissue sharpness and reducing noise and aliasing artifacts \cite{zhao2020smore}. For instance, BrainSR \cite{du2020brain} leverages dilated convolutions to expand the receptive field without increasing the number of parameters, thereby enabling richer contextual modeling for fine anatomical detail recovery. SpatioResNet \cite{lyon2024spatio} further extends CNN-based designs by integrating spatio-temporal residual learning, facilitating effective multi-contrast, 3D, and dynamic MRI SR. Collectively, these developments reinforce the central role of CNNs in state-of-the-art MRI SR pipelines.

\subsubsection{Long-Range Dependency Models} 
Methods that capture long-range dependencies play a key role in SR by going beyond local convolutional receptive fields. These can be grouped into three main families:

\textbf{Vision Transformers (ViTs):}  
ViTs rely on self-attention to explicitly model pairwise interactions across the entire feature map. SwinIR \cite{liang2021swinir}, for example, introduced hierarchical, window-based attention to reduce computational complexity while preserving global context. In MRI SR, ViT-style modules have been integrated for multi-contrast fusion \cite{lyu2023multicontrast}, cross-scale SR \cite{huang2023accurate}, and brain MRI \cite{liu2024mapanet}, demonstrating improved recovery of fine structures.

\textbf{State-Space Models (SSMs):}  
Recent alternatives to Transformers, such as Mamba \cite{ji2024deform, ji2024self}, replace explicit attention with structured state-space dynamics. These achieve linear scaling with sequence length, offering efficient global context modeling in MRI SR while maintaining competitive accuracy.

\textbf{Hybrid RNN–Attention Models:}  
RWKV \cite{yang2025restore} combines recurrent dynamics with attention-like updates, enabling efficient modeling of long sequences while retaining Transformer-level expressivity. In MRI SR, such hybrids improve robustness to noise and motion while lowering computational costs compared to full attention.

\textbf{Integration with CNNs:}  
In practice, long-range dependency modules are almost always combined with convolutional backbones. CNNs contribute strong local feature extraction and \textit{spatial inductive biases}—built-in assumptions about locality, translation invariance, and spatial structure—that ViTs, SSMs, and hybrids do not inherently provide. When used as modular components, long-range mechanisms complement CNNs by capturing global context while CNNs preserve fine local detail. This combination improves SR fidelity without substantially increasing architectural complexity, making it a dominant design choice in MRI SR, where anatomical accuracy must be balanced with computational efficiency.

\subsubsection{Generative Adversarial Networks (GANs)}
\label{GANs}
GANs consist of two components, a generator $\mathcal{G}(\cdot)$ and a discriminator $\mathcal{D}(\cdot)$, trained in a minimax game. The generator synthesizes HR images, while the discriminator learns to distinguish them from ground truth. This adversarial setup encourages perceptually realistic reconstructions, addressing the limitations of pixel-wise regression losses. The first application to SR was SRGAN \cite{ledig2017photo}, which combined adversarial and perceptual content losses to recover fine textures beyond conventional methods. ESRGAN \cite{wang2020enhanced} further improved visual fidelity through residual-in-residual dense blocks and a relativistic discriminator, setting new benchmarks in natural image SR.
Building on these advances, GANs have been widely adapted for MRI SR, where anatomical fidelity is critical. SOUP-GAN \cite{zhang2022soup} targeted through-plane slice-thickness SR using a perceptually tuned adversarial loss, enabling higher-quality reconstructions. DISGAN \cite{wang2023disgan} extended this to 3D volumetric data by introducing a wavelet-domain discriminator and residual-in-residual dense generators, jointly addressing denoising and SR with improved generalization across contrasts. To incorporate frequency priors, Xiao et al.~\cite{xiao2023novel} proposed a hybrid frequency-aware GAN that fused Fourier-domain processing with a complex-valued residual U-Net, achieving strong performance in both MRI and CT. More recent efforts integrate temporal and Transformer modules: for example, LSTM-Attention-GAN \cite{zhang2025super} combines recurrent units with attention mechanisms to improve temporal consistency in sequential brain and lung MR scans.
Overall, GAN-based approaches have substantially advanced MRI SR by enhancing perceptual realism and structural detail. However, they remain challenged by training instability, mode collapse, and computational overhead. Stabilization strategies—such as perceptual loss regularization, spectral normalization, and multi-task discriminators—are commonly employed to mitigate these issues.

\subsubsection{Diffusion Models (DMs)}
Diffusion models (DMs) were first applied to image SR in SR3 \cite{saharia2022image}, which demonstrated that iterative denoising can yield high-quality, perceptually realistic HR reconstructions. Unlike GANs, which generate images in a single forward pass, DMs synthesize outputs by reversing a stochastic diffusion process—gradually removing noise over multiple steps. This iterative design improves stability, controllability, and robustness, properties particularly valuable when structural fidelity is critical.
Recent work has adapted DMs for MRI SR with promising results. InverseSR \cite{wang2023inversesr} leverages a latent diffusion prior trained on UK Biobank brain MRI and introduces two inversion strategies: decoder-based and denoising diffusion implicit model (DDIM)-based, enabling flexible HR reconstruction across varying sparsity levels. To accelerate inference, partial diffusion models (PDMs) \cite{qu2024solving} proposed a latent alignment mechanism that reduces the number of sampling steps while maintaining accuracy. Building on the ResShift paradigm \cite{yue2023resshift}, Res-SRDiff \cite{safari2025mri} incorporated residual shifting to align voxel distributions between LR and HR MR volumes, achieving high-quality reconstructions in as few as four steps on ultra-high-field brain and prostate MRI datasets. While diffusion models are computationally demanding, innovations such as latent-space truncation, residual alignment, and accelerated sampling are making them increasingly viable. Compared to adversarial methods, DMs offer more stable training and stronger generalization in the presence of noise, motion, or low SNR, positioning them as a compelling alternative—or complement—to GANs in next-generation MRI SR.

\subsubsection{Implicit Neural Representations (INRs)}
INRs model data as continuous functions that map spatial coordinates—and optionally auxiliary features—directly to signal intensity values. Unlike conventional discrete grid representations, INRs encode entire images or volumes using neural networks (typically multi-layer perceptrons, MLPs) that learn the underlying signal as a continuous function:
\begin{equation}
    x_i = \mathcal{M}_\theta(\mathbf{c}_i, \mathbf{f}_i),
\end{equation}
where \( \mathbf{c}_i \in \mathbb{R}^{d} \) denotes the spatial coordinate (normalized to \( [0, 1] \) or \( [-1, 1] \)), \( \mathbf{f}_i \in \mathbb{R}^f \) is an optional feature vector (e.g., contrast type, modality, or subject embedding), and \( \mathcal{M}_\theta \) is the learnable mapping parameterized by \( \theta \). This formulation enables HR image reconstruction by querying the network at arbitrary coordinate locations, and allows the model to jointly leverage spatial and contextual information during optimization.

\begin{figure}[!t]
\centering
\includegraphics[width=0.49\textwidth,height=0.23\textwidth]{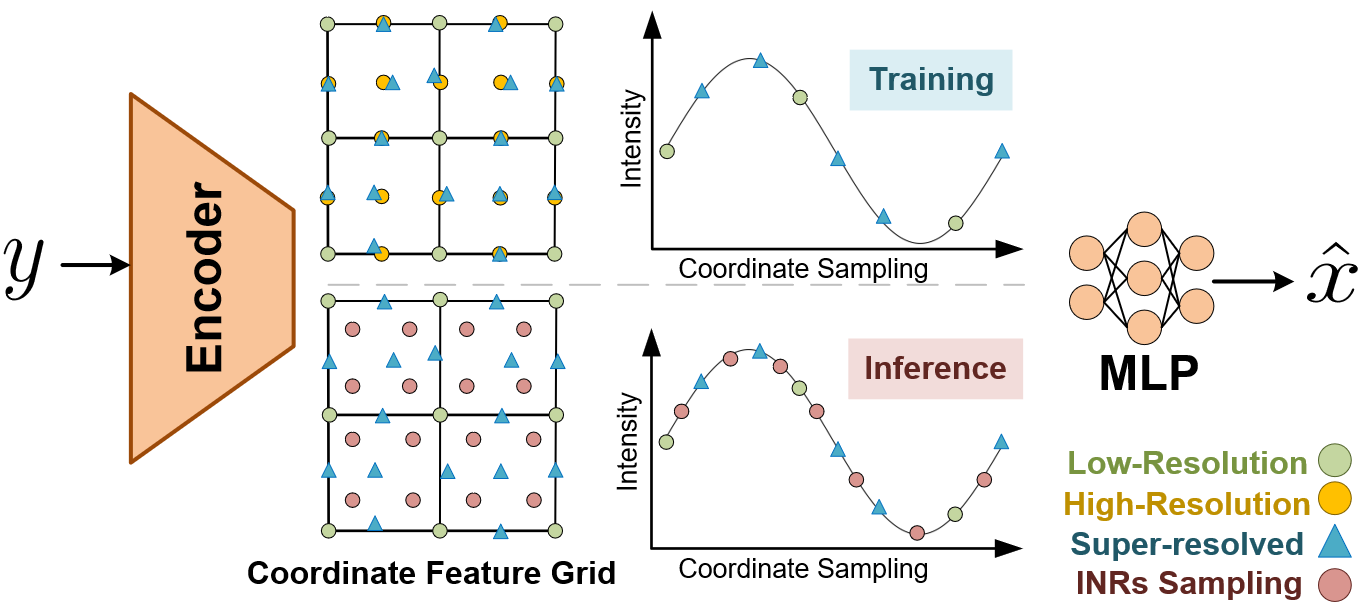}
\caption{Illustration of INR-based MRI super-resolution. An encoder maps the input image to a coordinate feature grid, from which spatial coordinates and corresponding features are sampled. During training, the network learns a continuous mapping from coordinates and features to signal intensities. At inference, the trained model can be queried at arbitrary coordinates to generate HR outputs. Inspired from \cite{li2024enhance}}
\label{Fig_INR}
\end{figure}

INRs offer several advantages for SR tasks: They eliminate the need for fixed-scale upsampling modules, reduce memory consumption, and naturally enable continuous spatial interpolation \cite{liu2024arbitrary}. In the context of MRI SR, INR-based methods have been effectively employed to model both individual slices and full volumetric data with high spatial fidelity.

Recent studies have significantly expanded the use of INRs across various MRI SR tasks. One of the earliest applications demonstrated volumetric SR via continuous coordinate-based representations~\cite{wu2022arbitrary}. INRs have since been employed in SVR, as in NeSVoR~\cite{xu2023nesvor}, and further enhanced through meta-learning strategies to improve cross-subject generalization in SVR settings~\cite{dannecker2025meta}. 
In cardiac MRI and 4D flow MRI, INR-based methods have enabled joint super-resolution and denoising, improving spatiotemporal coherence and data fidelity~\cite{lyu2025diffusion, saitta2024implicit}. For multi-contrast MRI, INRs have been used to learn continuous cross-contrast representations, enabling enhanced synthesis and preservation of contrast-specific features~\cite{wei2025multi, mcginnis2023multicontrast, mcginnis2023single, liu2025diff}. Personalized, patient-specific INR frameworks have also been introduced to capture individual anatomical variability~\cite{li2025patient}.
Cycle-INR~\cite{fang2024cycleinr} recently proposed a novel volumetric SR architecture incorporating local attention-based grid sampling and a cycle-consistency loss to address over-smoothing and inter-slice inconsistencies common in anisotropic 3D medical images. Furthermore, \cite{wu2025spatial} combined INRs with spherical harmonics to model continuous spatial-angular domains for SR of diffusion MRI, advancing representation fidelity in both spatial and directional dimensions.

\subsubsection{Gaussian Splatting (GS)}
\begin{figure}[!t]
    \centering
    \includegraphics[width=0.46\textwidth]{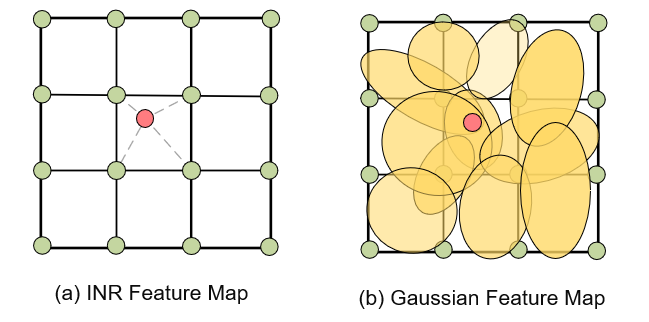}
    \caption{Comparison of feature representations between INR and GS. 
    (a) INRs model pixels as discrete point samples; 
    (b) GS represents each pixel as a self-adaptive continuous Gaussian field, allowing smooth and explicit evaluation of field values at arbitrary query locations~\cite{hu2025gaussiansr}.
    }
    \label{Fig_GS}
\end{figure}

GS models images as continuous fields by representing each pixel as a spatially adaptive Gaussian rather than a discrete sample. While grid-based methods such as INRs rely on coordinate-to-intensity mappings over fixed grids, GS captures local spatial continuity and uncertainty through overlapping Gaussian functions~\cite{hu2025gaussiansr, chen2025generalized}, producing smoother and potentially more detailed reconstructions (Fig.~\ref{Fig_GS}).

The signal intensity at a coordinate $\mathbf{p}$ is defined as:
\begin{equation}
    x_i = \mathcal{M}_\theta(\mathbf{c}_i, \mathbf{f}_i \mid \boldsymbol{\Sigma}_i, \xi_i),
\end{equation}
where $\mathbf{c}_i$ denotes the spatial coordinate, $\mathbf{f}_i$ the associated feature, $\boldsymbol{\Sigma}_i$ the covariance defining spatial extent, and $\xi_i$ the opacity controlling contribution strength. The Gaussian weighting function is expressed as:
\begin{equation}
    G_i(\mathbf{p} \mid \mathbf{c}_i, \boldsymbol{\Sigma}_i) =
    \frac{1}{2\pi |\boldsymbol{\Sigma}_i|^{1/2}}
    \exp\!\left[-\tfrac{1}{2}(\mathbf{p}-\mathbf{c}_i)^{T}
    \boldsymbol{\Sigma}_i^{-1}(\mathbf{p}-\mathbf{c}_i)\right],
\end{equation}
and the aggregated continuous representation is obtained as:
\begin{equation}
    F(\mathbf{p}) = \sum_i \sigma(\xi_i)\, G_i(\mathbf{p} \mid \mathbf{c}_i, \boldsymbol{\Sigma}_i)\, f_i,
\end{equation}
where $\sigma(\cdot)$ constrains opacity to $[0,1]$.

This formulation enables differentiable rendering, adaptive interpolation, and interpretable feature modeling. In super-resolution, GS provides a principled bridge between discrete sampling and continuous signal representation, achieving high-fidelity reconstruction across arbitrary spatial scales~\cite{hu2025gaussiansr}.
While GS has been applied to MRI reconstruction tasks~\cite{peng2025three}, it has not yet been introduced for MRI SR.

\subsubsection{Recurrent Neural Networks (RNNs)} 
RNNs, particularly long short-term memory (LSTM) and convolutional LSTM (ConvLSTM) architectures, have been employed in SR tasks where temporal or sequential dependencies are critical. Initially popularized in video SR \cite{li2019deepvolume}, these models capture correlations across frames to improve temporal consistency. Similar strategies have been extended to dynamic MRI, where temporal coherence is equally important. For example, Basty et al.\cite{basty2018super} applied SR to cardiac cine MRI, highlighting the benefits of modeling sequential dynamics. Qin et al.\cite{qin2018convolutional} used ConvLSTM blocks in interventional MRI, leveraging frame-to-frame coherence for real-time, high-fidelity reconstruction. More recently, Chatterjee et al.\cite{chatterjee2024ddos} introduced DDoS-UNet, a dual-channel recurrent UNet that conditions each frame on a high-resolution planning scan and the previous SR output, enabling temporally consistent enhancement of fetal and abdominal MRI. Similarly, Xu et al.\cite{xu2021stress} proposed STRESS, a self-supervised spatio-temporal SR framework that learns spatial and temporal dependencies without paired HR data, improving consistency in dynamic fetal MRI. These RNN-based approaches are particularly valuable in motion-sensitive or time-resolved MRI, where they enhance temporal continuity, reduce frame-to-frame artifacts, and improve overall reconstruction quality.

\subsubsection{Graph Convolutional Networks (GCNs)}  
GCNs extend convolution to graph-structured data, enabling the modeling of non-local anatomical relationships that conventional CNNs often miss. In MRI SR, nodes can represent voxels, patches, or regions, with edges encoding spatial or structural similarity, thereby integrating long-range context to improve structural fidelity and reduce artifacts. Isallari and Rekik \cite{isallari2021brain} proposed GSR-Net, a graph U-Net for brain connectivity graph super-resolution, demonstrating the feasibility of graph-based learning in neuroimaging. Ma and Cui \cite{ma2021hybrid} introduced a hybrid CNN–GCN framework for diffusion MRI, where CNNs captured local spatial features while GCNs modeled non-local relationships in the diffusion-encoding domain. More recently, Ma et al.~\cite{ma2024mri} developed GCESS, a hybrid network combining convolution and patch-based graph convolution to exploit self-similarity, effectively suppressing artifacts and enhancing detail in MRI reconstructions. While still less common than CNN- or GAN-based approaches, GCNs offer a principled means of embedding anatomical priors into SR pipelines.

\section{Foundation Models}
\label{sec:fm_mri_sr}
Foundation models (FMs) are large-scale models trained on diverse, heterogeneous datasets with the capacity to generalize across a wide spectrum of tasks. They can be efficiently adapted to downstream applications through fine-tuning, transfer learning, or prompting~\cite{bommasani2021opportunities}. In contrast to conventional MRI SR approaches—often designed for a single modality, anatomy, or acquisition protocol—FM-based methods exploit broad prior knowledge and rich feature representations learned from large-scale data. This confers improved adaptability, robustness, and scalability. Within medical imaging, FMs hold particular promise for MRI SR by enabling the incorporation of anatomical priors, semantic conditioning, and multimodal information (e.g., across imaging contrasts or between imaging and text), thereby unifying reconstruction, interpretation, and task-specific reasoning within a single framework.

\subsection{Recent Foundation Models for MRI Super-Resolution}
We highlight two representative FM paradigms for MRI SR, which are (i) anatomically guided multi-task enhancement and (ii) Graph-Based Multimodal Semantic Modeling.

\subsubsection{Anatomically Guided Multi-Task Enhancement (BME-X)}
BME-X~\cite{sun2025foundation} exemplifies the anatomically guided FM paradigm. A two-stage pipeline first infers soft tissue labels from low-quality inputs and then reconstructs high-resolution images conditioned on these priors, promoting anatomically plausible detail across motion correction, denoising, harmonization, and SR. Supervision is enabled via synthetic pairing: segmentation maps synthesize clean HR images, to which realistic degradations (e.g., downsampling, noise, motion blur, intensity bias) are applied to form LR counterparts with tight alignment. Trained on more than 13{,}000 brain MRI volumes from 19 public datasets spanning fetal to adult cohorts, BME-X demonstrates robustness to motion-corrupted infant scans and cross-domain scenarios (e.g., 3T$\rightarrow$7T SR). By coupling tissue-aware conditioning with controlled data generation it improves SR fidelity and yields more consistent downstream segmentation and registration.

\subsubsection{Graph-Based Multimodal Semantic Modeling (GraphMSR)}
In contrast to the anatomically guided paradigm of BME-X, GraphMSR~\cite{qin2025graphmsr} approaches MRI SR through graph-based multimodal reasoning. It constructs adaptive graphs over image patches, where nodes represent local anatomical regions and edges capture contextual or structural relationships. A hierarchical graph attention mechanism then models both local detail and global topology. Distinctively, GraphMSR integrates semantic priors derived from a large medical language model: clinically relevant descriptors (e.g., anatomical regions, quality attributes, pathological terms) are encoded as text embeddings and fused with image features via cross-attention. This allows the model to emphasize diagnostically important regions while suppressing artefactual content during reconstruction.

\subsection{Comparison and Direction} 
BME-X and GraphMSR represent two distinct FM strategies for MRI SR: one relies on explicit tissue priors for broad enhancement, while the other uses semantic–graph reasoning to emphasize perceptual and diagnostic fidelity. Both highlight the unifying potential of foundation models. Table~\ref{tab:foundation_models_comparison} summarizes their design and capabilities. It should also be noted, however, that reliance on foundation models carries a broader concern: errors or biases present in a base model may be inherited by all downstream models that build upon it, leading to widespread propagation of the same flaws.

\begin{table*}[ht]
\centering
\caption{Comparison of MRI Foundation Models for Super-Resolution}
\label{tab:foundation_models_comparison}
\begin{tabular}{lcc}
\toprule
\textbf{Aspect} & \textbf{BME-X}~\cite{sun2025foundation} & \textbf{GraphMSR}~\cite{qin2025graphmsr} \\
\midrule
Architecture & Tissue-aware dual CNN & Graph neural network with attention \\
Knowledge Priors & Tissue segmentation maps & Semantic prompts (e.g., text embeddings) \\
Multimodal Input & No & Yes (text + image) \\
Enhancement Tasks & SR, denoising, motion correction, harmonization & SR (spatial detail and perceptual enhancement) \\
Generalization Scope & Lifespan, scanners, pathologies & Diverse anatomical regions and LR or artifact-affected MRIs \\
Downstream Tasks & Segmentation, registration, diagnosis & Perceptual SR, diagnostic fidelity \\
\bottomrule
\end{tabular}
\end{table*}

\section{Generative AI}
\label{generative_ai}

Generative artificial intelligence (GenAI) has emerged as a transformative paradigm for MRI SR, enabling the synthesis of HR images by modeling the underlying distribution of anatomical structures. Beyond enhancing perceptual quality, GenAI approaches can integrate prior knowledge and operate in paired or unpaired learning settings.  

Prominent GenAI frameworks include GANs, variational autoencoders (VAEs), diffusion models (DMs), and normalizing flows (NFs). Each has distinct strengths and limitations: GANs produce sharp, perceptually realistic outputs but may introduce hallucinations and suffer from training instability; VAEs capture uncertainty and enable controllable generation but often yield overly smooth textures; NFs focus on density estimation with diverse sampling but require deep, memory-intensive architectures, making them less practical for large-scale applications; and, DMs achieve state-of-the-art fidelity and robustness, albeit with high computational cost;
Among these, diffusion models have rapidly gained prominence for MRI SR due to their ability to model complex anatomical distributions, support flexible conditioning, and integrate seamlessly into inverse problem frameworks without retraining.

\subsection{Diffusion Models (DMs) for Super-Resolution} 
DMs are generative frameworks that generate high-quality data by reversing a gradual noising process~\cite{ho2020denoising, he2025diffusion}. Originally developed for image generation, they now achieve state-of-the-art performance in image synthesis \cite{dayarathna2024deep}, and inverse imaging problems like SR~\cite{ saharia2022image}. DMs model complex data distributions and produce high-quality outputs through iterative refinement. All variants share a two-stage structure:

\begin{itemize}
    \item[(I)] \textbf{Forward process}: Gradually corrupts a clean sample \(x_0\) into noisy samples \(x_t\) over a sequence of timesteps \(t = 0, \ldots, T\), eventually producing the final noisy sample \(x_T\).

    \item[(II)] \textbf{Reverse process}: A neural network learns to denoise \( x_T \), step by step, to recover the original data distribution. In SR, this process is conditioned on the observed LR image \( y \) to ensure fidelity to the input.
\end{itemize}

\begin{figure}[!th]
\centering
\includegraphics[width=0.49\textwidth]{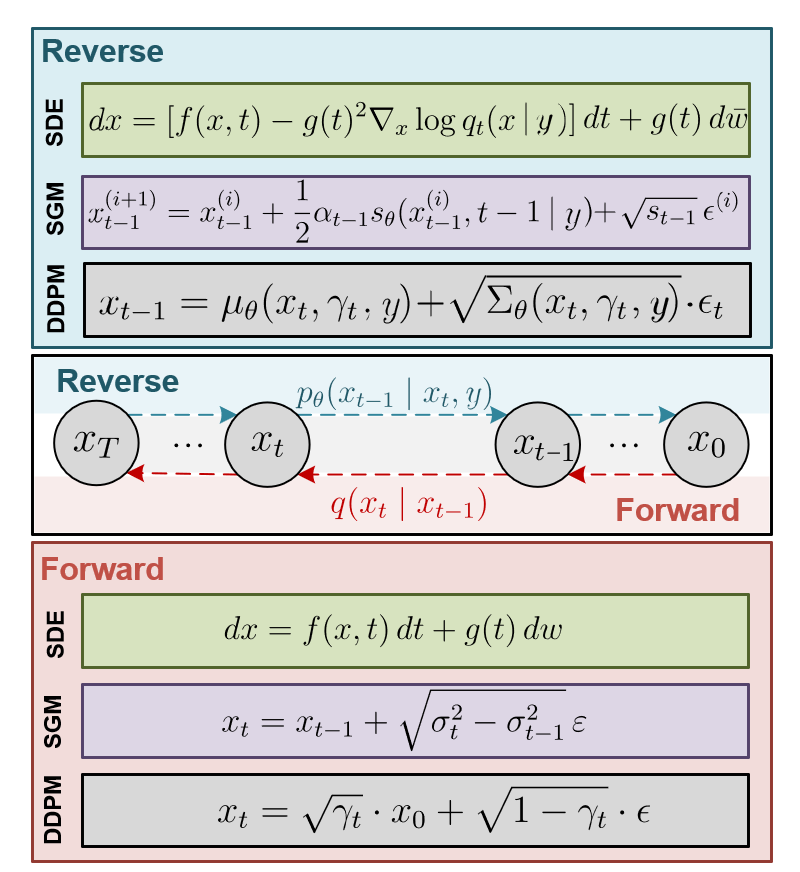}
\caption{
Summary of forward and reverse processes in three major diffusion model formulations: DDPMs, SGMs, and SDE-based models. 
The bottom row illustrates the forward processes (in red), where clean data \( x_0 \) is progressively corrupted using Gaussian noise or stochastic dynamics. 
The top row shows the corresponding reverse processes (in blue): DDPMs perform noise prediction and Gaussian denoising; SGMs apply Langevin dynamics based on learned score functions; and SDE-based models use reverse-time trajectories guided by continuous-time score estimates. 
Adapted from \cite{moser2024diffusion}.}
\label{fig_diffusion_formulations}
\end{figure}

There are several formulations of DMs used in MRI SR, including denoising diffusion probabilistic models (DDPMs)~\cite{ho2020denoising}, score-based generative models (SGMs)~\cite{song2019generative}, stochastic differential equation (SDE)–based diffusion models~\cite{song2020score}, denoising diffusion implicit models (DDIMs)~\cite{song2020denoising}, and latent diffusion models (LDMs)~\cite{rombach2022high}—each with distinct mathematical formulations and inference procedures. Their respective forward and reverse processes are summarized in Fig.~\ref{fig_diffusion_formulations} to highlight the connections with the models discussed below.

\subsubsection{Denoising Diffusion Probabilistic Models (DDPMs)}

DDPMs model generation as the reversal of a fixed noising process. The forward step adds Gaussian noise to a clean image \( x_0 \):
\begin{equation}
q(x_t \mid x_{t-1}) = \mathcal{N}(x_t;\, \sqrt{1 - \alpha_t} x_{t-1},\, \alpha_t \mathbf{I}),
\end{equation}
with the closed-form:
\begin{equation}
x_t = \sqrt{\gamma_t} x_0 + \sqrt{1 - \gamma_t} \epsilon,\quad \epsilon \sim \mathcal{N}(0, \mathbf{I}).
\end{equation}

\noindent The reverse process is modeled as:
\begin{equation}
p_\theta(x_{t-1} \mid x_t, y) = \mathcal{N}(x_{t-1};\, \mu_\theta(x_t, y, t),\, \Sigma_\theta(x_t, y, t)),
\end{equation}
where a neural network predicts noise-conditioned parameters using the LR input \( y \). Instead of recovering \( x_0 \) directly, the network estimates the noise:
\begin{equation}
\mathcal{L}_{\mathrm{CDM}} = \mathbb{E}_{x_0, \epsilon, t} \left[ \left\| \epsilon - \epsilon_\theta(x_t, t \mid y) \right\|^2 \right].
\end{equation}
This loss simplifies training and improves sampling. Conditioning on \( y \) enforces consistency with the measured data.

In MRI SR, conditional DDPMs have been adapted for various reconstruction tasks. Wu~\etal~\cite{wu2023super} proposed a brain MRI SR framework improving structural fidelity, later extending it with multi-scale self-attention, channel reduction for faster training, and achieving state-of-the-art perceptual quality on T1- and diffusion-weighted MRI. Zhao~\etal~\cite{zhao2024mri} introduced the partial diffusion model (PDM), which initializes from an LR-derived intermediate to cut denoising steps and inference time without quality loss. Ma~\etal~\cite{ma2025diffusion} developed a cross-attention-based DDPM that enhances fidelity to LR inputs while preserving fine anatomical details, and Qiu~\etal~\cite{qiu2024dbsr} proposed DBSR, a quadratic conditional diffusion model for blind cardiac MRI SR that predicts blur kernels from LR inputs and incorporates them into reconstruction via a cascaded residual attention network to improve robustness against motion blur.

For multi-contrast SR, Chang~\etal~\cite{chang2024high} leveraged conditional diffusion priors for detail restoration in MRI synthesis. Li~\etal~\cite{li2024rethinking} proposed DiffMSR, performing diffusion in a compact latent space to produce high-frequency priors decoded by a prior-guided large window Transformer (PLWformer) for efficient reconstruction. Chen~\etal~\cite{chen2025edge} introduced Eg-Diff, an edge-guided conditional diffusion model that injects contrast-adaptive edge features and applies adaptive multi-modality fusion to preserve anatomical boundaries. Mao~\etal~\cite{mao2023disc} presented DisC-Diff, a disentangled conditional diffusion framework that separates shared and modality-specific features, enhancing robustness and fidelity across contrasts.

\subsubsection{Score-Based Generative Models (SGMs)}
SGMs synthesize data by following the score function, i.e., the gradient of the log-density \( \nabla_{x_t} \log p_t(x_t) \) \cite{song2019generative}. The forward process perturbs clean data with Gaussian noise:
\begin{equation}
x_t = x_0 + \sigma_t \epsilon, \quad \epsilon \sim \mathcal{N}(0, \mathbf{I}),
\end{equation}
while sampling is performed via the reverse-time SDE:
\begin{equation}
dx_t = [f(x_t, t) - g^2(t)\nabla_{x_t} \log p_t(x_t)]\,dt + g(t)\,d\bar{w}_t,
\end{equation}
where the score \( \nabla_{x_t} \log p_t(x_t) \) is approximated by a neural network \( s_\theta(x_t, t \mid y) \), conditioned on the observed LR image.

Training is done via denoising score matching, where the model learns to approximate the true score from noisy inputs:
\begin{equation}
\mathcal{L}_{\mathrm{DSM}} = \mathbb{E}_{x_0, \epsilon, t} \left[ \left\| s_\theta(x_t, t \mid y) + \frac{1}{\sigma_t} \epsilon \right\|^2 \right].
\end{equation}

SGMs allow flexible sampling using Langevin dynamics or probability-flow ODEs. Conditioning on \( y \) ensures that the reconstructed HR outputs are consistent with the measured LR data.
Liu~\etal~\cite{liu2025score} addressed the challenge of limited cardiac MR data with tailored score learning. Cao~\etal~\cite{cao2024high} focused on high-frequency recovery to accelerate sampling. Wu~\etal~\cite{wu2023wavelet} employed wavelet-domain decomposition for artifact suppression, while Hou~\etal~\cite{hou2023fast} proposed FRSGM, integrating ensemble denoisers and spatial self-consistency for fast, robust MRI reconstruction.

\subsubsection{Stochastic Differential Equation (SDE)-Based Diffusion Models}

SDE-based models generalize SGMs by formulating the forward process as a continuous-time stochastic differential equation \cite{song2020score}:
\begin{equation}
dx = f(x, t)\,dt + g(t)\,dw_{t},
\end{equation}
where \( f \) is the drift term, \( g \) controls the noise scale, and \( w \) denotes standard Brownian motion.

The reverse-time SDE is given by:
\begin{equation}
dx = [f(x, t) - g^2(t)\nabla_x \log p_t(x \mid y)]\,dt + g(t)\,d\bar{w}_{t},
\end{equation}
where the score function \( \nabla_x \log p_t(x \mid y) \) is approximated by a neural network conditioned on the LR image \( y \).

Like SGMs, training is done via score matching, where the network learns to approximate the conditional score. Sampling can be performed using numerical solvers (e.g., Euler–Maruyama) or deterministic probability-flow ODEs, offering flexibility and stability during inference. While less common in current practice, SDE-based models provide a principled framework for incorporating uncertainty, adaptive priors, and continuous-time regularization—valuable for robust and interpretable MRI reconstruction.

\subsubsection{Denoising Diffusion Implicit Models (DDIMs)}

\begin{figure}[!th] 
\centering
\includegraphics[width=0.47\textwidth]{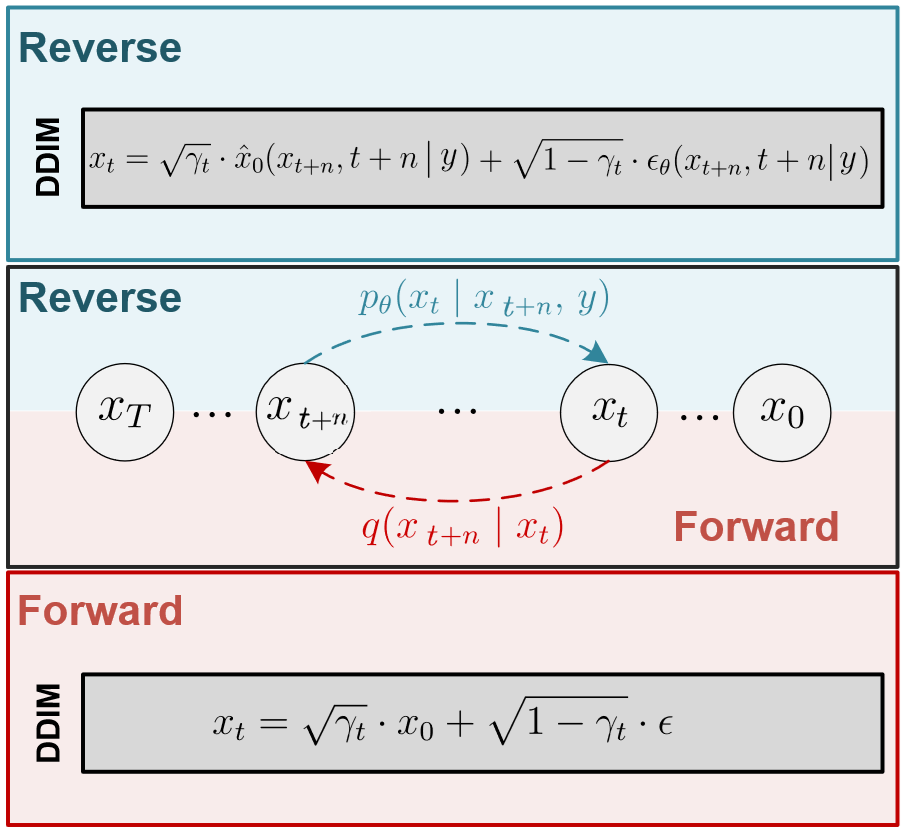}
\caption{
Illustration of DDIM sampling: instead of traversing all steps, the model maps \( x_{t+n} \rightarrow x_t \) using a learned deterministic (or semi-stochastic) function conditioned on \( y \). This enables accelerated inference by skipping intermediate states while preserving fidelity.
}
\label{fig_DDIM}
\end{figure}

DDIMs~\cite{song2020denoising} introduce a non-Markovian and often deterministic sampling strategy that preserves the marginal distributions of DDPMs under a fixed noise schedule. Unlike DDPMs, DDIMs allow \emph{step skipping}, mapping \( x_{t+n} \rightarrow x_t \) directly over a sparse subset of timesteps (skip factor \( n \)), controlled by a stochasticity parameter \( \eta \), where \( \eta = 0 \) yields a fully deterministic process (see Fig.~\ref{fig_DDIM}).

A typical update rule is:
\begin{equation}
x_{t} = \sqrt{\gamma_t} \cdot \hat{x}_0 + \sqrt{1 - \gamma_t} \cdot \epsilon_\theta(x_{t+n}, t+n),
\end{equation}
where \( \hat{x}_0 \) is the predicted clean image, and \( \gamma_t \) is the cumulative noise factor.

In MRI SR, DDIMs are widely adopted as inference-time accelerators. They are particularly effective in handling distribution shifts via diffusion prior adaptation~\cite{chung2024deep}. Examples include InverseSR~\cite{wang2023inversesr}, which performs DDIM sampling in latent space for faster 3D brain SR, and PDM~\cite{zhao2024mri}, which skips early steps using LR-derived latents. Res-SRDiff~\cite{safari2025mri} further improves efficiency via residual-based initialization, achieving high-fidelity results in just a few steps.

\subsubsection{Latent Diffusion Models (LDMs)}
LDMs~\cite{rombach2022high} perform diffusion in the latent space of a pretrained autoencoder, substantially reducing computational cost compared to pixel-space generation. A frozen autoencoder consists of an encoder $\mathcal{E}$, which maps an input image $x$ to a compact latent representation $z = \mathcal{E}(x)$, and a decoder $\mathcal{D}$, which reconstructs the image as $\tilde{x} = \mathcal{D}(z)$. Diffusion is applied only in the latent space.

In MRI SR, the latent formulation enables efficient 3D reconstruction. Wang~\etal~\cite{wang2023inversesr} introduced \emph{InverseSR}, training an LDM prior on UK Biobank brain data for high-quality 3D SR inference. Yoon~\etal~\cite{yoon2024latent} showed that latent diffusion substantially improves Alzheimer’s and mild cognitive impairment classification by preserving diagnostic features. Mármol-Rivera~\etal~\cite{marmol2025latent} extended LDM-based SR to arbitrary zoom factors, enabling flexible and consistent upscaling across resolution levels.  

Compared to pixel-space accelerations such as DDIM, which reduce inference time by skipping denoising steps, LDMs achieve efficiency at the representation level and can be combined with DDIM-style step reduction for further speed gains.

\subsubsection{Frequency-Domain Diffusion Models}
Instead of operating in the pixel or latent space, frequency-domain diffusion models apply noise corruption and denoising in the frequency space (e.g., wavelet or Fourier domain). This approach is particularly well suited for MRI SR, where high-frequency components contain critical diagnostic information.

Wu~\etal~\cite{wu2023wavelet} proposed a wavelet-domain DDPM that suppresses noise and artifacts while preserving fine structures. Similarly, Cao~\etal~\cite{cao2024high} introduced a frequency-aware SGM that processes high-frequency bands separately, improving perceptual quality and sampling efficiency. These methods exploit the structured sparsity of MRI frequency content and provide robustness against noise and aliasing.
Frequency-based models can also be combined with latent or DDIM accelerations, making them flexible for clinical deployment.

\begin{figure}[!th] 
\centering
\includegraphics[width=0.49\textwidth]{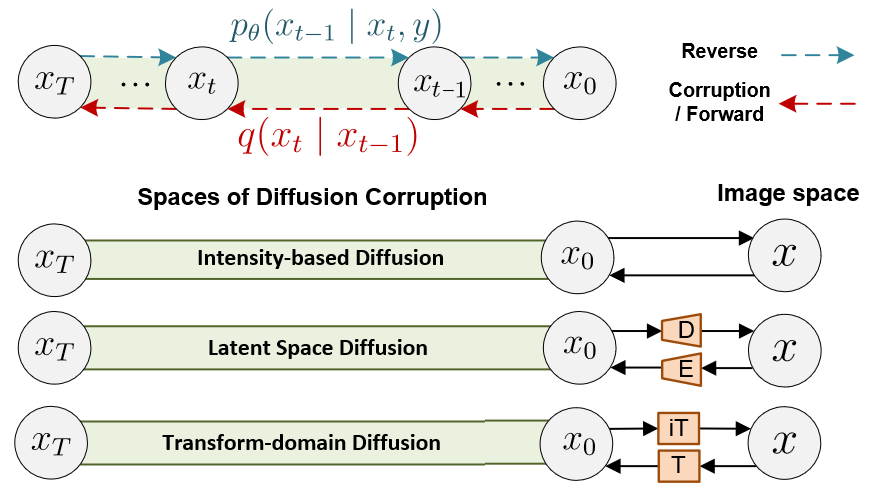}
\caption{
Overview of diffusion corruption spaces in MRI SR. 
\textbf{Top:} Forward (red) and reverse (blue) denoising steps. 
\textbf{Bottom:} Three types of domains in which the diffusion process occurs—(i) pixel-based diffusion, operating directly on image pixels; (ii) latent space diffusion, operating in a latent representation obtained via an encoder (E) and mapped back through a decoder (D); and (iii) frequency-based diffusion, operating in a transformed domain ($T$) and its inverse ($iT$), such as wavelet or Fourier. Adapted from \cite{moser2024diffusion}.
}
\label{fig_corruption_domain}
\end{figure}

\subsubsection{Mixture of Diffusion Models (MDMs)}
Mixture-of-experts (MoE) diffusion models route inputs to specialized denoisers to better capture anatomical variability and acquisition differences. In MRI SR, Wang~\etal~\cite{wang2025moediff} introduce a region-aware gating mechanism that assigns anatomy-specific diffusion experts, enhancing fine-detail preservation and robustness across contrasts and sites. This specialization mitigates single-model brittleness and homogenization but adds complexity in expert calibration and potential boundary artefacts. MDMs thus offer a scalable approach to SR by leveraging diverse priors without retraining a monolithic model.

\subsubsection{Diffusion Models for Imaging Inverse Problems via Posterior Estimation}
In many real-world inverse problems, such as super-resolution (Eq.~\ref{eq_1}), only the degraded observation \(y\) is available, and paired LR-HR datasets are not available for supervised training. Diffusion models address this challenge by serving as powerful generative priors for posterior sampling, enabling accurate image reconstruction without retraining. Specifically, they approximate the posterior \(p(x \mid y)\) by combining a denoising prior with a fidelity term, following the identity:
\begin{equation}
    \nabla_x \log p(x \mid y) = \nabla_x \log p(x) + \nabla_x \log p(y \mid x).
\end{equation}
This decomposition underlies iterative sampling schemes that alternate between a denoising step—using the learned score \(\nabla_x \log p(x)\)—and a measurement consistency step—driven by the likelihood gradient \(\nabla_x \log p(y \mid x)\). Such strategies, commonly referred to as posterior sampling, enable solving inverse problems without retraining.
 
Representative approaches include ILVR~\cite{choi2021ilvr}, which aligns low-frequency components during sampling; DDRM~\cite{kawar2022denoising}, which leverages closed-form posterior means for linear measurements; and DPS~\cite{chung2022diffusion}, which injects explicit likelihood gradients into the diffusion trajectory. These methods have demonstrated strong results across a wide range of imaging tasks. For a broader taxonomy and analysis of diffusion-based inverse solvers, refer to the survey by Daras \etal~\cite{daras2024survey}.

%%%%%%%%%%%%%%%%%%%%%%%%%%%%%%%%%
\subsection{Alternative Generative Models for Super-Resolution}

While diffusion models currently dominate generative SR, alternative paradigms—variational autoencoders (VAEs) and normalizing flows (NFs)—offer complementary trade-offs in inference speed, interpretability, and likelihood estimation.

\subsubsection{Variational Autoencoders (VAEs)}
For SR, VAEs adopt a probabilistic latent-variable framework in which an encoder $q_\phi(z\!\mid\!x,y)$ approximates the posterior distribution over latent variables $z$ given the HR target $x$ and its LR counterpart $y$. A decoder $p_\theta(x\!\mid\!z,y)$ then reconstructs the HR image from the latent representation:
\begin{equation}
    z \sim p(z), \quad x \sim p_\theta(x\mid z,y),
\end{equation}
where $p(z)$ is typically a standard Gaussian prior.  
Training maximizes the \emph{evidence lower bound} (ELBO), which balances reconstruction accuracy and latent regularization:
\begin{equation}
    \mathcal{L}_{\text{VAE}} =
    \mathbb{E}_{q_\phi(z\mid x,y)}[\log p_\theta(x\mid z,y)]
    \;-\;
    \mathrm{KL}\!\left(q_\phi(z\mid x,y)\,\|\,p(z)\right).
\end{equation}

The reconstruction term enforces HR fidelity, while the KL divergence regularizes the latent posterior toward the prior, enabling sampling and smooth interpolation~\cite{kingma2013auto}. VAEs provide fast inference and structured latent spaces, but Gaussian decoders often produce over-smoothed textures~\cite{larsen2016autoencoding,blau2018perception}. 
In MRI SR, Andrew~\etal~\cite{andrew2021super} proposed a lightweight autoencoder framework for brain MRI SR that surpasses conventional CNN-based approaches in PSNR, SSIM, and computational efficiency. Kapoor~\etal~\cite{kapoor2023multiscale} developed a multiscale VAE for 3D brain MRI, enhancing anatomical variability and perceptual realism.
Despite these advances, VAEs generally lag behind diffusion-based methods in perceptual quality and fine-detail reconstruction.

\begin{figure}[!th]
\centering
\includegraphics[width=0.49\textwidth]{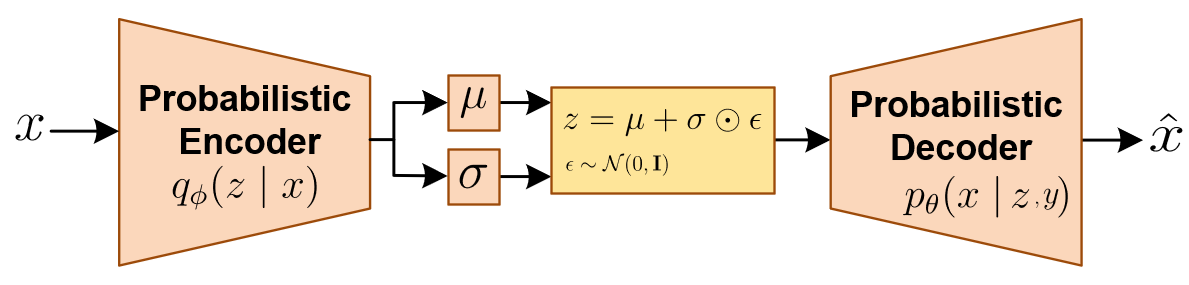}
\caption{
Illustration of the VAE-based MRI super-resolution framework. 
The encoder $q_\phi(z \mid x)$ approximates the posterior over latent variables given HR ground truth $x$ and LR input $y$, while the decoder $p_\theta(x \mid z, y)$ reconstructs the HR image.
Training maximizes the ELBO to balance fidelity and latent regularization.
\label{fig_VAE}
}
\end{figure}

\subsubsection{Normalizing Flow Models (NFMs)}
Normalizing flows transform a simple base distribution (e.g., a standard Gaussian) into a complex target distribution via a sequence of invertible mappings:
\begin{equation}
    x = f_K \circ \cdots \circ f_1(z), \quad z \sim \mathcal{N}(0, \mathbf{I}),
\end{equation}
where each $f_k$ is bijective and differentiable.  
This property allows the exact computation of the data log-likelihood using the change-of-variables formula:
\begin{equation}
    \log p(x) = \log p(z) + \sum_{k=1}^K \log \left| \det \left( \frac{\partial f_k}{\partial h_{k-1}} \right) \right|,
\end{equation}
where $h_{k-1}$ is the intermediate representation before $f_k$.  

For SR, \emph{conditional} flows model $p(x\mid y)$ by incorporating LR features $y$ into each transformation, enabling both exact likelihood estimation and the generation of diverse HR outputs by sampling different $z$. In MRI SR, Ko~\etal~\cite{ko2023mriflow} introduced MRIFLow, a conditional flow model producing anatomically consistent and diverse reconstructions, while Dong~\etal~\cite{dong2025flow} proposed a multi-scale invertible network to improve fidelity across contrasts. Despite their flexibility, NFMs often require deep architectures to be expressive, which can increase memory and computational cost.

\begin{figure}[!th] 
\label{fig_GAN}
\centering
\includegraphics[width=0.49\textwidth]{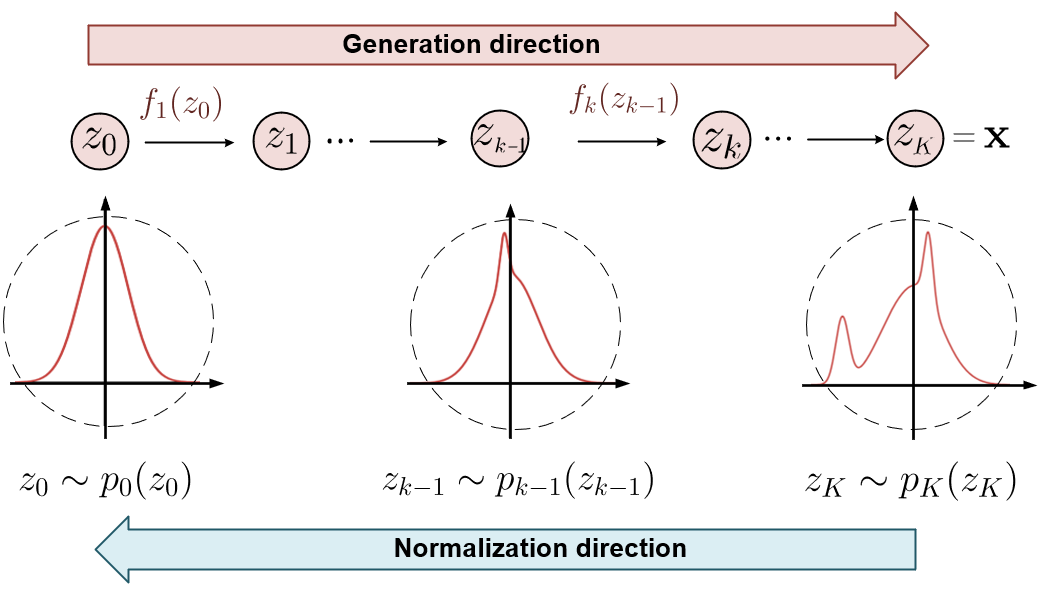}
\caption{
Illustration of a normalizing flow model. A simple base distribution \( z_0 \sim p_0(z_0) \), typically standard Gaussian, is transformed through a sequence of invertible functions \( f_1, \ldots, f_K \) into a complex target distribution \( x = z_K \). The forward (generation) path increases distribution complexity, while the reverse (normalization) path enables exact likelihood computation and inference. Adapted from~\cite{shen2023psrflow}.
}
\end{figure}

\subsubsection{Generative Adversarial Networks (GANs)}
\label{GenAI_GANs}

\noindent GANs were first introduced in Section~\ref{GANs} from a network architecture standpoint. In this section, we revisit them through the lens of generative modeling, highlighting their adversarial training dynamics and applications in MRI super-resolution.

GAN-based SR methods consist of a generator \(\mathcal{G}_{\theta}(\cdot)\) that produces HR images \( \hat{x} = \mathcal{G}_{\theta}(y) \) from LR inputs \( y \), and a discriminator \(\mathcal{D}_{\phi}(\cdot)\) that assesses whether the generated images are real or generated. The training involves a min-max game with the following objective functions:
\begin{multline}
    \min_{\mathcal{G}_\theta} \max_{\mathcal{D}_{\phi}} \mathcal{L}_{\text{adv}}(\mathcal{G}_{\theta}, \mathcal{D}_{\phi}) = \mathbb{E}_{x \sim p(x)} [ \log 
    \mathcal{D}_{\phi}(x) ] + \\ \mathbb{E}_{y \sim p(y)} [ \log (1 - \mathcal{D}_{\phi}(\mathcal{G}_{\theta}(y))) ],
\end{multline}

\noindent where \( p(x) \) denotes the distribution of real HR images, and \( p(y) \) is the distribution of LR inputs. Additionally, perceptual or content losses are often used to ensure fidelity to the ground truth. GAN-based SR methods have been extensively studied in MRI~\cite{chen2018efficient,chen2020mri,zhang2022soup} prior to the emergence of diffusion models, due to their one-pass generation (hence, fast), memory efficiency, and ability to produce perceptually realistic images.

\begin{figure}[!th] 
\centering
\includegraphics[width=0.49\textwidth]{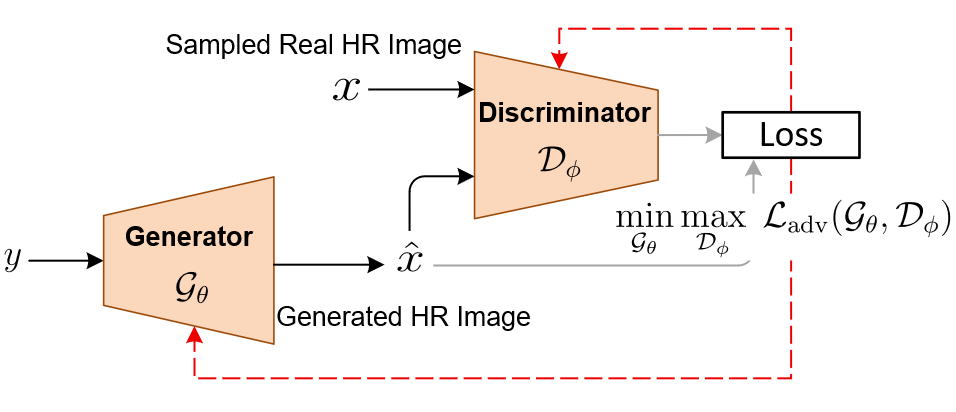}
\caption{
Overview of GAN-based super-resolution. The generator \( \mathcal{G}_\theta \) maps the LR image \( y \) to an estimated HR image \( \hat{x} \), while the discriminator \( \mathcal{D}_\phi \) distinguishes real HR images from generated ones. The training is adversarial, encouraging the generator to produce perceptually realistic outputs aligned with the HR data distribution. Adapted from \cite{zhao2023generative}.
\label{fig_GAN}
}
\end{figure}

\subsection{Challenges with GenAI}

\noindent \textbf{Hallucinations and Latent Space Limitations:} Generative models may produce anatomically implausible or unrealistic content—commonly termed “hallucinations”—that do not correspond to true tissue structures. Such artifacts pose critical concerns in clinical applications, where diagnostic reliability and anatomical accuracy are paramount.

\noindent \textbf{GANs} provide a compact and often interpretable latent space, enabling intuitive sampling and interpolation. However, they are prone to mode collapse and may inadequately cover the full data distribution. This can result in samples that appear realistic but deviate from physiological plausibility.

\noindent \textbf{VAEs} learn a smooth, probabilistic latent representation that promotes broader mode coverage compared to GANs. Yet, their reliance on pixel-wise reconstruction often leads to overly smoothed or blurry outputs, limiting their utility in tasks demanding high-frequency detail.

\noindent \textbf{DMs} avoid using compact latent embeddings. Instead, they generate images by progressively denoising high-dimensional noise over multiple steps. This iterative process enables excellent coverage of the data distribution and minimizes hallucination risk. However, because DMs do not rely on a structured latent space, their internal representations are less interpretable and offer limited direct control over the generation process—potentially posing challenges for clinical integration where transparency and fine-grained control are important.

\noindent \textbf{2D vs.\ 3D Super-Resolution:} Most generative models for SR are designed for 2D data, which simplifies computation but overlooks cross-slice dependencies. Extending these approaches to 3D SR poses challenges due to the increased memory requirements and the need to maintain spatial continuity across volumetric dimensions.

\begin{table*}[h]
\renewcommand{\arraystretch}{0.2} % Adjusts the space between rows consistently
\caption{Image Quality Assessment Metrics in Super Resolution}
\centering
\begin{tabularx}{\linewidth}{m{4.5cm} >{\raggedright\arraybackslash}m{6.5cm} m{6cm}} % Ensure all columns have the same vertical alignment, with left-aligned formulas
\toprule
\textbf{Metric} & \textbf{Formula} & \textbf{Description} \\
\midrule
\textbf{Peak Signal-to-Noise Ratio (PSNR)}
& 
$\textbf{PSNR} = 10 \cdot \log_{10}\left(\frac{{\text{MAX}^2}}{{\text{MSE}}}\right)$
& Full-reference  \newline Measures distortions (dB)  \newline \textbf{Range:} $[0, \infty)$, typically $[20, 40]$ dB \\
\midrule
\textbf{Structural Similarity Index (SSIM)}
& 
$\textbf{SSIM}=(x, \hat{x}) = \left[\textbf{C}_l(x, \hat{x})\right]^\alpha \left[\textbf{C}_c(x, \hat{x})\right]^\beta \left[\textbf{C}_s(x, \hat{x})\right]^\gamma$
& Full-reference  \newline Measures perception \newline \textbf{Range:} $[0, 1]$ \\
\midrule
\textbf{Visual Information Fidelity (VIF)}
& 
$\textbf{VIF} = \frac{\sum I(\text{Ref}; \text{Distorted}|\text{HVS})}{\sum I(\text{Ref}; \text{Ideal})}$
& Full-reference  \newline Measures perceptual information preserved \newline \textbf{Range:} $[0, 1]$ \\
\midrule
\textbf{Contrast-to-Noise Ratio (CNR)}
& 
$\textbf{CNR} = \frac{\left|\mu_{\text{ROI}} - \mu_{\text{background}}\right|}{\sigma_{\text{noise}}}$
& No-reference  \newline  Measures contrast against noise \newline \textbf{Range:} $[1, \infty)$, typically $> 1$ \\
\midrule
\textbf{Task-based Evaluation}
& 
\text{Evaluates performance in downstream tasks} 
& Full or No-reference \newline Measures perception/distortion/fidelity \newline \textbf{Range:} Variable, depending on task and metric \\
\midrule
\textbf{Mean Opinion Score (MOS)}
& 
\text{Aggregates human ratings of image quality} 
& Full or No-reference  \newline Measures perception \newline \textbf{Range:} $[1, 5]$ (poor to excellent) \\
\bottomrule
\end{tabularx}
\label{table:image_quality_metrics}
\end{table*}

\section{Performance Evaluation}

\subsection{Image Quality Assessments}
Image quality assessment (IQA) is crucial to ensure the accuracy and reliability of HR images generated by MRI SR algorithms for clinical and research applications. IQA methods are generally classified into subjective and objective approaches. Subjective methods rely on expert evaluation, closely aligning with human perception but are time-consuming and costly. In contrast, objective methods use computational algorithms to efficiently quantify distortions, reducing cost, bias, and inconsistencies from human evaluation. However, objective methods often fail to fully capture the nuanced aspects of human visual perception that are critical in clinical settings. IQA can be conducted using full-reference methods, which compare generated HR images to a reference, or no-reference methods, which assess image quality without requiring a reference. This section explores both subjective and objective IQA metrics, highlights the need for learning-based IQA approaches in MRI SR, and discusses the critical trade-offs between perceptual quality and distortion evaluations.

\subsubsection{Peak Signal-to-Noise Ratio (PSNR)} The PSNR is a full-reference IQA metric that measures the quality of a reconstructed image compared to the original reference image. It is expressed in decibels (dB) and is calculated using the following formula:
\begin{equation}
\label{eq_psnr}
\text{PSNR} = 10 \cdot \log_{10}\left(\frac{{\text{MAX}^2}}{{\text{MSE}}}\right)
\end{equation}
Where \text{MAX} is the maximum possible pixel value of the image (e.g., 255 for an 8-bit grayscale image). And \text{MSE} is the mean squared error, which represents the average of the squared differences between the original and reconstructed pixel values over all pixels in the image. Although PSNR is commonly used as a quality metric in SR, it does not fully align with human visual perception. Techniques that focus on optimizing voxel- /pixel-level correlations, which often lead to higher PSNR values, can produce outputs that are overly smooth and lack intricate details. In contrast, GANs, which are designed to enhance fine details, may result in lower PSNR scores, highlighting the discrepancy between PSNR values and human perception of image quality \cite{saharia2022image}.

\subsubsection{Structural Similarity Index (SSIM)} The SSIM is a full-reference IQA metric designed to assess the structural similarity between images, taking into account the characteristics of the human visual system \cite{wang2004image}. It evaluates images based on their luminance ($C_l(x,\hat{x})$), contrast ($C_c(x,\hat{x})$), and structural ($C_s(x,\hat{x})$) contents as follows:
\begin{subequations}\label{eq:SSIM}
\begin{equation}
C_l(x, \hat{x}) = \frac{2\mu_x\mu_{\hat{x}} + c_1}{\mu_x^2 + \mu_{\hat{x}}^2 + c_1}, \label{eq:SSIMa}
\end{equation}
\begin{equation}
C_c(x, \hat{x}) = \frac{2\sigma_x\sigma_{\hat{x}} + c_2}{\sigma_x^2 + \sigma_{\hat{x}}^2 + c_2}, \label{eq:SSIMb}
\end{equation}
\begin{equation}
C_s(x, \hat{x}) = \frac{\sigma_{x\hat{x}} + c_3}{\sigma_x\sigma_{\hat{x}} + c_3}, \label{eq:SSIMc}
\end{equation}
\end{subequations}
here, $\mu_x$ and $\sigma_x$ denote the mean and standard deviation of image $x$, and $\sigma_{x\hat{x}}$ represents the covariance between images $x$ and $\hat{x}$. $c_1$, $c_2$, and $c_3$ are small positive constants introduced for stability in the calculations. The SSIM score is defined by, 
\begin{equation}
\label{eq_ssim}
\text{SSIM}(x, \hat{x}) = [C_l(x, \hat{x})]^\alpha [C_c(x, \hat{x})]^\beta [C_s(x, \hat{x})]^\gamma,
\end{equation}
where $\alpha>0$, $\beta>0$, and $\gamma>0$ are control parameters that offer flexibility to adjust the relative importance of luminance, contrast, and structural information in the overall SSIM measurement. It is worth noting that SSIM operates on small windows, typically with dimensions like $11\times11$, and the final reported value is derived from the mean SSIM scores across these windows.

Multi-scale SSIM \cite{wang2003multiscale} was then introduced to better capture image quality across different resolutions by analyzing the image at multiple scales. This approach provides enhanced sensitivity to variations in viewing conditions, making it more robust in applications where image quality assessment needs to account for multiple levels of detail.

% \subsubsection{Information Fidelity Criterion (IFC)}  
% The IFC is a full-reference IQA metric that measures shared statistical information between reference and distorted images \cite{sheikh2005information}. It calculates mutual information across sub-bands using:
% \begin{equation}
% I(C_N; D_N \mid S_N = s_N) = \sum_{i=1}^{N} \frac{1}{2} \log_2 \left(1 + \frac{g_i^2 s_i^2 \sigma_U^2}{\sigma_V^2}\right),
% \end{equation}
% where \( g_i \) and \( s_i \) are model and scale parameters, and \( \sigma_U^2 \), \( \sigma_V^2 \) are variances of the reference and noise signals. The total IFC is obtained by summing over all sub-bands:
% \begin{equation}
% \text{IFC} = \sum_{k \in \text{subbands}} I(C_{Nk}, D_{Nk} \mid s_{Nk}),
% \end{equation}
% where \( C_{Nk} \) and \( D_{Nk} \) are sub-band coefficients. IFC ranges from 0 (no fidelity) to \( \infty \) (perfect fidelity), emphasizing fidelity over distortion.

\subsubsection{Visual Information Fidelity (VIF)}
The VIF is a full-reference IQA metric that quantifies the amount of visual information preserved in a distorted image relative to a reference, based on statistical models of natural scenes and the human visual system (HVS)~\cite{sheikh2006image}. The reference image is modeled using Gaussian scale mixtures in the wavelet domain, and distortions are assumed to result from signal attenuation and additive noise. The HVS is approximated by a set of local frequency-selective channels followed by a contrast masking mechanism.
VIF is calculated as the ratio of two mutual information terms:
\begin{equation}
\text{VIF} = \frac{\sum_{k \in \text{subbands}} I(\text{Reference}; \text{Distorted} \mid \text{HVS})}{\sum_{k \in \text{subbands}} I(\text{Reference}; \text{Ideal HVS})}
\end{equation}
Here, the numerator estimates the information that a human observer (modeled by the HVS) can extract from the distorted image. The denominator represents the maximum extractable information by an idealized human visual system with perfect perception and no distortion. VIF scores typically range from 0 to 1, where higher values indicate better perceptual quality. Compared to PSNR or SSIM, VIF correlates more strongly with subjective human evaluations and is gaining adoption in perceptual assessments of MRI SR~\cite{sun2025foundation}.

\subsubsection{Contrast-to-Noise Ratio (CNR)}
The CNR is a no-reference IQA metric that quantifies the contrast between a region of interest (ROI) and the background, relative to the level of noise in the image \cite{bernstein2004handbook}, with the following calculation:
\begin{equation}
\label{eq_CNR}
\text{CNR} = \frac{\left|\mu_{\text{ROI}} - \mu_{\text{background}}\right|}{\sigma_{\text{noise}}}
\end{equation}
where \(\mu_{\text{ROI}}\) is the mean intensity of the region of interest, \(\mu_{\text{background}}\) is the mean intensity of the background, and \(\sigma_{\text{noise}}\) is the standard deviation of the noise. A higher CNR value indicates better differentiation between the ROI and the background, which is crucial for detecting fine structures in MRI.

\subsubsection{Task-based Evaluation}
SR models are frequently assessed based on their impact on downstream tasks, particularly in medical imaging. Original and reconstructed HR images are input into pre-trained models, and reconstruction quality is inferred from changes in task performance. This is particularly relevant for applications such as tumor segmentation \cite{zhou2022super}, detection \cite{kelkar2021task}, and segmentation \cite{wu2022arbitrary}, where the effectiveness of SR models is determined by improvements in key clinical outcomes. Recent work has also demonstrated that SR can influence downstream morphometric analyses. For example, Nian et al.~\cite{nian2023toward} used a 3D-EDSR model to generate multiresolution MRI volumes and systematically evaluated how spatial resolution affects cortical thickness estimation across multiple processing pipelines.

\subsubsection{Mean Opinion Score (MOS)} The MOS is an IQA technique that relies on qualitative evaluation methods. In this approach, human raters are asked to assess the perceptual quality of images, providing scores that typically range from 1 (poor) to 5 (excellent). The mean value of these scores is used to represent the overall performance. \cite{rudie2022clinical, terada2022clinical} Although MOS provides a perceptual assessment through human judgment, it is prone to biases, non-linear perception scales, and inconsistencies in rating criteria \cite{wang2020deep}.

\subsubsection{Advancing MRI-Specific IQA} Accurate and clinically meaningful evaluation of super-resolved MRI requires domain-specific IQA techniques. This section reviews recent progress, emerging needs, and future directions for MRI-tailored IQA.

\textbf{MRI-specific Metrics:} Recent work has focused on MRI-specific no-reference IQA metrics, addressing contexts where ground truth is unavailable. For example, Stępień et al. \cite{stkepien2023no} developed a deep learning IQA model that correlates strongly with radiologists' perceptual ratings. Meanwhile, Rubert et al. \cite{rubert2021data} proposed the brain mask inlier fraction, a task-specific metric for evaluating super-resolved fetal brain MRI reconstructions. This metric demonstrated high concordance with expert ratings and enabled adaptive scan termination—facilitating more efficient and reliable clinical workflows. Clinical validation of knee MRI SR is presented in~\cite{vosshenrich2025clinical}, whereas its impact on spinal cord imaging—specifically regarding lesion detection sensitivity, inter-rater variability, and clinical utility—is assessed in~\cite{wintergerst2025measuring}. Multi-center fetal MRI SR is addressed in~\cite{sanchez2025automatic}, where a quality-aware generative model is proposed to handle acquisition variability and improve both visual and segmentation outcomes.

\textbf{Need for Learning-Based Perceptual IQA:}
In computer vision, learning-based IQA methods such as NIMA \cite{talebi2018nima} and LPIPS \cite{zhang2018unreasonable} assess perceptual quality using deep feature representations. Adapting these approaches to MRI SR may better reflect clinical relevance than traditional metrics like PSNR or SSIM. Recent work \cite{kastryulin2023image} benchmarked 35 IQA metrics using 14,700 expert scores across MRI tasks, showing that metric performance varies with network architecture and that some perceptual metrics align closely with radiologist evaluations—highlighting the importance of developing MRI-specific IQA tools.

\textbf{Assessing Synthetic MRI Images:}
Recent work \cite{deo2025metrics} highlights the limitations of commonly used no-reference IQA metrics in evaluating generative models for medical imaging. Using brain MRI data, the study shows that many metrics poorly reflect clinical relevance, particularly in detecting localized anatomical errors or assessing suitability for downstream tasks. It underscores the need for task-aware, clinically validated evaluation frameworks when assessing synthetic medical images.

\textbf{Emerging Directions: Multi-Modal Language Models} 
In computer vision, multi-modal language models have been explored for image quality assessment by integrating visual and textual information \cite{you2025depicting}, offering a promising direction for future MRI IQA development.

\subsubsection{Perception-Distortion Trade-off} 
\begin{figure}[!t]
\centering
\includegraphics[width=0.32\textwidth,height=0.25\textwidth]{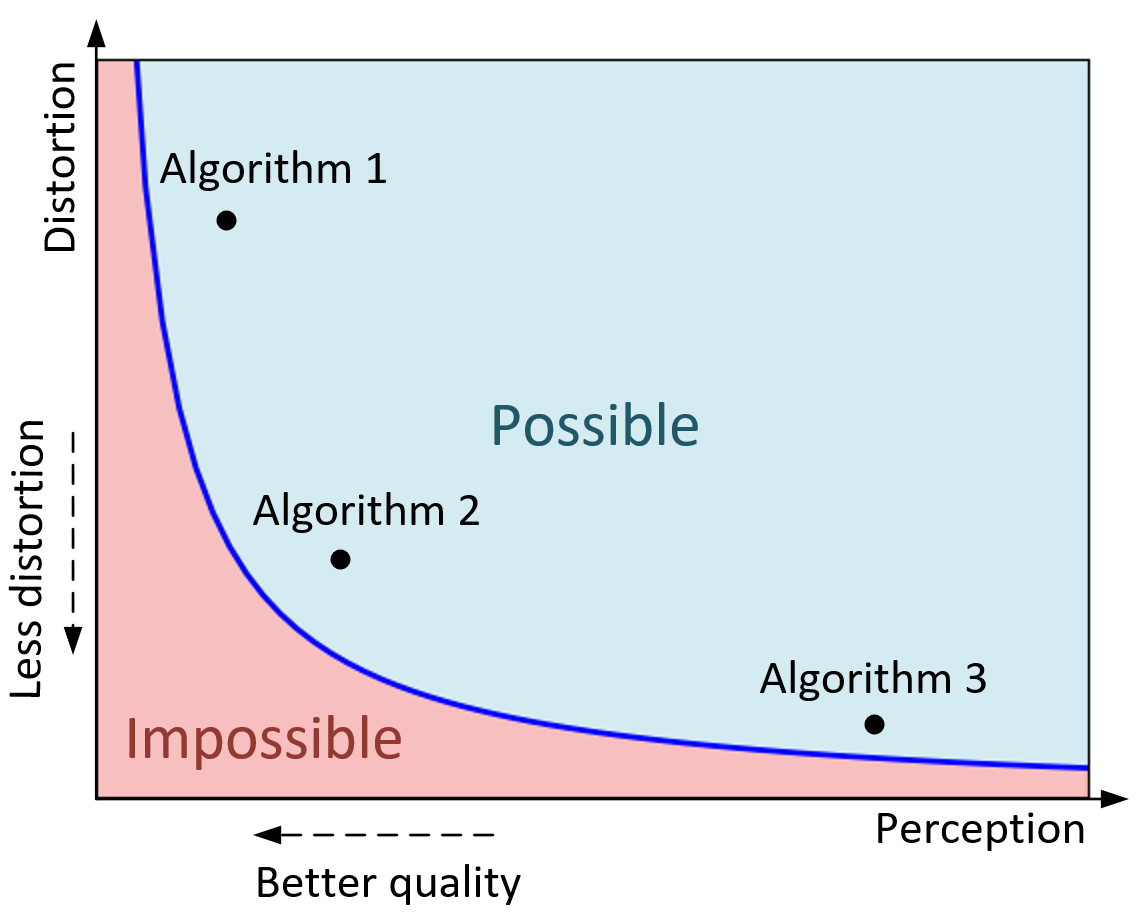}
\caption{The perception-distortion trade-off in image restoration algorithms reveals a fundamental challenge: Regardless of algorithmic differences, there is a region in the perception-distortion plane that remains unattainable. When algorithms approach this region, the choice emerges: improve either distortion or perceptual quality, not both simultaneously \cite{blau2018perception, cohen2024looks}.}
\label{perception_distortion_trade_off}
\end{figure}

Evaluating SR image quality is inherently challenging due to the trade-off between distortion-based metrics (e.g., PSNR) and perceptual metrics (e.g., expert ratings). While distortion metrics assess pixel-level fidelity, they often fail to capture perceptual realism. Conversely, optimizing for perceptual quality may increase pixel-wise errors. This fundamental trade-off implies that improving one often degrades the other (see Fig.~\ref{perception_distortion_trade_off}). Consequently, a balanced evaluation strategy—combining both metric types—is essential. In scientific and clinical contexts, unlike generic computer vision tasks, greater emphasis is often placed on distortion fidelity, though perceptual quality remains a critical consideration.

\subsection{Benchmark Datasets}

This section organizes benchmark datasets based on imaging modalities, anatomical focus (brain, body), and field strength, as summarized in Table~\ref{tab:datasets}. The imaging modalities discussed include standard MRI sequences \cite{mcrobbie2017mri}:

\begin{itemize}
    \item \textbf{Structural MRI (sMRI)}: Includes T1-weighted (T1w), T2-weighted (T2w), PD, and FLAIR sequences used to characterize brain and anatomical structures.
    \item \textbf{Proton Density-weighted (PD)}: Offers high anatomical detail by minimizing T1 and T2 contrast.
    \item \textbf{Fluid-Attenuated Inversion Recovery (FLAIR)}: Suppresses cerebrospinal fluid (CSF) signals to highlight lesions.
    \item \textbf{Diffusion-Weighted Imaging (DWI)} and \textbf{Diffusion Tensor Imaging (DTI)}: Assess microstructural tissue properties and white matter tracts.
    \item \textbf{Magnetic Resonance Angiography (MRA)}: Visualizes blood vessels using MRI.
    \item \textbf{Functional MRI (fMRI)}: Measures brain activity using blood-oxygen-level-dependent (BOLD) contrast, including resting-state and task-based protocols.
\end{itemize}

\begin{table*}[ht]
\caption{Benchmark MRI Datasets Used in Super-Resolution Studies}
\centering
\scriptsize
\renewcommand{\arraystretch}{1.2}
\setlength{\tabcolsep}{4pt}
\begin{tabular}{|l|l|c|c|l|l|}
\hline
\textbf{Dataset Name} & \textbf{Body Part} & \textbf{\# Subjects} & \textbf{Field Strength (T)} & \textbf{Purpose} & \textbf{Modality} \\ \hline
\multirow{2}{*}{fastMRI \cite{fastMRI, zbontar2018fastmri}} & Knee & 1,398 & 1.5, 3 & \multirow{2}{*}{MRI Acceleration} & sMRI (T2, PD) \\ \cline{2-4} \cline{6-6}
 & Brain & 6,970 & 1.5, 3 &  & sMRI (T1, T2, FLAIR) \\ \hline
NAMIC \cite{NAMICData} & Brain & 20+ & 1.5, 3 & Segmentation, SR & sMRI (T1, T2), dMRI, fMRI, PD, MRA \\ \hline
IXI \cite{ixiDataset} & Brain & 581 & 1.5, 3 & Brain Imaging & sMRI (T1, T2, PD), dMRI, MRA \\ \hline
HCP \cite{hcpData, van2013wu} & Brain & 1,200 & 3, 7 & Connectomics & sMRI (T1, T2, FLAIR), fMRI, dMRI \\ \hline
dHCP \cite{dhcp2017data} & Neonatal Brain & 500+ & 3 & Early Development & sMRI (T1, T2), fMRI \\ \hline
% FeTA \cite{fetaSoftware} & Fetal Brain & 80 & 1.5, 3 & Tissue Segmentation & sMRI (T2) \\ \hline
CRL \cite{gholipour2017normative} & Fetal Brain & 80 & 3 & Brain Atlas & sMRI (T2) \\ \hline
EPI \cite{luo2017vivo} & Fetal Brain & 111 & 3 & EPI Time Series & fMRI \\ \hline
BraTS \cite{bakas2018identifying, lloyd2017high, menze2014multimodal} & Brain (Tumor) & 350+ & 1.5, 3 & Tumor Segmentation & sMRI (T1, T1c, T2, FLAIR) \\ \hline
\end{tabular}
\label{tab:datasets}
\end{table*}

\subsubsection{fastMRI}
The fastMRI dataset provides fully sampled and retrospectively undersampled multi-coil k-space data for knee and brain MRI \cite{fastMRI, zbontar2018fastmri}. It includes PD and PD with fat suppression (PD-FS) sequences for the knee, as well as T1w, T2w, and FLAIR sequences for the brain. The dataset is designed to support machine learning research in MRI reconstruction, acceleration, and super-resolution.

\subsubsection{IXI} 
The IXI dataset contains nearly 600 MRI scans from healthy adult volunteers across three hospitals in London, acquired at 1.5T and 3T field strengths. Each subject includes T1w, T2w, PD, DWI, and MRA sequences \cite{ixiDataset}.

\subsubsection{HCP} 
The Human Connectome Project (HCP) provides high-resolution MRI data from approximately 1,200 healthy adults, covering both structural and functional imaging. It includes T1w, T2w, fMRI (resting-state and task-based), and DWI modalities acquired at 3T; a subset of participants also underwent 7T imaging \cite{hcpData, van2013wu}. 

\subsubsection{dHCP}  
The Developing Human Connectome Project (dHCP) provides high-resolution structural and functional MRI data of neonates. The latest release includes imaging data from 558 subjects, with T1w, T2w, and resting-state fMRI scans acquired at 3T using motion-robust protocols and a dedicated neonatal head coil \cite{dhcp2017data}.

\subsubsection{NAMIC} 
The NAMIC dataset, developed under the National Alliance for Medical Image Computing, provides multi-modal brain MRI data from over 20 healthy subjects. It includes T1w, T2w, PD, DWI, fMRI, and MRA modalities, acquired at 1.5T and 3T. The dataset supports research in segmentation, registration, and super-resolution \cite{NAMICData}.

% \subsubsection{FeTA}
% The FeTA dataset enables segmentation of fetal brain tissues from in utero T2w MRI scans. It includes manual annotations for multiple tissue classes, including cortex, ventricles, and cerebellum \cite{fetaSoftware}.

\subsubsection{CRL}
The Computational Radiology Lab (CRL) fetal brain atlas dataset provides in utero T2w MRI scans of fetuses with gestational ages ranging from 21 to 38 weeks. The data were acquired at 3T using motion-robust protocols and were used to construct a normative spatiotemporal atlas of fetal brain development \cite{gholipour2017normative}.

\subsubsection{EPI}
The EPI dataset contains fetal echo-planar imaging (EPI) time series from 111 in utero MRI sessions, spanning gestational ages from 25 to 35 weeks. It was developed to study spontaneous fetal brain activity and supports motion correction and fetal fMRI research \cite{luo2017vivo}.

\subsubsection{BraTS}
The Brain Tumor Segmentation (BraTS) dataset is curated for the development and benchmarking of glioma segmentation algorithms. It comprises multimodal MRI scans, including T1w, T1-weighted post-contrast (T1c), T2w, and FLAIR sequences, from patients with high- and low-grade gliomas. The scans are acquired at 1.5T and 3T and include expert-annotated tumor subregions \cite{bakas2018identifying, lloyd2017high, menze2014multimodal}.

\subsubsection{Other Public Repositories Suitable for MRI SR}

While the datasets in Table~\ref{tab:datasets} are commonly used when evaluating MRI SR methods, several large public repositories host MRI data that are highly suitable for SR studies but remain underutilized. These sources provide diverse scanners, protocols, and resolutions, enabling rigorous evaluation of generalization, domain shift, and reliability. Some examples include: 

\textit{OpenNeuro (formerly OpenfMRI):}
A large archive of BIDS-compliant datasets across sMRI, fMRI, and dMRI, including multi-site 3T and 7T acquisitions \cite{markiewicz2021openneuro, OpenNeuro_ds000248}.

\textit{Medical Segmentation Decathlon (MSD):}
MSD included a ten-task collection of 2,633 3D volumes across diverse anatomies/modalities. Designed for segmentation, but its MRI subsets (Brain, Hippocampus, Cardiac, Prostate) with heterogeneous protocols and anisotropic voxels make it a practical SR testbed, enabling standardized downstream evaluation (e.g., segmentation). \cite{antonelli2022medical, MSDWebsite} 

\textit{Alzheimer’s Disease Neuroimaging Initiative (ADNI):} A multi-site, longitudinal Alzheimer’s disease cohort with standardized MRI at 1.5/3T (T1w, T2w, FLAIR, dMRI, fMRI, and ASL; PET is also available). ADNI distributes raw MRI and automated derivatives (e.g., FreeSurfer). \cite{ADNI_DataPage}

\textit{The Cancer Imaging Archive (TCIA):} A large-scale repository of de-identified medical images, including numerous MRI collections spanning brain, breast, prostate, and other cancers. TCIA provides multi-institutional, multi-sequence data (e.g., T1w, T2w, DWI), often with accompanying clinical and genomic information, making it highly valuable for SR studies involving oncology and cross-domain generalization. \cite{clark2013cancer}

\section{Global Access and Clinical Applications}

\subsection{Technology Distribution, Economic Significance, and the Importance of Super-Resolution}

According to OECD statistics, there are an estimated 65{,}000~MRI scanners worldwide, corresponding to roughly 7 units per million population~\cite{liu2021low, nature2021mri}.  
Access is highly unequal: high-income countries average over 25 scanners per million people, while low- and middle-income countries have around 1 scanner per million, and approximately two-thirds of the global population has little or no access to MRI~\cite{murali2024bringing, jalloul2023mri}.  
In terms of field strength, the majority of systems in high-income countries are mid- to high-field (1.5~T and 3~T), whereas a substantial proportion of scanners in some countries (e.g., $\sim$50\% in China, $\sim$42\% in Japan) operate below 1.5~T~\cite{jalloul2023mri}. Ultra-high-field (7~T and above) systems remain confined to research centers, while ultra-low-field and portable MRI are emerging as cost-effective options for point-of-care imaging in resource-limited environments, where super-resolution techniques can help mitigate lower SNR and resolution~\cite{liu2021low, ljungberg2025characterization}.

\begin{table*}[htbp]
\caption{MRI Scanner Field Strength and Applications}
\centering
\renewcommand{\arraystretch}{1.15}
\resizebox{\textwidth}{!}{%
\begin{tabular}{@{}p{2.4cm}p{12.5cm}@{}}
\toprule
\textbf{Field Strength} & \textbf{Typical Use / Regions} \\
\midrule
Ultra-low ($<$0.1T)     & Portable and bedside imaging used in emergency rooms, rural clinics, and low-resource settings. \\
Low-field ($<$1.0T)     & Basic imaging in low- and middle-income Countries; cost-effective in infrastructure-limited regions. \\
Mid-field (1.0--1.5T)   & Standard diagnostic workhorse globally; both developing and developed countries. \\
High-field (3.0T)       & High-resolution imaging; widely adopted in high-income countries. \\
Ultra-high ($\geq$7.0T) & Used mainly for research and specialized clinical cases in high-income centers. \\
\bottomrule
\end{tabular}%
}
\label{tab:mri-distribution}
\end{table*}

\textbf{Democratized MRI:} 
Democratizing MRI through affordable portable systems, especially in low-income countries, can greatly enhance healthcare access by providing advanced imaging without the high costs and infrastructure needs of traditional MRI. Ultra-low-field MRI systems, like  0.055T, paired with SR techniques, can improve image quality by reconstructing HR images like those acquired by higher-field MRI \cite{lau2023pushing, figini2024evaluation, baljer2025ultra, baljer2025gambas}. This allows for more accurate diagnoses using cost-effective equipment, making high-quality MRI accessible in under-resourced areas and supporting equitable healthcare \cite{baljer2024multi}.

\subsection{Applications of MRI Super-Resolution}
MRI super-resolution (SR) has been applied across a broad spectrum of medical imaging domains.  
By computationally enhancing spatial resolution without increasing acquisition time, SR improves anatomical visualization, diagnostic accuracy, and quantitative analysis in both human and animal studies.  
Table~\ref{tab:mri_sr_applications} summarizes representative application areas where MRI SR contributes to improved image quality, supporting both clinical practice and research.

\setlength{\tabcolsep}{6pt}
\renewcommand{\arraystretch}{1.3} 
\begin{table*}[ht]
\centering
\caption{Representative Applications of MRI Super-Resolution Across Medical Imaging}
\label{tab:mri_sr_applications}
\begin{tabularx}{\textwidth}{@{}l p{10cm} l@{}}
\toprule
\textbf{Application Area} & \textbf{Description / Benefit} & \textbf{References} \\
\midrule

Image-guided surgery \& Interventions & Enhances anatomical detail for planning and MRI-guided therapy. & \cite{yuan2019fast, grover2024super} \\
Neurodegenerative disorders & Enables early detection of Alzheimer’s disease and cognitive decline. & \cite{huang2023mr, grigas2024positive} \\
Pediatric \& Fetal imaging & Allows faster, motion-robust scans for pediatric and fetal patients. & \cite{gholipour2010robust, tsai2023robust} \\
Robotics \& Navigation & Improves visual feedback for MRI-guided robotic procedures. & \cite{martinez2021super, wang2024stereo} \\
Point-of-Care / Emergency & Enhances portable MRI usability in resource-limited environments. & \cite{donnay2024super, islam2023improving} \\
Atlas reconstruction & Builds HR anatomical atlases for research and clinical studies. & \cite{gholipour2017normative, lee2024super} \\
Oncologic imaging & Improves tumor delineation, treatment planning, and therapy monitoring. & \cite{zhou2022super} \\
Cardiovascular imaging & Provides better visualization of myocardial structures and blood flow. & \cite{lin2020efficient, qiu2024dbsr} \\
Spinal \& Musculoskeletal imaging & Supports spinal cord and soft-tissue assessment with improved detail. & \cite{dewey2024super, chaudhari2018super} \\
Specialized modalities & Enables CEST-MRI and MPI for HR molecular imaging. & \cite{wu2023comet, gungor2022transms} \\
Functional \& Metabolic imaging & Enhances spatial resolution in fMRI and MR spectroscopy. & \cite{ota2022super} \\
Radiomics \& AI diagnostics & Improves feature extraction and ML-based classification accuracy. & \cite{iqbal2019super, hou2023deep} \\
Knee osteoarthritis & Supports early detection of cartilage degeneration. & \cite{wang2025super} \\
Enhanced segmentation & Enables self-supervised SR for resource-efficient 3D MRI segmentation. & \cite{song2025rehrseg} \\
4D flow MRI of neurofluids & Provides HR velocity fields of cerebrospinal and cerebral blood flow. & \cite{patel2025super} \\
Tissue microstructure mapping & Estimates HR tissue microstructure maps from low-quality dMRI. & \cite{qin2021multimodal} \\
Ultra-high-field fMRI & Localizes fine-scale motion-selective sites in early visual cortex. & \cite{li2024resolution} \\
Diffusion \& Perfusion Imaging & Enhances microstructural and vascular characterization in dMRI and perfusion MRI. & \cite{wu2025spatial, sun2025conditional} \\
Neuroimaging in Animal Models & Provides HR imaging for rodent and primate neuroscience research. & \cite{liang2023mouse, dzyubachyk2014interactive} \\
Tractography Validation & Improves microstructure mapping and fiber connectivity estimation. & \cite{isallari2021brain, ordinola2025super} \\
Anatomical Quantification & Enables precise morphological and volumetric analysis. & \cite{suwannasak2024deep} \\
Quantitative Imaging & Improves dynamic mapping and water quantification in MRI. & \cite{lee2024deep, thomas2024vivo} \\
% Atlas Generation & Supports detailed anatomical atlas creation for research studies. & \cite{lee2024super} \\
Bone and Kidney Imaging & Facilitates whole-body or organ-specific MRI reconstruction. & \cite{dzyubachyk2014interactive} \\
\bottomrule
\end{tabularx}
\end{table*}

\section{Theoretical and Conceptual Foundations}

\subsection{Theoretical Limits of Bandwidth Extrapolation in Super-Resolution}
Classical signal processing theory defines fundamental limits on how much high-frequency information can be recovered in SR. Youla and Webb~\cite{youla1978generalized} established the feasibility of image reconstruction as the intersection of multiple physical constraint sets using alternating orthogonal projections, providing a general geometric framework for inverse problems. Building on this, Donoho~\cite{donoho1992superresolution} derived rigorous stability bounds for bandwidth extrapolation, showing that reliable recovery of frequencies beyond the diffraction or sampling limit is fundamentally impossible without strong priors: even minimal noise causes unbounded amplification. Under sparsity assumptions, partial super-resolution becomes theoretically attainable, but stability degrades rapidly with increasing resolution factors. These foundational results underscore that classical SR is fundamentally constrained by the physics of sampling and noise. In contrast, modern learning-based methods sidestep these limitations by synthesizing perceptually plausible—though not strictly accurate—high-frequency details, reflecting the perception–distortion trade-off~\cite{blau2018perception}.

\subsection{Denoising, Deblurring, Inpainting, and Super-Resolution: A View of Inverse Imaging}

Denoising, deblurring, inpainting, and SR are classical inverse imaging problems, unified under a common goal: recovering a high-fidelity image \( x \) from a degraded observation \( y \). As introduced in Section~\ref{Section_SR_formulation}, the SR problems can be described by,
\[
y = \mathcal{H}_\delta(x) = (\Gamma(x) \circledast \kappa) \downarrow_s + n,
\]
where the degradation operator \( \mathcal{H}_\delta \) incorporates spatial transformation \( \Gamma \), blur kernel \( \kappa \), downsampling \( \downarrow_s \), and additive noise \( n \). The specific inverse problem is determined by which of these degradation components are active:

\begin{itemize}
    \item \textbf{Denoising:} 
    \[
    y = x + n \quad \text{with} \quad \Gamma = I,\ \kappa = \delta,\ s = 1
    \]
    Remove additive noise from otherwise clean images.

    \item \textbf{Deblurring:} 
    \[
    y = x \circledast \kappa + n \quad \text{with} \quad \Gamma = I,\ s = 1,\ \kappa \neq \delta
    \]
    Restore image sharpness degraded by blurring.

    \item \textbf{Inpainting:} 
    \[
    y = M \odot x + n \quad \text{with} \quad M \in \{0,1\}^{W \times H}
    \]
    Recover missing or occluded image regions using a binary mask.

    \item \textbf{Super-Resolution:} 
    \[
    y = (\Gamma(x) \circledast \kappa) \downarrow_s + n
    \]
    Recover spatial resolution from downsampled, blurred, and noisy observations with possible misalignment.
\end{itemize}

\noindent Although the degradations differ, these tasks are governed by a shared mathematical structure and are often addressed using similar strategies—such as MAP estimation, regularization, plug-and-play priors, and deep unfolding. Denoising \cite{iskender2023red, romano2017little, wu2025self, xiang2023ddm, milanfar2025denoising}, in particular, is foundational and frequently used as a module within broader frameworks (e.g., PnP \cite{venkatakrishnan2013plug}, equivariant imaging\cite{chen2023imaging}). Deblurring enhances sharpness and structural fidelity \cite{lim2020deblurring}, while inpainting can recover the missing data \cite{zhu2024advancing} or under-sampled acquisitions \cite{kang2021deep}. Among these, SR is generally the most ill-posed due to its compounded degradation effects.

\subsection{Uncertainty Quantification in Super-Resolution}

Uncertainty quantification is critical in super-resolution, particularly for medical imaging applications where reliability is essential. Bayesian approaches have been used to estimate predictive uncertainty~\cite{tanno2021uncertainty, liu2023spectral, kar2021fast}, enhancing interpretability and trust in SR outputs. GAN-based methods have also incorporated uncertainty modeling to improve robustness under ambiguous inputs~\cite{ma2024uncertainty}. Recent work on equivariant bootstrapping~\cite{tachella2023equivariant} and comprehensive reviews~\cite{abdar2021review, chen2023deep} highlight both methodological advances and open challenges in integrating uncertainty into deep learning-based SR pipelines.

\section{Technical \& Methodological Considerations}

\subsection{Challenges with Real Data and Solutions}

Real-world clinical MRI data present numerous challenges for SR models, ranging from missing or imperfect ground truth to variability in image quality and scale. These issues hinder the direct application of supervised learning and call for alternative strategies. Table~\ref{tab:realdata_challenges} summarizes key challenges and corresponding solutions provided by DL-based strategies—including reference-free training, misalignment-aware models, domain adaptation, and continuous-scale SR techniques—that have been proposed to address the limitations of real-world data in MRI SR pipelines.

\begin{table*}[!t]
\centering
\caption{Key Challenges in MRI Super-Resolution with Real-World Data and Representative Solutions}
\label{tab:realdata_challenges}
\renewcommand{\arraystretch}{1.3}
\begin{tabular}{p{3.2cm} p{5.3cm} p{6.2cm}}
\toprule
\textbf{Challenge} & \textbf{Description} & \textbf{Representative Solutions} \\
\midrule
\textbf{Absence of References} & Paired HR images are rarely available in clinical settings, making supervised training impractical. & Reference-free strategies: unsupervised learning~\cite{liu2024unsupervised, zhou2022blind, iglesias2023synthsr}, self-supervised learning~\cite{zhao2020smore, shocher2018zero, xu2021stress}, and deep plug-and-play methods~\cite{romano2017little, chen2023imaging, tamir2019unsupervised, monga2021algorithm, terris2024equivariant}. \\
\addlinespace
\textbf{Imperfect References} & Available HR images may contain noise, or scanner-induced distortions. & Training with imperfect supervision via “no-clean” reference is possible and discussed in several studies~\cite{lehtinen2018noise2noise, tachella2024unsure, khateri2024noclean}. \\
\addlinespace
\textbf{Misregistration} & Spatial misalignments between LR and HR pairs, especially in multi-contrast, longitudinal, or SVR contexts. & Robust training with misalignment-aware networks or registration preprocessing~\cite{wei2024misalignment, lei2025robust}. \\
\addlinespace
% \textbf{Limited Data Availability} & High-quality MRI datasets remain scarce across tasks and contrasts. & Transfer learning, data augmentation, domain adaptation, and self-/unsupervised or zero-shot learning~\cite{spieker2023deep}. \\
\addlinespace
\textbf{Arbitrary Scaling Factors} & Real-world LR-HR pairs often involve non-integer resolution ratios. & Meta-learning, continuous representation models, and implicit neural networks~\cite{wu2022arbitrary, li2023rethinking, tan2020arbitrary, han2024arbitrary, tu2024synergizing, zhang2023self, zhao2025arbitrary, pang2025nexpr}. \\
\bottomrule
\end{tabular}
\end{table*}

\subsection{Deviations from General Deep Learning Practices}

While DL-based SR models often follow conventional DL methodologies, certain practices—such as batch normalization and dropout—require domain-specific reconsideration.

\subsubsection{Batch Normalization (BN)} 
BN is widely used in DL to accelerate and improve stability of training by reducing internal covariate shift \cite{ioffe2015batch}. Early SR models incorporated BN for these benefits. However, subsequent research revealed that BN can impair the preservation of fine textures and introduce artifacts, particularly in pixel-level tasks like SR. As a result, many recent SR architectures have omitted BN to retain image fidelity and reduce memory usage, allowing for deeper or wider network designs \cite{lepcha2023image}.

\subsubsection{Dropout} 
Dropout is a regularization technique that mitigates overfitting by randomly deactivating neurons during training \cite{srivastava2014dropout}. However, its stochastic nature can disrupt the learning of fine spatial details, making it rarely used in SR models. Kong et al.~\cite{kong2022reflash} showed that applying dropout selectively—e.g., in the final layers—can improve performance under multi-degradation settings. In contrast, Xu et al.~\cite{xu2025adaptive} found that standard dropout degrades high-frequency detail reconstruction and proposed an adaptive variant tailored for SR tasks.

\subsection{Efficiency}
Balancing model performance with computational complexity is essential for practical applications of MRI SR. Key factors for evaluating efficiency include model size, computational cost (e.g., multiplication-addition operations), memory usage, and inference time. These aspects directly impact deployment feasibility, particularly on resource-constrained clinical systems \cite{li2109beginner, anwar2020deep}.

\subsubsection{Model Size}
Model size reflects the number of trainable parameters and determines storage and memory requirements. Larger models may offer higher performance but are often impractical for deployment on devices with limited computational resources. Designing compact and lightweight architectures is therefore critical for real-time and embedded applications.

\subsubsection{Multiplication-Additions (Mult-Adds)}
Multiplication-addition operations (Mult-Adds) serve as a fundamental metric for estimating a model's computational complexity. FLOPs (floating point operations) are often approximated as twice the number of Mult-Adds. Since inference time scales proportionally with Mult-Adds, minimizing them is essential for achieving fast and efficient performance.

\subsubsection{Memory Usage}
Memory consumption during inference affects both speed and device compatibility. High memory usage can restrict deployment on edge devices or increase latency. Techniques such as mixed-precision training—leveraging both float32 and float16 arithmetic—have shown to reduce memory footprints significantly without compromising model accuracy \cite{micikevicius2017mixed}.

\subsubsection{Inference Time}
Inference time refers to the wall-clock time required for a model to process a given input. It is influenced by model size, Mult-Adds, and memory efficiency. In latency-sensitive settings such as interactive image-guided surgery or emergency diagnostics, minimizing inference time is critical for usability and safety.

\subsection{Preprocessing}
Preprocessing is often necessary before training or applying SR to collected MRI data to improve anatomical consistency, normalize intensity statistics, mitigate common MRI artifacts, and reduce nuisance variability between images. While public datasets may come with a degree of preprocessing already applied, this is not universally true and will not be the case for newly acquired or clinical data. The exact preprocessing steps required depend on MRI acquisition details, the anatomy of interest, and the specific SR objective. Table~\ref{tab:preprocessing} lists some common operations and a non-exhaustive set of some widely-used tools.

\begin{table*}[ht]
\caption{Common Preprocessing Steps and Selected Tools for MRI}
\label{tab:preprocessing}
\small
\renewcommand{\arraystretch}{1.3}
\setlength{\tabcolsep}{6pt}
\begin{tabularx}{\linewidth}{@{}p{3cm}p{6cm}>{\footnotesize}p{7cm}@{}}
\toprule
	\textbf{Preprocessing Step} & \textbf{Purpose} & \textbf{\footnotesize Representative Tools} \\
\midrule
Image Volume Registration & Align scans across timepoints, modalities, or subjects. &
	ool{MCFLIRT}~\cite{jenkinson2002improved} (FSL), \tool{antsRegistrationSyN}~\cite{avants2011reproducible} (ANTs), \tool{Elastix}~\cite{klein2010elastix} (ITK), \tool{mni\_autoreg}~\tool{collins1994auto,collins1995auto} (MINC) \\
Slice-wise Motion Correction & Reduce intra-scan motion artifacts, partially important in fetal, neonatal, and elderly populations. &
\tool{EDDY} (FSL), \tool{SVRTK}~\cite{uus_deformable_2020}   \\
Bias Field Correction & Corrects spatially varying non-uniformities due MRI coil physics. &
\tool{N4ITK}~\cite{tustison2010n4itk} (ANTs or ITK), \tool{FAST} \cite{zhang_segmentation_2001} (FSL), \tool{nu\_correct} \cite{vincent_minc_2016} (MINC), \tool{dwibiascorrect} (MRtrix3) \\
Intensity Normalization & Standardizes image intensities. &
	ITK/simpleITK functions, intensity-normalization (python library), \tool{dwiintensitynorm} (MRtrix3), \tool{inormalize} (MINC) \\
% Skull Stripping/Brain Extraction & Removes non-brain tissue. &
% \tool{BET} (FSL)~\cite{smith2002fast}, \tool{HD-BET}~\cite{isensee2019automated}, \tool{3DSkullStrip} (AFNI), \tool{SynthStrip}~\cite{hoopes2022synthstrip}, \tool{BEasT} (MINC) \\
Susceptibility and/or Eddy Current Distortion Corrections & Mitigates geometric distortions common in EPI used for fMRI and DWI. &
	\tool{TOPUP} (FSL), \tool{EDDY} (FSL), \tool{DR-BUDDI} (AFNI), \tool{3dQwarp} (AFNI), \tool{SynB0-DISCO} \\
Denoising & Reduces thermal noise influence &
	\tool{dwidenoise} (MRtrix3), \tool{patch2self,nlmeans,localpca} (DIPY), \tool{DenoiseImage} (ANTs), \tool{SANLM} (CAT12) \\
Gibbs Unringing & Reduces discrete Fourier transform ringing artifacts. &
	\tool{mrdegibbs} (MRtrix3),  \tool{gibbs\_removal} (DIPY)\\
\bottomrule
\end{tabularx}
\end{table*}

Additionally, pipelines such as fMRIPrep~\cite{esteban2019fmriprep}, sMRIPrep~\cite{esteban_smriprep}, QSIPrep~\cite{cai2021dmriprep}, FreeSurfer's \tool{recon-all}, FSL's \tool{fsl\_anat}, AFNI's \tool{afni\_proc}, MRtrix3's \tool{dwipreproc}, MINC's \tool{bic-pipelines}, TORTOISE~\cite{pierpaoli2010tortoise} and others provide preconfigured workflows validated within their respective regimes. When used in SR studies, each step should be considered in the context of the study goal. For example, aggressive denoising or smoothing may erase details that SR is meant to reconstruct; leaving bias fields, misalignments, or distortions may hinder training or may be addressed in SR model design; segmentation steps in existing pipelines may be undesirable on LR input and improved by inserting SRR steps; some intensity normalization approaches are only applicable to the brain; some software tools may have performance issues on SR.

\subsection{Open-Source and Proprietary Software}
A curated list of open-source and proprietary tools relevant to MRI super-resolution, including implementations, datasets, and official repositories, is available in our GitHub\footnote{\scriptsize \url{https://github.com/mkhateri/Awesome-MRI-Super-Resolution}}.

%%%%%%%%%%%%%%%%%%%%%%%%%%%%%%%%%%%%%%%%%%%%%%%
\section{Critical Outlook and Future Directions}
In this survey, we reviewed DL-based MRI SR from the perspectives of computer vision, computational imaging, inverse problems, and MR physics. We analyzed key scenarios spanning end-to-end, physics-driven, and image-to-image translation paradigms; proposed a taxonomy covering supervision level, architectural backbone, and physics coupling; and consolidated advances in learning strategies, IQA, benchmark datasets, and deployment considerations. We also curated open-source resources to facilitate further research. While categorizing existing approaches provides a structured view of the field, translating MRI SR into routine clinical practice will require coordinated advances across the following fronts.

\subsection{Learning Approach}
MRI SR in real-world settings is often constrained by the lack of paired LR–HR data, absence of reliable HR ground truth, limited training samples, and distribution shifts across sites, vendors, and protocols. Progress will require greater emphasis on unsupervised and self-supervised paradigms capable of leveraging unpaired data and adapting to unseen domains. Data-efficient strategies, such as transfer learning and few-/zero-shot adaptation, offer a promising direction for addressing these challenges without the need for large paired datasets, yet they have received limited attention.

\subsection{Network Design}
Loss functions and architectural innovation remain central to SR performance, yet recent progress has been incremental. Most work has prioritized optimizing PSNR/SSIM, which may not reflect clinical utility. Future designs should balance fidelity, robustness, efficiency, and interpretability, incorporating domain-specific inductive biases and physics-aware components to improve both accuracy and reliability. Networks should be optimized not only for perceptual and quantitative quality but also for downstream clinical tasks, potentially through task-aware objectives and expert-in-the-loop feedback.

An important yet underexplored issue is the well-known regression-to-the-mean effect that arises when training with $\ell_p$-norm losses, the most widely used objective functions. In ambiguous regions where multiple plausible high-frequency structures exist, $\ell_p$ optimization favors their average, leading to overly smoothed reconstructions and the potential loss of clinically relevant fine details \cite{delbracio2023inversion}. Addressing this limitation will require new objective functions and architectural strategies that preserve structural variability. Promising directions include distributional modeling approaches that explicitly capture uncertainty \cite{delbracio2021projected}, or adaptive step-size strategies in diffusion-based frameworks \cite{delbracio2023inversion}.

\subsection{Generative AI}
Generative AI has shown superb results in 2D MRI SR. Extending these gains to 3D remains difficult due to the higher data dimensionality, large GPU memory demands, and the need to maintain spatial coherence across slices. Another major challenge is \emph{hallucinations}—synthetic but non-existent anatomical features—which threaten diagnostic reliability in clinical use. This tension between perceptual realism and structural fidelity defines a central challenge for the field~\cite{he2024one,zhu2024oftsr}.
Future progress will require methods that: (1) ensure volumetric consistency; (2) improve computational efficiency via MRI-specific latent diffusion; (3) integrate generative realism with physics-based constraints through imaging inverse problem frameworks~\cite{darassurvey, zhu2023denoising, martin2024pnp}; (4) enable domain adaptation, such as steerable or robust diffusion models~\cite{barbano2025steerable}; (5) develop reliable synthetic HR MRI pipelines for augmentation in rare-disease and data-scarce settings; and (6) leverage multi-modal generation to synthesize well-aligned complementary data (e.g., multi-contrast, multi-field strength) when direct acquisition or precise matching is not feasible.

\subsection{Foundation Models}
While progress in applying foundation models to MRI SR has been limited, their potential is considerable. Trained on large, diverse datasets, these models can capture rich anatomical and modality-aware representations, enabling robust generalization across sites, vendors, and acquisition protocols. Although their direct application may not always yield optimal results, coupling them with few- or zero-shot adaptation offers a powerful means to rapidly adapt to new data with minimal retraining.

\subsection{Interpretability}
The black-box nature of DL can hinder trust in safety-critical settings. Developing interpretable approaches aligned with MRI physics, coupled with calibrated uncertainty estimation and failure detection, can enhance transparency, mitigate risk, and support regulatory approval.

\subsection{Data}
\subsubsection{Data Acquisition}
Acquiring paired LR–HR MRI data under controlled conditions—such as breath-hold scans, artifact-minimized acquisitions, or specialized protocols—can provide high-quality references for model training. Such datasets enable learning of realistic degradation patterns that go beyond simplistic downsampling or $k$-space truncation, capturing the complex noise, motion, and system-specific effects present in real acquisitions. Incorporating diverse anatomical regions, pathologies, and scanner settings during acquisition further enhances generalization and clinical relevance.

\subsubsection{Data Curation}
Beyond simply collecting data, effective curation shapes its clinical value. Vision–language methods linked to radiology reports can spotlight disease-relevant regions, ensuring that SR models focus where it matters most. Each dataset should undergo rigorous annotation, quality control, and standardization, with detailed metadata, scanner settings, acquisition context, and pathology labels, enabling precise training, fair evaluation, and robust adaptation across diverse clinical environments.

\subsection{Benchmarking}
Despite significant progress in MRI SR, fair comparison remains challenging. Even when the same benchmark datasets are used, differences in preprocessing pipelines often lead to discrepancies in reported performance. Progress will require standardized benchmarking frameworks with high-quality, publicly available datasets and consistent evaluation protocols. Such benchmarks should span diverse anatomies, pathologies, scanners, and acquisition settings, with harmonized preprocessing and realistic, well-documented degradation models to ensure comparability and reproducibility.

\subsection{MR Physics in Forward Model}
An important aspect often overlooked in the forward degradation model in Equation \ref{eq_2} used for MRI SR is the incorporation of the MRI slice profile. The slice profile characterizes the sensitivity variation across the slice thickness due to non-ideal radiofrequency (RF) excitation pulses and slice-selective gradient fields. In practice, this profile is rarely an ideal rectangular function; imperfections in RF pulse shape, gradient non-linearities, and system bandwidth constraints lead to tapered or asymmetric sensitivity profiles that extend beyond the nominal slice thickness. These deviations cause through-plane blurring and partial volume effects, reducing effective spatial resolution and introducing anisotropy. 

To account for this, the blur component $\kappa$ in the degradation operator $H_{\delta}(\cdot)$ can be decomposed into an in-plane point spread function and a through-plane slice profile kernel, often modeled as a 1D convolution along the slice-selection axis. This modification more accurately represents the physics of volumetric MR acquisition. Incorporating the slice profile into the forward model enables SR algorithms to predict and reverse slice-profile--induced blurring, yielding reconstructions that are more physically consistent with the underlying acquisition. This is particularly critical in through-plane SR and slice-to-volume reconstruction scenarios, where ignoring slice profile effects can limit achievable resolution, alter tissue contrast, and introduce artifacts. Accurate modeling of slice sensitivity supports better network generalization across acquisition protocols and contributes to more reliable quantitative and qualitative outcomes in clinical and research settings.

\subsection{Application to Diffusion MRI (dMRI)}
SR has particular significance in dMRI, where increasing spatial resolution, angular (q-space) resolution, and coverage across multiple $b$-values leads to substantially longer scan times and greater sensitivity to motion and artifacts. While only a handful of SR studies have explicitly addressed dMRI (e.g., \cite{wu2025spatial, lyon2024spatio}), a recent review \cite{karimi2024diffusion} highlights growing efforts in image enhancement, interpolation, and super-resolution of dMRI.  

Future directions for dMRI SR should explicitly target the preservation of diffusion-specific information. Promising avenues include: (1) physics-aware models that respect the mathematical constraints of the diffusion signal, account for noise statistics, and model the effects of motion and eddy currents; (2) task-driven objectives that optimize SR reconstructions not only for visual fidelity but also for downstream tasks such as tractography, connectomics, and microstructural biomarker estimation; and (3) cross-domain integration, leveraging structural MRI, anatomical priors, or biophysical models (e.g., tensor or multi-compartment models) to constrain and regularize SR.

\subsection{Current and Future Clinical applications}
\subsubsection{In-Plane Resolution Enhancement}
All major MRI vendors now offer DL-based SR solutions that enable higher resolution imaging from standard acquisitions without increasing scan time. These techniques use neural networks trained on paired low and high resolution datasets to synthesize fine anatomical details, effectively doubling in plane matrix sizes while preserving edge sharpness and tissue contrast. Raw data consistency checks ensure diagnostic reliability by preventing artifactual hallucinations. The technology is clinically implemented for neuroimaging applications including small vessel visualization and musculoskeletal MRI where submillimeter resolution is critical.

\subsubsection{Fetal MRI Super Resolution}
Fetal MRI presents unique challenges because random fetal movement prevents reliable generation of standard axial, coronal and sagittal views during acquisition. SR reconstruction addresses this limitation by computationally combining motion-corrupted 2D slices from multiple orientations to create isotropic 3D volumetric datasets. This AI-driven technique is now clinically established for visualizing critical neurodevelopmental features, including cortical folding patterns, brainstem anatomy, and subtle structural abnormalities. The resulting isotropic resolution enables radiologists to reconstruct diagnostically useful images in any plane while maintaining high spatial resolution, overcoming a fundamental constraint of conventional fetal MRI protocols.

\subsubsection{Portable and Low-Field MRI} Portable MRI systems are transforming healthcare by enabling accessible, cost-effective imaging in diverse settings, from rural clinics to emergency rooms. They facilitate rapid diagnosis and treatment decisions without the infrastructure demands of high-field scanners, expanding access to advanced imaging in underserved areas. Super-resolution techniques can substantially enhance the clinical utility of portable MRI by compensating for the lower SNR and spatial resolution of low-field systems, enabling diagnostically accurate, high-resolution reconstructions \cite{iglesias2022quantitative, islam2023improving, man2023deep, li2024fast, baljer2025ultra}. 
Several portable and low-field MRI systems have already received FDA clearance for clinical use, underscoring their growing translational impact. Recent reviews further highlight the evolving regulatory landscape and clinical potential of these devices \cite{sheth2021portable, cooley2023lowfield, arnold2023lowfield}.

\subsubsection{Future Directions}
Future advancements may enable multi-contrast SR from single acquisitions and reduce reliance on motion correction through real-time adaptive reconstruction. The integration of federated learning could expand anatomical diversity in training datasets, while open innovation platforms may accelerate customization for niche applications, including fetal cardiac MRI and oncology. The convergence of higher resolution outputs with faster scan times promises to redefine diagnostic thresholds in MRI.

\subsection{Competitions}
Despite the success of SR competitions in computer vision—such as NTIRE~\cite{agustsson2017ntire, ren2024ninth} and PIRM~\cite{ignatov2018pirm}—no dedicated platform exists for MRI SR. This absence limits standardized evaluation, reproducibility, and coordinated technical progress in a domain with distinct clinical requirements.

A community-driven MRI SR competition could be organized into specialized tracks, each addressing a key bottleneck:

\begin{itemize}
\item \textbf{Method Development:} Advancing algorithms purpose-built for MRI-specific challenges, covering diverse SR scenarios outlined in Section.~\ref{sec: MRISR_scenarios}.

\item \textbf{Image Quality Assessment:} Innovating evaluation methods beyond PSNR/SSIM to capture perceptual and clinical relevance, particularly in settings without HR ground truth.

\item \textbf{Benchmark Datasets:} Curating and standardizing diverse, clinically representative datasets spanning anatomy, pathology, vendor, and field-strength variations to enable fair and reproducible comparisons.

\item \textbf{Preprocessing Standards:} Establishing harmonized preprocessing pipelines, tailored to each anatomy and dataset type, covering spatial registration, motion correction, and intensity normalization to ensure comparability across methods.

\end{itemize}

\noindent Running these tracks in parallel would enable progress in algorithms, evaluation, and datasets while collectively advancing MRI SR toward clinical translation. Such competitions could also yield lasting resources—reference implementations, standardized pipelines, and reproducible baselines—lowering barriers to entry and accelerating adoption in practice.

\subsection{Clinical Validation}
Most current MRI SR studies rely on PSNR and SSIM, which fail to capture modality-specific nuances or the true clinical utility of reconstructed images. Future validation should incorporate MRI-specific and task-relevant metrics, expert radiologist assessments, multimodal evaluation using vision--language models, and performance on downstream clinical tasks. For generative AI outputs, it is necessary to implement safeguards that can detect hallucinations, anatomical inconsistencies, and clinically irrelevant details, ideally by using hybrid pipelines that combine automated screening with expert review.

Most current SR algorithms are primarily evaluated on synthetic data created by downsampling high-resolution images, with limited testing on real data. For a meaningful assessment of their performance, it is essential to evaluate these algorithms on real-world data—and ideally on clinical data that includes relevant pathologies.

Clinical validation of MRI SR methods should adopt a rigorous and multi-layered strategy that addresses both technical performance and clinical applicability. Robust protocols can include multi-center reader studies where trained radiologists evaluate SR outputs across representative patient populations, imaging protocols, and scanner types. These studies should be blinded, include inter-rater and intra-rater agreement analyses, and assess image fidelity for clinically important endpoints such as accurate delineation of anatomical structures, lesion conspicuity, and diagnostic confidence.

Evaluation should also be carried out in downstream AI-enabled diagnostic tasks, including tumor segmentation, brain volumetry, and radiomics-based prediction, to quantify the real-world impact of SR models on clinical decision-making. Coupling image quality metrics with such task-based endpoints ensures that improvements in visual appearance correspond to measurable diagnostic benefits.

For models that may hallucinate or generate non-physical structures, validation frameworks should integrate automated artifact and hallucination detection methods, for example, structural consistency mapping or physics-informed plausibility checks, followed by targeted expert review of flagged cases. These safety measures are essential for regulatory approval and clinical trust.

The development of standardized and publicly accessible clinical validation benchmarks is a critical next step. Such resources should provide reference datasets with paired LR-HR MRI, expert annotations, and well-defined evaluation protocols to ensure reproducibility, fairness in comparison, and continuous improvement of SR methods.

\bibliographystyle{ieeetr}
\bibliography{main.bib,for_srrR.bib}

\end{document}